%% file: noInter_final.tex
\documentclass[longauth,traditabstract]{aa} 
\usepackage{graphicx}
\usepackage{lscape}
\usepackage{booktabs}
\usepackage{longtable}
\usepackage{multicol}
\usepackage{multirow}
\usepackage{txfonts}
\usepackage{natbib}
\usepackage{subfigure}
\usepackage{color}
\usepackage{url}
\usepackage[pdfpagemode=UseThumbs,colorlinks=true,bookmarks=true]{hyperref}
\hypersetup{citecolor={blue},linkcolor={blue},urlcolor={blue}}
\bibpunct{(}{)}{;}{a}{ }{,}

\begin{document} 
    \title{Kinematic alignment of non-interacting CALIFA galaxies}
    \subtitle{Quantifying the impact of bars on stellar and ionised gas velocity field orientations}

   \author{J.~K.~Barrera-Ballesteros
          \inst{1,2}\fnmsep\thanks{\email{jkbb@iac.es}}
          \and
          J.~Falc\'{o}n-Barroso\inst{1,2}
      	  \and
          B.~Garc\'{\i}a-Lorenzo\inst{1,2}
	  \and
	  G.~van~de~Ven\inst{3} 
          \and
          J.~A.~L.~Aguerri\inst{1,2}
          \and
          J.~Mendez-Abreu\inst{1,2,4}
          \and
          K.~Spekkens\inst{5}
          \and
          S.~F.~S\'{a}nchez\inst{6,7,8} 
          \and
          B.~Husemann\inst{9,10}
          \and
          D.~Mast\inst{6,8} 
          \and
          R.~Garc\'{\i}a-Benito\inst{6}
          \and 
          J.~Iglesias-Paramo\inst{6,8} 
          \and
          A.~del~Olmo\inst{6}
          \and
          I.~M\'{a}rquez\inst{6}
          \and
          J.~Masegosa\inst{6}
          \and
          C.~Kehrig\inst{6}
          \and
          R.~A.~Marino\inst{11}
          \and
          L.~Verdes-Montenegro\inst{6}
          \and
          B.~Ziegler\inst{12}
          \and
          D.~H.~MacIntosh\inst{13}
          \and 
          J.~Bland-Hawthorn\inst{14}
          \and
          C.~J.~Walcher\inst{10}
          \and
          the CALIFA Collaboration
          }

   \institute{Instituto de Astrof\'{i}sica de Canarias, Calle V\'{i}a Lact\'{e}a s/n, E-38205, Spain
         \and
         Departamento de Astrof\'{i}sica, Universidad de La Laguna (ULL), E-38200 La Laguna, Tenerife, Spain
         \and
          Max-Planck Institute for Astronomy, K\"{o}nigstul 17, 69117 Heidelberg, Germany    
         \and
         School of Physics and Astronomy, University of St Andrews, North Haugh, St Andrews, KY16 9SS, UK
         \and
	  Department of Physics, Royal Military Collage of Canada, PO Box 17000, Station Forces, Kingston, Ontario, K7K7B4, Canada
	 \and
          Instituto de Astrof\'{i}sica de  Andaluc\'{i}a (CSIC), Glorieta de la Astronom\'{i}a S/N, 18008 Granada, Spain
          	 \and
	 Instituto de Astronom\'{\i}a, Universidad Nacional Auton\'{o}ma de Mexico, A.P. 70-264, 04510, M\'exico,D.F.
         \and
	  Centro Astron\'{o}mico Hispano Alem\'{a}n de Calar Alto (CSIC-MPG), C/ Jes\'{u}s Durb\'{a}n Rem\'{o}n 2-2, 4004 Almer\'{\i}a, Spain.
	  \and
	 European Southern Observatory, ESO Headquarters, Karl-Schwarzschild-Str. 2, 85748 Garching b. M\"{u}nchen, Germany
	 \and
	 Leibniz-Institut f\"{u}r Astrophysik Potsdam (AIP), An der Sternwarte 16, 14482, Potsdam, Germany
	 \and 
	 CEI Campus Moncloa, UCM-UPM, Departamento de Astrof\'{\i}sica y CC. de la Atm\'{o}sfera, Facultad de CC. F\'{\i}sicas, Universidad Complutense de Madrid, Avda. Complutense S/N, 28040 Madrid, Spain
	 \and
	 University of Vienna, Departament of Astrophysics, T\"{u}rkenschanzstr. 17, 1180 Vienna, Austria
	 \and
	 Department of Physics \& Astronomy, University of Missouri-Kansas City, 5110 Rockhill Road, Kansas City, MO, USA
	 \and
	 Sydney Institute for Astronomy, School of Physics A28, University of Sydney, NSW 2006, Australia  }

\abstract{We present 80 stellar and ionised gas velocity maps from the Calar
Alto Legacy Integral Field Area (CALIFA) survey in order to characterize the kinematic orientation of non-interacting galaxies. The study of galaxies in isolation is a key step towards understanding how fast-external processes, such as major mergers, affect kinematic properties in galaxies. We derived the global and individual (projected approaching and receding sides) kinematic position angles (PAs) for both the stellar and ionised gas line-of-sight velocity distributions. When compared to the photometric PA, we find that morpho-kinematic differences are smaller than 22 degrees in 90 \%  of the sample for both components; internal kinematic misalignments are generally smaller than 16 degrees. We find a tight relation between the global stellar and ionised gas kinematic PA consistent with circular-flow pattern motions in both components. This relation also holds generally  in barred galaxies across the bar and galaxy disk scales. Our findings suggest that even in the presence of strong bars, both the stellar and the gaseous components tend to follow the gravitational potential of the disk. As a result, kinematic orientation can be used to assess the degree of external distortions in interacting galaxies.}

\maketitle

\section{Introduction}

To first order, galaxies are rotationally-supported disk systems
\cite[for a historical review see e.g.,][]{2001ARA&A..39..137S}. The
observed line-of-sight velocity distributions are expected to show a regular
rotation pattern in the form of the so-called ``spider diagram'', with the
position angle (PA) of the minimum and maximum aligned with the optical major
axis. However, several physical processes can disturb this regular behaviour. On
the one hand, kinematic perturbations can be induced by external factors such as
ram-pressure stripping \citep[e.g.,][]{2008A&A...483..783K} or galactic mergers
\citep[e.g.,][]{2003ApJ...597..893N}. On the other hand,
kinematic departures from regular rotation can be due to internal instabilities
such as spiral arms or bars  \citep[][]{2008gady.book.....B,
2013pss5.book..923S}. To disentangle the perturbations due to external processes
it is necessary first to study in detail the possible kinematic perturbations
due to internal processes in both stars and gas.

Integral-field spectroscopy (IFS) is a unique tool to study the kinematic
properties across the entire galaxies. Even though 3D spectroscopic surveys are now feasible for a large set of galaxies  \citep[e.g.: GHASP, GH$\alpha$FaS, ATLAS$^{\mathrm{3D}}$,  SAMI; ][]{2008MNRAS.388..500E, 2008PASP..120..665H, 2011MNRAS.413..813C, 2012MNRAS.421..872C}, these surveys have usually focused on the study of the kinematics of  early-type systems \citep[e.g.: SAURON, ATLAS$^{\mathrm{3D}}$; ][]{2004MNRAS.352..721E, 2011MNRAS.414.2923K} or on specific wavelength range like H$\alpha$ range in spiral galaxies in different environments \citep[e.g., GHASP, and GH$\alpha$FaS][]{2009ApJ...704.1657F, 2008MNRAS.388..500E}.

The CALIFA survey \citep{2012A&A...538A...8S} provides an unique opportunity to
study simultaneously the stellar and ionised gas global kinematics in a sample
of galaxies that covers a wide range of morphological types, stellar masses and
environments. The spectral and spatial coverage provided by the CALIFA IFS data
allow us to study the effect of structures such as bars in non-interacting
galaxies on the global kinematics. Bars are frequent elliptical-like structures
present in disk galaxies. About 40$-$50 \% of nearby disk galaxies
observed in optical wavelengths show bars features
\citep[e.g.,][]{2007AAS...211.9712M,2008ApJ...675.1194B,2009A&A...495..491A}.
This fraction is even higher (about 60$-$70 \%) for near-infrared
observations \citep[e.g.,][]{2000AJ....119..536E, 2000ApJ...529...93K,
2007ApJ...657..790M}. Hydrodynamical simulations of these systems show that the
bar component contributes only between 10 and 20 \% of the disk potential
\citep[e.g., ][]{2010MNRAS.407L..41S}, and yet it can produce important changes
in the gas dynamics. Bars are thought to be very efficient in redistributing
angular momentum, energy and mass in both luminous and dark matter components
\citep[e.g.,][]{1985MNRAS.213..451W, 1998ApJ...493L...5D,
2000ApJ...543..704D,2003MNRAS.341.1179A,2006ApJ...637..214M,
2006ApJ...637..567S, 2006ApJ...639..868S,2009ApJ...707..218V}. The effects of
this angular momentum redistribution is different in gas and stars due to their
different properties \citep[e.g.,][]{1982MNRAS.199..151T}. For example, flows
are observed within the co-rotation radius towards the galaxy centre only in the
gas component \citep[e.g.,][]{2002A&A...386...42R}.

The aim of this paper is to characterize the stellar and the ionised gas kinematics for a sample of non-interacting galaxies by means of their global orientation. In a companion paper (Barrera-Ballesteros et al., in preparation) we will use the results of the present study to quantify how much the kinematic of a sample of interacting galaxies differentiates with respect to the kinematics of isolated ones. In other words, the result from this work will yield a yardstick describing kinematic distortions due to internal processes with which to compare to kinematic disturbances induced by major merger events. 

The paper is organized as follows. In Section 2, we present the sample as well as the methods for extracting the stellar and ionised gas kinematics. The procedure used to derive directly the kinematic PA from the velocity maps is presented in Section 3. Section 4 presents the kinematic characterization of the non-interacting galaxies: morpho-kinematic and kinematic misalignments are compared against different bar strengths as well as at different radii of barred galaxies. The impact of bars in the velocity field is discussed in Section 5. Our findings and main conclusions are summarized in Section 6.

\section{Sample and observations}

\subsection{Sample selection}
\label{sec:sample}

The sample presented in this work is drawn from the CALIFA mother sample
\citep[][]{2012A&A...538A...8S} observed until November 2013. This mother sample includes 939 galaxies, selected from the SDSS DR7 \citep{2009ApJS..182..543A}. The selection criteria is such that galaxies included are in the nearby Universe with redshifts  0.005\,$<z<$\,0.03 and their isophotal diameter in the SDSS $r$-band is 
45$^{\prime\prime}$\,$\lesssim D_{25}\lesssim$\,80$^{\prime\prime}$.

Using the SDSS $r$-band images of the observed galaxies we select 80
galaxies without evident signatures of interaction such as tidal
tails, bridges, rings, shells or any other morphological distortion caused by
a merging. Moreover, we consider galaxies isolated from companions
within a physical radius of 250\,kpc, a systemic velocity difference
smaller than 1000\,km\,s$^{-1}$ and a difference in magnitude in the SDSS
r-band images larger than 2\,mag (position and systemic velocities were taken from NED\footnote{NASA/IPAC Extragalactic database. http://ned.ipac.caltech.edu/} ). We use a conservative physical radius for rejecting close companions with respect to pairs-survey criteria \citep[e.g.,][]{2008AJ....135.1877E}. The systemic velocity selection criteria is supported by recent cosmological simulations \citep{2013MNRAS.tmp.2336M}. Finally, we also selected those galaxies for which stellar and ionised gas signal-to-noise (S/N) ratios allow us a reliable estimation of their kinematic properties (see section \ref{sec:vel_fields}).

It is important to note that these selection criteria does not exclude from our sample galaxies
with minor companions or galaxies located in the outskirts of groups. However, we estimate the perturbation caused by tidal forces induced by companions by means of the f-value \citep{2004A&A...420..873V}. Values larger than -2 are required to produce sizeable effects on a disk galaxy. We were able to measure in 71 objects the f-value. We found that 95\% of these galaxies have f-values smaller than -2 with a mean value of -4.0. This suggests that our sample is a fairly
representation of isolated galaxies. In fact, 10 galaxies from our
sample are also part of the AMIGA sample \citep[Analysis of the interstellar
Medium of Isolated GAlaxies, ][ CALIFA id: 2, 30, 131, 152, 275, 743, 748, 777, 779, 856]{2005A&A...436..443V}. AMIGA sample is defined based on strict isolation criteria, and shows different physical properties at all wavelengths studied so far than galaxies in denser environments (even field galaxies). These 10 galaxies
are located in the region of lower tidal forces and number density of AMIGA full sample (Fig. 6 in \cite{2007A&A...472..121V}).

Our non-interacting sample includes different morphological types, bar strengths and a relatively wide range of  luminosities (-24\,$\lesssim$\,$M_r$\,$\lesssim$\,$-19$; where $M_r$ is the absolute magnitude in SDSS $r$-band). Morphological types and bar strengths were obtained by visual inspection of the SDSS $r$-band images by different members of the collaboration (Walcher et al., in preparation, see also table \ref{table_morph}). Even more, the sample is dominated by Sbs and Sc morphological types, which are also the dominant population of the AMIGA sample,
hence characteristic of isolated galaxies. Finally, from a visual inspection of the velocity maps we exclude from our sample the early-type galaxy NGC\,5623. This galaxy presents a stellar counter-rotating disk, presumably originated from an early encounter, and thus it will be included in the sample of interacting galaxies of Barrera-Ballesteros et al. (in prep.).

\begin{figure*}[!htb]
\begin{centering}
\includegraphics[width=\linewidth,angle=0,clip=true]{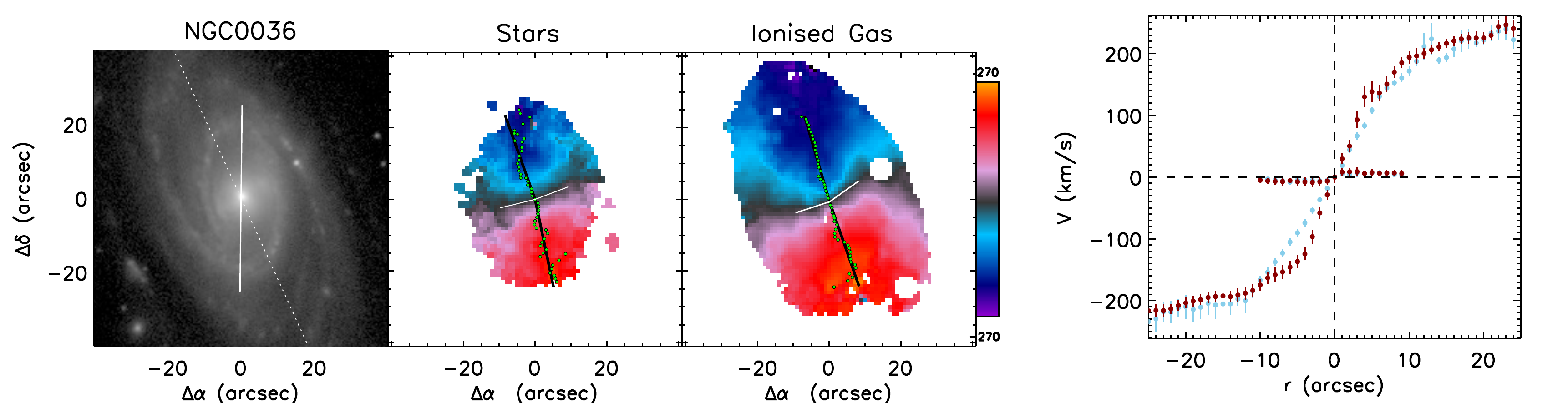}
\caption{\label{kin_example}Example of the method used to determine projected kinematic position angles (PAs) from the velocity maps of the galaxy \object{NGC\,36}. Left most panel shows the SDSS $r$-band image of the galaxy. White solid line represents the photometric PA measured at the same galactocentric distance as the kinematic PA ($\mathrm{PA_{morph}}$, measured from the $r$-band image; see section \ref{sec:Robust_Kinematic}). Dashed line represents the photometric PA measured at the outermost isophotes of the image (PA$\mathrm{_{morph}^{out}}$, see section \ref{sec:Impact}). Middle-left and middle-right panels show the stellar and ionised gas velocity maps, respectively. Green points highlight the locations where the maximum velocity is located at given radius, determined from the position-velocity diagram. Black lines for each kinematic side represent the average kinematic PA (PA$_{\mathrm{kin}}$), while black thin lines along the zero-velocity curve show the average minor kinematic PA. Right panel shows the distance from the galactic centre versus the maximum for the stellar (blue points) and ionised gas (red points) components. The curve along  0\,km\,s$^{-1}$ represent the velocities along the zero-velocity curve.  The error bars represent the uncertainty in velocity determined from Monte Carlo simulations.}
\end{centering}
\end{figure*}

\subsection{CALIFA velocity maps}
\label{sec:vel_fields}
Observations were carried out using the PPAK instrument at Calar Alto
Observatory \citep{2005PASP..117..620R}. Its main component consists of 331
fibers of 2\farcs7 diameter each, concentrated in a single hexagon bundle
covering a field-of-view of $74\arcsec\times64\arcsec$, with a filling factor of
$\sim$ 60\%\,. Three dithered pointings are taken for each object to reach a filling factor of 100\% across the entire FoV. The final data cube consists of more than 4000 spectra at a sampling of $1\arcsec\times1\arcsec$ per spaxel \cite[see details in][]{Husemann_12}. The CALIFA survey has two spectral setups, V500 and V1200, low and intermediate resolution, respectively. Objects included in this work have been observed in the V500 setup. This setup has a nominal resolution of $\lambda$/$\Delta\lambda$\,$\sim$\,850 at $\sim$5000\AA\ and its nominal wavelength range is 3745$-$7300\AA.However, the final data cube has an homogenized spectral resolution (FWHM) over the entire wavelength range of 6.0\AA\ and the wavelength sampling per spaxel is 2.0\AA. The total exposure time per pointing is fixed for all the observed objects to 45 min. The data reduction is performed by a pipeline designed specifically for the CALIFA survey. Detailed reduction process is explained by \cite{2012A&A...538A...8S} and improvements on this pipeline are presented by \cite{Husemann_12}. 

The stellar kinematic extraction method for the CALIFA survey will be presented in Falc\'{o}n-Barroso et al. (in prep.). Here we highlight the main steps. Spaxels with continuum S/N\,$<$\,3 in the original V500 cube were considered unreliable and therefore not considered in further analysis. To achieve a minimum S/N of 20, we used a Voronoi-binning scheme for optical IFS data implemented by \cite{2003MNRAS.342..345C}. We will refer to these Voronoi bins as ``voxels''. From this selection in the continuum S/N, a large fraction of the voxels ($\sim$80\%) have the same size as the spaxels of the cube. The remaining fraction of the voxels are located in the outer regions of the galaxies and include a range between three or five spaxels. To derive the line-of-sight velocity maps, for each cube, a non-linear combination of a subset of stellar templates from the Indo-U.S. library \citep{2004ApJS..152..251V} is fit to each voxel using the penalized pixel-fitting method \citep[pPXF,][]{2004PASP..116..138C}.  Errors for each voxel, determined via Monte Carlo simulations, range from 5 to 20\,km\,s$^{-1}$ for inner to outer voxels, respectively. Note that these uncertainties are smaller than the spectral sampling per spaxels from the data cubes. Since each spaxel offers a wide range of absorption lines to fit its line-of-sight velocity, the spectral location of these absorption features can be determine even at scales smaller than the spectral sampling.  Figure\,\ref{kin_example} (middle-left panel) shows an example of a stellar kinematic maps for one of our galaxies

The detailed description of the ionised-gas kinematic extraction method for the CALIFA survey is provided by Garc\'{\i}a-Lorenzo et al. (2014, submitted). Briefly, to obtain the ionised-gas emission in each spaxel we subtracted the stellar continuum spectra derived from the best stellar pPXF fit in its corresponding voxel. No binning was done for the ionized gas.  We assumed that the stellar populations in each voxel are rather smooth and do not change from spaxel to spaxel within each voxel. A cross-correlation (CC) method is used to measure the velocity of the ionised-gas \citep[see ][for details on the method]{2013MNRAS.tmp..534G}. The method compares the spectrum in each spaxel in a given wavelength range  with a template that includes the H$\alpha$+[NII]$\lambda\lambda$6548,6584 emission lines (6508$-$6623\AA). The template corresponds to a Gaussian model for each emission line in the given wavelength range, shifted to the systemic velocity reported in NED and assuming a velocity dispersion equals to the instrumental resolution ($\sigma$ $\sim$ 90\,km\,s$^{-1}$ at the H$\alpha$ emission line). We selected spaxels with S/N \,$>$\, 8  in this emission line for the ionised gas velocity maps. Garc\'{\i}a-Lorenzo et al. (2014, submitted) present examples on the kinematic fitting per spaxel and the extraction of the ionized gas component. Fig.~\ref{kin_example} (middle-right panel) shows an example of an ionised-gas velocity maps. Estimated uncertainties in the location of the maximum of the CC function are $\sim$10\,km\,s$^{-1}$. Both the pPXF and the CC methods are able to determine velocity
dispersion maps  however, due to the low spectra resolution of the V500 data we
did not attempt any further analysis of these maps. They are also not required for the work presented here.

\section{Kinematic position angles from velocity maps}
\label{sec:Robust_Kinematic}

Appendix~\ref{sec:maps} shows the complete set of stellar and ionised gas
velocity maps for the galaxies in this study. Some of them are also presented 
in Falc\'{o}n-Barroso et al. (in prep., for stellar kinematics in the V1200 setup) and
Garc\'{\i}a-Lorenzo et al. (2014, submitted for the ionised gas).

In the last years, several analytic methods have been developed to determine
kinematic properties from two-dimensional velocity distributions. These models
consider symmetric velocity distributions, either by assuming a thin disk
geometry \citep[e.g., ][]{2003AJ....125.1164B, 2008MNRAS.388..500E},
pure-circular motion on elliptical galaxies \citep{2006MNRAS.366..787K} or
symmetrical radial distortions with respect to some axis
\citep{2007ApJ...664..204S}. Even though these methods
yield good results for symmetric velocity maps, they may provide (depending on
its use) biased or uncertain results for galaxies with strong departures from
order motions. 

The aim of this paper is to establish the observed (i.e., projected) typical ranges of kinematic parameters expected for non-interacting galaxies. From those kinematic values, we will be able to shade some light on the possible kinematic distortions induced by an interaction. For this reason, we provide in this study a quantitative characterization of the kinematics of galaxies through parameters derived \textit{directly} from their stellar and ionised gas velocity
distributions. No assumption on the behaviour of the galactic components is
made. We will use the same method on our sample of interacting galaxies in order 
to make a consistent comparison. A pure-rotational velocity field can be caracterized by several parameters: kinematic centre, systemic velocity, global orientation of the field (i.e., kinematic PA), and its inclination with respect to the sky's plane. In this study we restrict ourselves only to kinematic projected properties of the sample. We consider that an estimation of the inclination via the velocity distribution would require assuming an intrinsic behavior on the kinematic component \citep[e.g., tilted-ring modeling, ][]{2007A&A...468..731J}. Although, this could be the case of some of the non-interacting galaxies presented in this study (restricted even only to the nebular component), it might not be the case for highly asymmetric velocity fields, such as those exhibit by interacting/merging galaxies. In the work presented here we focus mainly on the characterzation of the velocity fields via  their different components kinematic PAs. 

In order to determine the global orientation of the velocity maps first it is necessary to set the kinematic centre. For an ideal pure rotating disk the kinematic centre should be at the position with the largest velocity gradient (or gradient peak, GP hereafter) and in agreement with optical nucleus.  The reader is refered to Garc\'ia-Lorenzo et al. (2014, submitted) for details of the method used to determine the velocity gradient in the CALIFA ionized gas velocity fields.  Briefly, in a region of $10\arcsec\times10\arcsec$ centred in the optical nuclei we determine the velocity gradient  in each spaxel with respect to the surrounding spaxels. Then, we select the positions where  the velocity gradient is larger than its average inside this box. Finally, the GP is estimated from the weighted average location of the selected positions, using as weights the velocity gradient at each location.

 For an observed velocity field the GP does not always coincided with the optical nucleus (ON hereafter). Taking the size of the original fiber as the minimum distance to report an offset between the GP and ON (i.e., 2.7$\arcsec$). We find that a large fraction of the sample in their stellar component do not present a significant offset between these two positions (69/80) with a mean offset of 1.5 and standard deviation of 1.1 arcsecs.  In the rest of the stellar velocity fields this offset is due to possible SNe in the central region (e.g.,  NGC~7478) or the possibility in barred galaxies that the GP is located at the end of bar (e.g.  NGC~6004). The ionized gas present similar offsets (mean offset of 1.3 and standard deviation of 0.9 arcsecs). However the fraction of galaxies with small offsets is larger (74/80). The galaxies with large offsets are either edge on galaxies (e.g., UGC~841) or patchy in emission (e.g., NGC~4956).   Finally, as  kinematic centre we choose either ON or GP, depending on the best symmetry of the rotational curve (see details below). In all the cases with large offset in the GP (in both components) the ON yields a more symmetric velocity curve than the GP. This yield that in 21/80  and 5/80 velocity fields the ON was selected as the kinematic centre for the stellar and ionized gas components, respectively.   As we note above most of the offsets are smaller than the original spatial resolution of the instrument, in particular, for the stellar component where spatial binning  is done (although small for the nuclear region) this yield a larger number of fields where the ON was selected as kinematic centre. In tables \ref{table_Skin} and \ref{table_Gkin} we list the gradient peak location with respect to the optical centre when  the choice for the kinematic centre was the  GP, and the systemic velocity derived for both components as the mean value within a 2.7$\arcsec$ radius of the kinematic centre.

 \begin{figure}[!htb]
 \includegraphics[width=\linewidth]{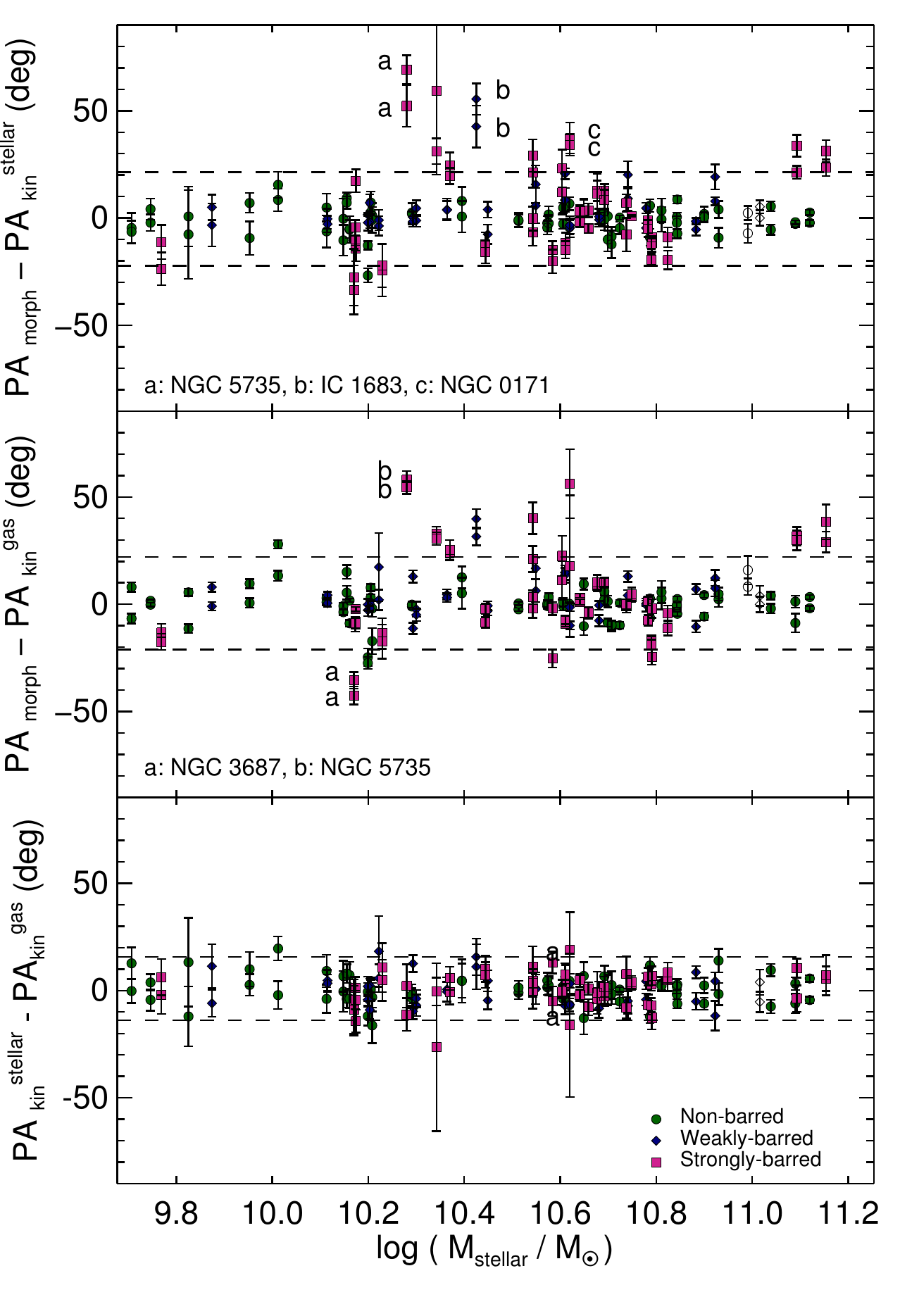} 
 \caption{Difference between the morphological PA ($\mathrm{PA_{morph}}$) and
 kinematic PA ($\mathrm{PA_{kin}}$) against the total stellar mass for the
 stellar (top) and the ionised gas (middle), as well as the stellar versus ionised-gas kinematic PA (bottom). Both kinematic sides of the velocity maps were compared  with $\mathrm{PA_{morph}}$ (top panels) and between each component (bottom panel). In each panel, the green-filled circles represent the non-barred
 galaxies, blue diamonds are weakly-barred galaxies and violet squares correspond
 to barred galaxies. For early-type galaxies we use open circles. Dashed lines
 represent the 2$\sigma$ dispersion of the sample. Labels indicate the
 3$\sigma$ outliers. Error bars are determined from Monte Carlo simulations.} 
 \label{PA_morph} 
 \end{figure}

The major (projected) kinematic axis (PA$_{\mathrm{kin}}$, hereafter) provides a measurement of the global kinematic orientation. It can be determined directly from the positions of the spaxels defining the kinematic lines of nodes \citep{1992ApJ...387..503N}. In practice, we plot the radial velocity for all the spaxels up to a given radius in a position-velocity diagram centred in the kinematic centre previously assigned (see right panel of Fig.~\ref{kin_example}). We then choose those spaxels with the maximum (minimum) projected velocity at the receding (approaching) side at different radii. Finally, for each side, we identify the selected spaxels in the velocity map. The average of their polar coordinates provides an estimation of PA$_{\mathrm{kin}}$ and their standard deviation ($\delta$PA$_{\mathrm{kin}}$) measures the scatter of these points around the straight line defining the kinematic PA. Following a similar procedure, we also determine the minor (receding and approaching) kinematic PA by selecting the spaxels with the lowest velocity differences respect to the kinematic center. To compare the stellar and the ionised gas kinematic PA we pick the common maximum radius where both components can be measured. Although PA$_{\mathrm{kin}}$ is usually defined as the angle between the north and the receding side of the velocity field \citep[e.g., ][]{1997MNRAS.292..349S}, for the sake of homogeneity, we report the approaching and receding kinematic PAs independently (PA$_{\mathrm{kin,app}}$ and PA$_{\mathrm{kin,rec}}$) between 0 and 180 degrees from North to East. Typical errors in PA$_{\mathrm{kin}}$ and $\delta$PA$_{\mathrm{kin}}$ are of the order of 7 $^\circ$ obtained from Monte Carlo simulations. We have also investigated the uncertainties introduced by the Voronoi binning in the stellar velocity maps (by artificially increasing the level of binning in a few of our galaxies) and established that they could bias the determination of PA$_{\mathrm{kin}}$ at most 4$^\circ$.\looseness-2 

In this study, for comparison, we also estimate the photometric orientations
$\mathrm{PA_{morph}}$ of the galaxies by fitting an ellipse model using the
standard task \textit{ellipse} of IRAF\footnote{ IRAF is distributed by the
Optical Astronomy Observatory, which is operated by the Association of
Universities for research in Astronomy (AURA) under cooperative agreement with the National Science Fundation} on the isophotes of the SDSS $r$-band image at two galactocentric distances: 1) at the same galactocentric distance where PA$_{\mathrm{kin}}$  are derived ($\mathrm{PA_{morph}}$, see solid white line in left panel of Fig.~\ref{kin_example}) and 2) at the outer most isophotes of the image (PA$\mathrm{_{morph}^{out}}$, see dotted white line in left panel of Fig.~\ref{kin_example}). We use the ellipticity of the latter  ($\epsilon$) as an homogeneous proxy for the inclination of the galaxies in our sample (see Table \ref{table_morph}).  All the position angles used in this work, as well as the radial distances where they have been computed, are presented in Tables~\ref{table_Skin} and \ref{table_Gkin}.
 \begin{figure*}[t]
 \minipage{0.5\textwidth}
 \includegraphics[width=\linewidth]{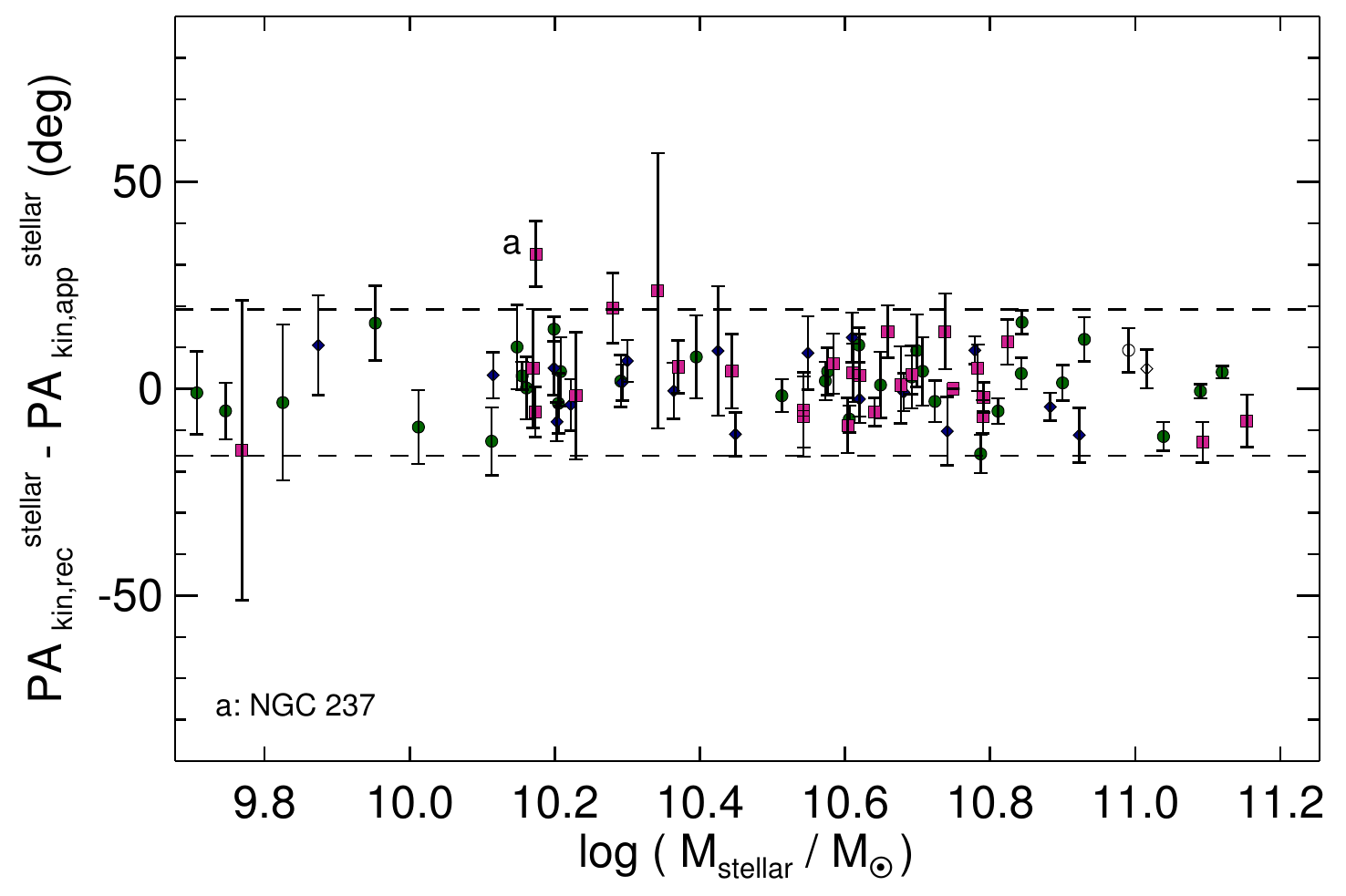}
 \endminipage\hfill
 \minipage{0.5\textwidth}
 \includegraphics[width=\linewidth]{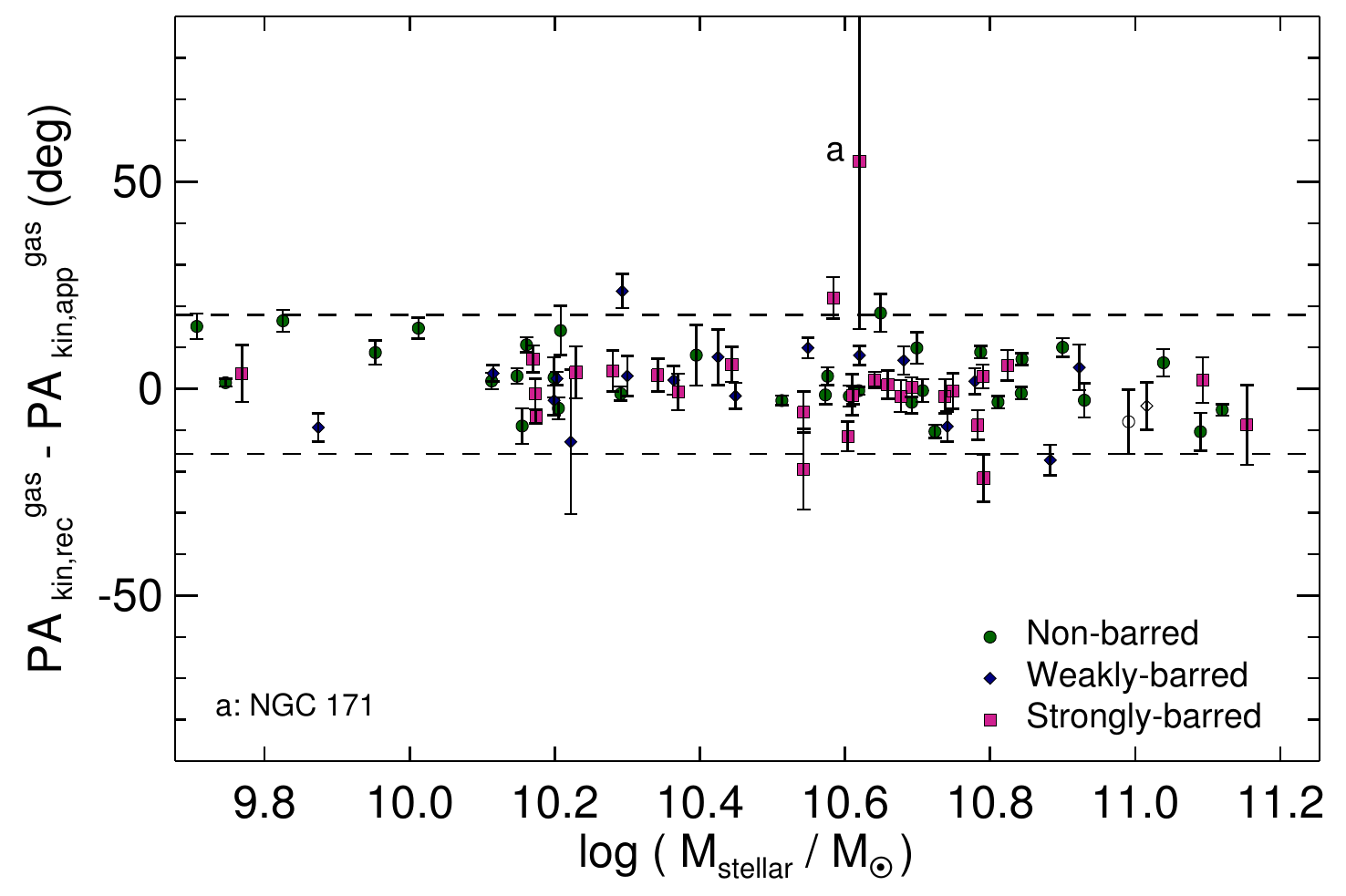}
 \endminipage\hfill
 \caption{Internal kinematic misalignment between the receding PA
 ($\mathrm{PA_{kin,rec}}$) and approaching ($\mathrm{PA_{kin,app}}$) kinematic
 position angle against the stellar mass. In each panel the green-filled circles
 represent the non-barred galaxies, blue diamonds correspond to weakly-barred
 galaxies and violet squares represent the barred galaxies. Open circles
 represent the early-type galaxies. Dashed lines represent the 2$\sigma$
 dispersion of the sample ($\sim$15$^\circ$). Labels indicate the 3$\sigma$
 outliers. Error bars at the left-top of each panel represent the typical error determined from Monte Carlo simulations.}
 \label{PA_kin} 
 \end{figure*}
%
%
\section{Kinematic alignment of galaxies}
\label{sec:alignments}
\subsection{Global morpho-kinematic misalignments}
\label{sec:mk_mis}

In Fig.~\ref{PA_morph} we show the morpho-kinematic PA misalignment (i.e.,
$\mathrm{PA_{morph}}-\mathrm{PA_{kin}}$) for the stellar and ionised-gas
components (top and middle panels, respectively) against the total stellar mass
of the galaxies (from Walcher et al., in preparation). Since there is no
preference between the kinematic approaching and receding PA, both of them were
compared with respect to $\mathrm{PA_{morph}}$. We do not find any trend between
the stellar mass and the morpho-kinematic PA misalignment, in any component, for
our sample of galaxies.

\begin{figure*}[t]
 \minipage{0.5\textwidth}
 \includegraphics[width=\linewidth]{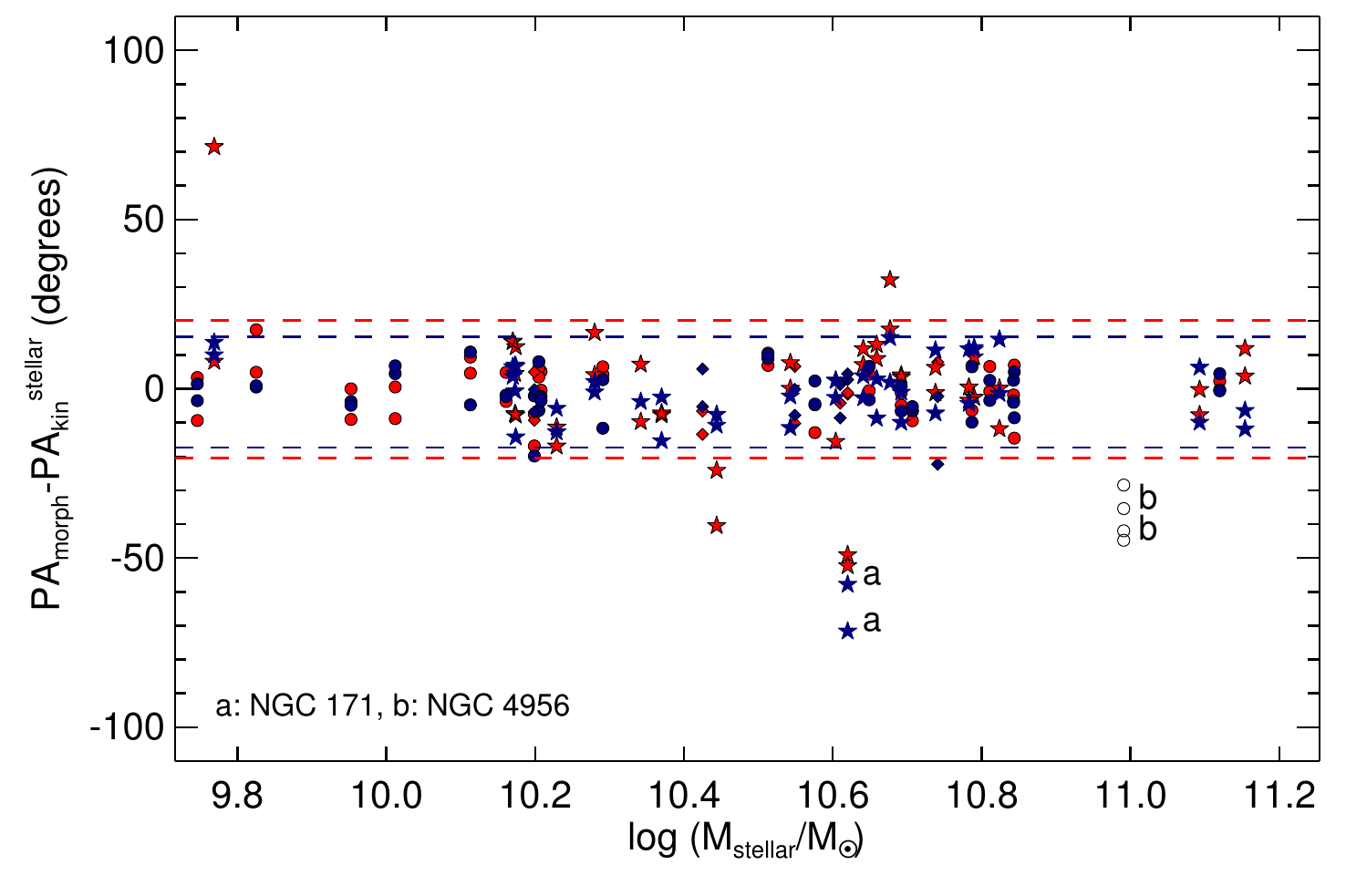}
 \endminipage\hfill
 \minipage{0.5\textwidth}
 \includegraphics[width=\linewidth]{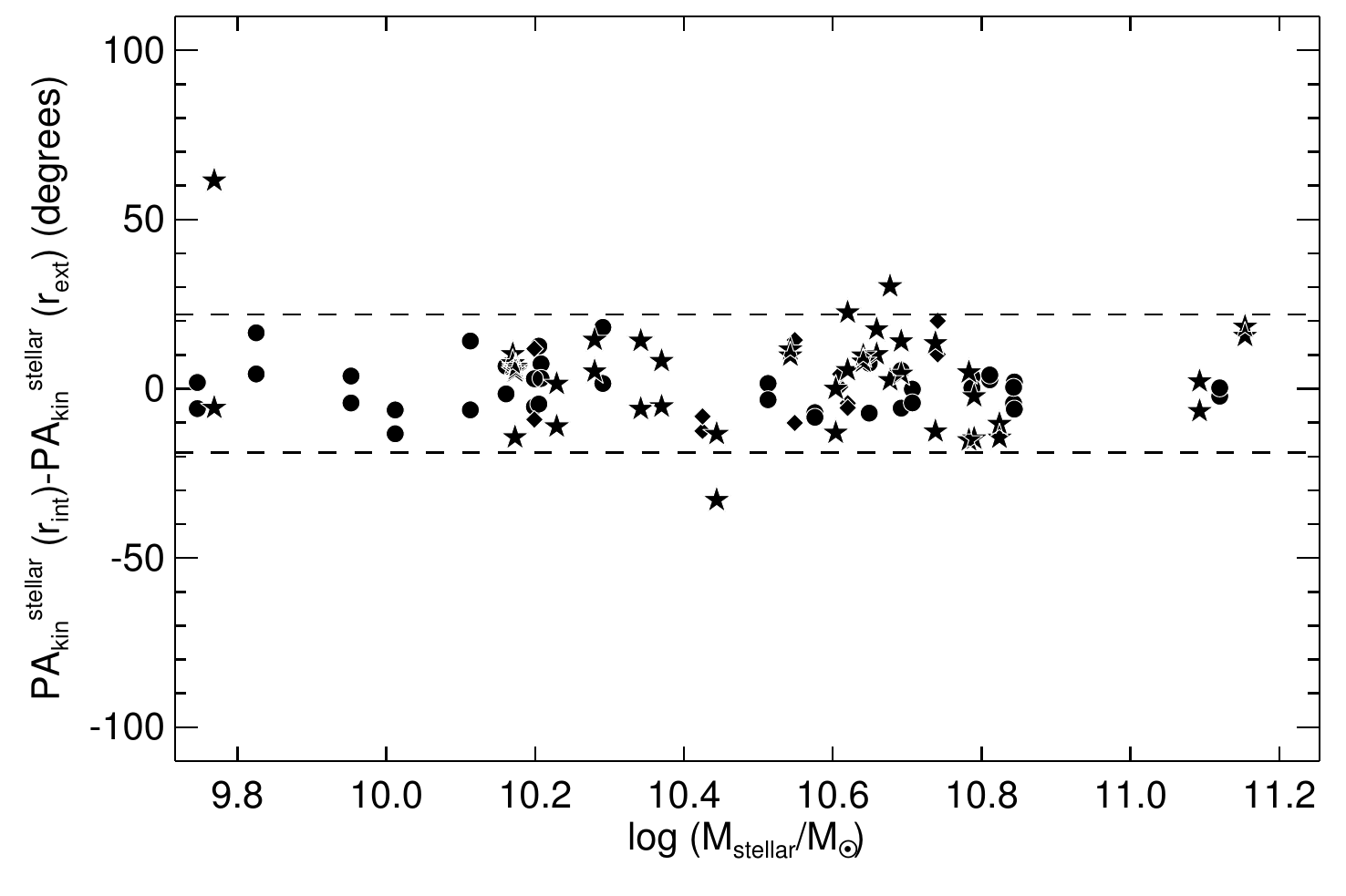}
  \endminipage\hfill
  \minipage{0.5\textwidth}
  \includegraphics[width=\linewidth]{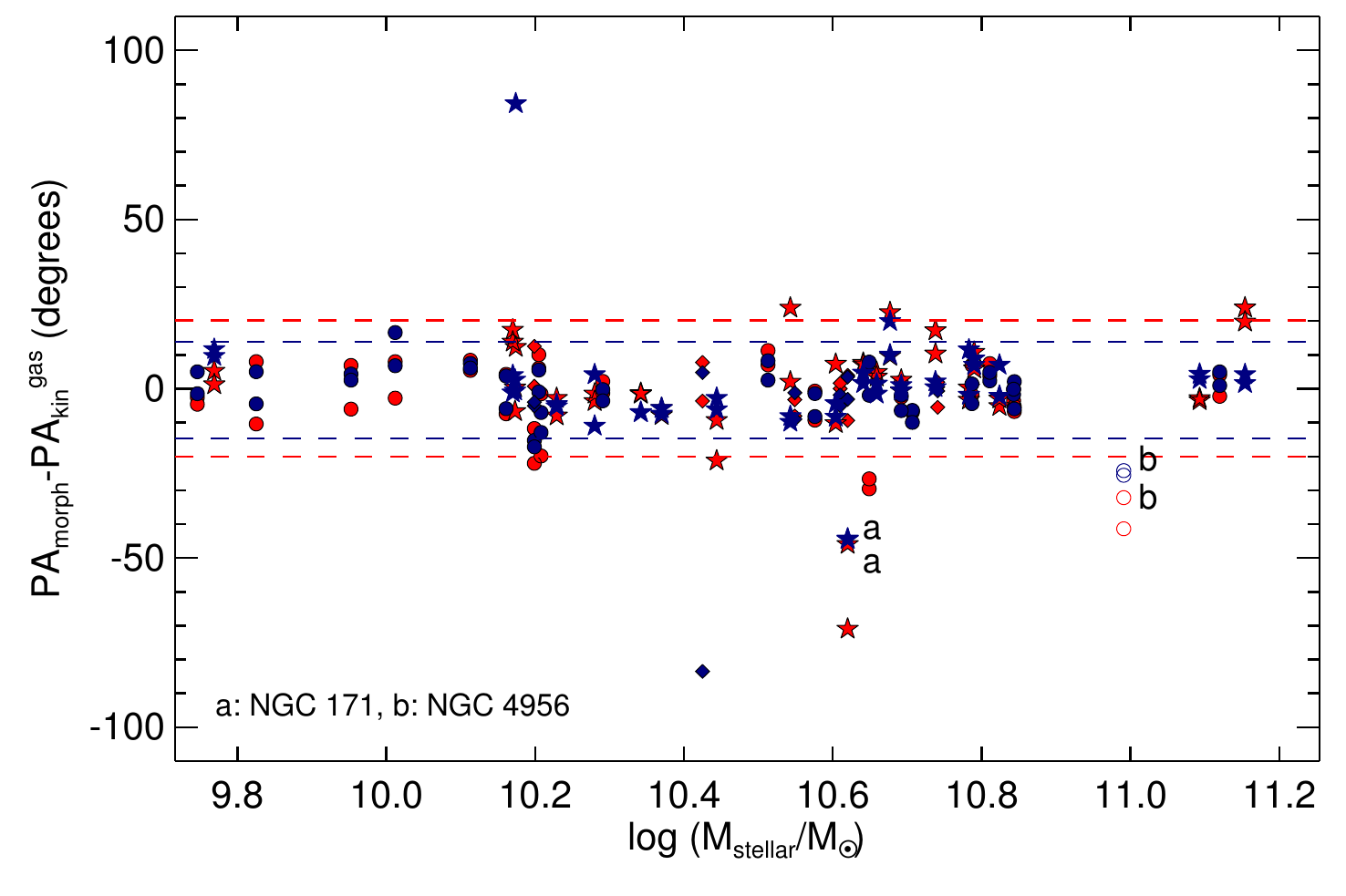}
  \endminipage\hfill
  \minipage{0.5\textwidth}
  \includegraphics[width=\linewidth]{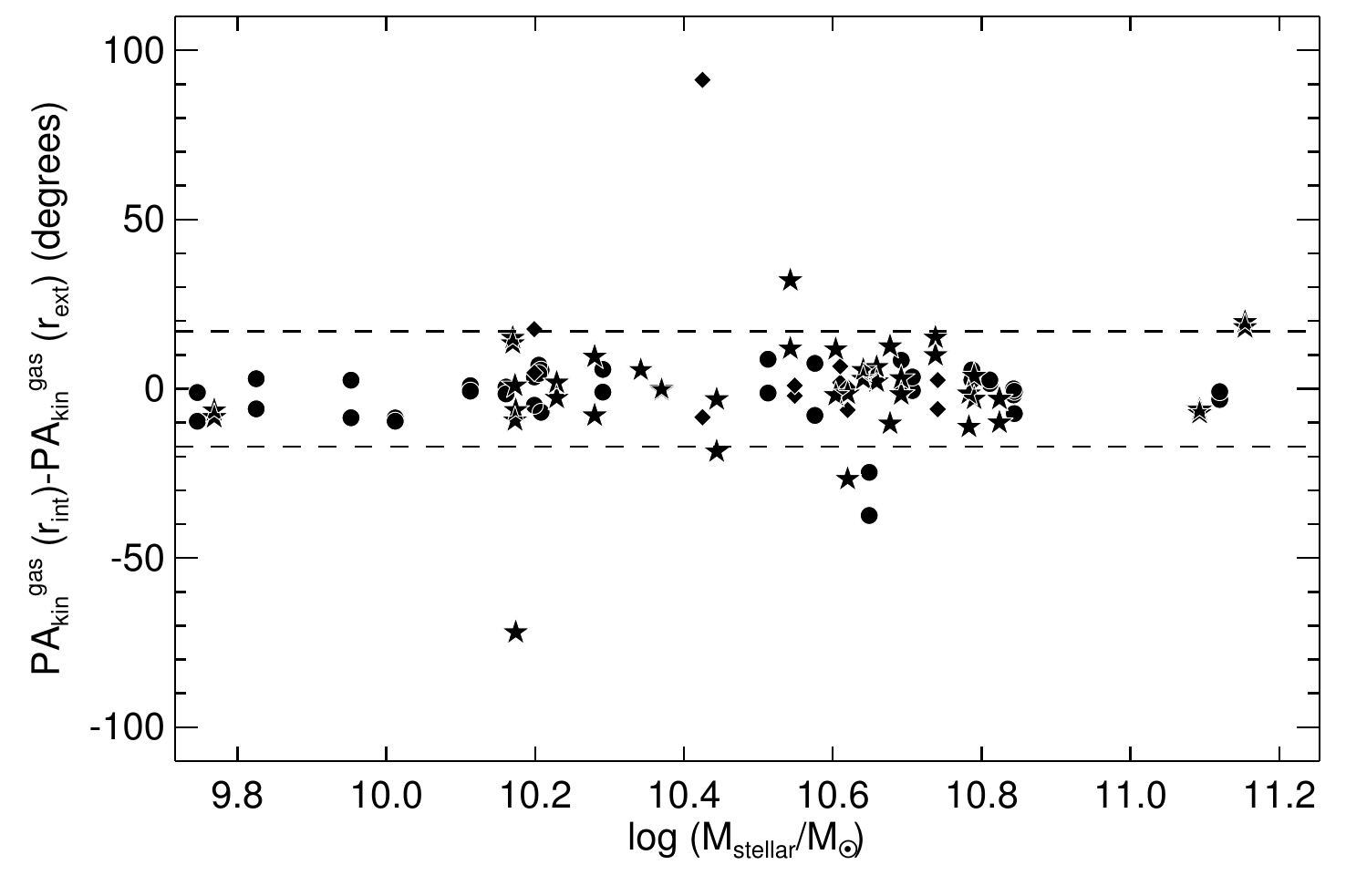}
  \endminipage\hfill
 \caption{Left panels: Morpho-kinematic PA misalignments at different regions of
 the galaxy against the stellar mass for both the stellar (top panel) and the
 ionised gas (bottom panel) components in a subsample of low-inclined galaxies. 
 In each of these panels, the red points represent the morpho-kinematic PA
 misalignment between PA$\mathrm{_{morph}^{out}}$ (see table \ref{table_morph})
 and  PA$\mathrm{_{kin}}$ at r$\mathrm{_{in}}$. Blue points represent the
 morpho-kinematic misalignment between PA$\mathrm{_{morph}^{out}}$ and
 PA$\mathrm{_{kin}}$ up to r$\mathrm{_{out}}$ (see section \ref{sec:Impact} for
 details). Filled circles, squares and stars represent non-barred, weakly-barred
 and strongly-barred galaxies, respectively. Right panels: Difference between the
 kinematic PA derived at r$\mathrm{_{in}}$ and r$\mathrm{_{out}}$ against the
 stellar mass for both the stellar (top panel) and the ionised gas (bottom panel)
 components. In the panels we highlight those objects with larger misalignments.}
 \label{PA_bars}
 \end{figure*}

About 90\% of the galaxies have morpho-kinematic misalignments smaller than 21(stellar)/ 22 (ionised gas) degrees. The mean misalignment and standard deviation are $\sim$0$^\circ$ and $\sim$10$^\circ$, respectively  in both components.  This indicates a global agreement between morphology and kinematics regardless the morphological type or bar strength. Barred galaxies present in both components the larger morpho-kinematic difference and the larger error bars (see tags in Fig.~\ref{PA_morph}). On the one hand, larger differences in the stellar component are observed in  IC~1683, NGC~5735, and NGC~171. For the two former galaxies the kinematic PA is similar to PA$\mathrm{_{morph}^{out}}$ rather than  $\mathrm{PA_{morph}}$. However $\mathrm{PA_{kin}}$ of NGC~171 is not similar to any of the photometric PA presented here (see Fig. \ref{PA_bars}). On the other hand for the ionised component, NGC~3687 and NGC~5735 present dissagrements with the photometric and kinematic PA. Similarly as in the stellar component, this galaxies seems to be more aligned to PA$\mathrm{_{morph}^{out}}$ than $\mathrm{PA_{morph}}$.
In Sec.~\ref{sec:Impact}, we study the impact of deriving the kinematic PA at different radius, in particular for barred galaxies. 

\cite{2011MNRAS.414.2923K} found that for 90\% of the ATLAS$^{\mathrm{3D}}$
galaxies displayed stellar morpho-kinematic misalignments smaller than
16$^\circ$ in their stellar component. From a sample of 24 early-type bulges,
\cite{2006MNRAS.369..529F} found that 80\% showed stellar morpho-kinematic
differences smaller than 20$^\circ$. Our results, based on a sample of
non-interacting galaxies with a wide range of morphological types, are
consistent with those findings. For the ionised-gas, the Fabry-Perot survey of
late-type spirals and irregular galaxies, GHASP \citep[][]{2008MNRAS.388..500E},
reported a 85\% of the objects with morpho-kinematic misalignments smaller than
15$^\circ$. These results are also in agreement with our findings. Similar
results have also been found for smaller samples \citep[e.g.][]{2008A&A...488..117K}.

Although we present along this study projected kinematic properties of non-interacting galaxies, in Appendix \ref{sec:inclination_effects} we study the possible impact of the inclination for the determination of kinematic properties. In Fig.~\ref{PA_morph_e} we plot the previous morpho-kinematic misalignments with respect to the photometric ellipticity as a proxy for the inclination ($\epsilon$, see Sec.~\ref{sec:Robust_Kinematic}). As is expected, larger misalignments are found at low ellipticities whereas at high ellipticities the morpho-kinematic misalignements get reduced due to the projection effects. However, we suggest that even at low ellipticities this misalignments reduced significantly (see Sec.~\ref{sec:Impact}).

Our results suggest that for a non-interacting galaxy the misalignment between the photometric and global (stellar- and/or ionised gas-) kinematic orientation is small for a wide range of morphologies and stellar masses (smaller than 20$^\circ$, with a mean value of  $\sim$ 0$^\circ$). This also held for galaxies with  strict isolation critera (i.e., AMIGA subsample), where the average misalignment for these galaxies is $\sim$1$^\circ$ with a standard deviation of 8$^\circ$ and 5$^\circ$ for the stellar and ionized gas component, respectively. As we noted previously only a small fraction of galaxies (2/80, barred) show large morpho-kinematic misalignments.

In the bottom panel of Fig.~\ref{PA_morph} we show the difference
between the stellar and ionised gas kinematic PA (i.e.$\mathrm{PA_{kin}^{stellar}}-\mathrm{PA_{kin}^{gas}}$). As in previous plots, we separate our sample according to their different bar strengths. In this stellar mass range, we do not find any trend. Barred and non-barred galaxies present similar alignments between their components. In fact, the mean difference for these components orientations is $\sim$1$^\circ$ with a standard deviation of $\sim$7$^\circ$. For 90\% of the galaxies, the global orientation of the stellar and the ionised gas velocity fields agrees, with differences smaller than 16$^\circ$. The only galaxy that seems to present large kinematic misalignment between both components is the late-type barred galaxy NGC~171. The large uncertainties in  the ionized gas component as well the fact that the locations of maximum velocity  present a large deviation from a straight line in the gas component suggest that the strong bar is the driver that distorted the ionised gas velocity field. From the SDSS $r$-band image, this object presents a bar as large as the FoV of the instrument with an extended ring. This is the only galaxy with these features in the sample presented here. In Fig.~\ref{PA_kin_e1} we plot these misalignments respect to the ellipticity. For a wide range of inclinations the alignment between both components is consistent. Summarizing, our measurements reveal small morpho-kinematics misalignments for both component as well as small stellar versus gas kinematic misalignments for barred galaxies, comparable to those for non-barred galaxies. At the spatial and spectral resolution of CALIFA survey, both components thus seem to follow a similar kinematic pattern even in the presence of bars.

\subsection{Internal kinematic PA misalignments}
\label{sec:int_kin}

In Sec.~\ref{sec:Robust_Kinematic} we estimated kinematic PA from the approaching
and receding side as well as their departures from a straight line
($\mathrm{PA_{kin,rec}}$, $\mathrm{PA_{kin,app}}$,
$\delta\mathrm{PA_{kin,rec}}$, $\delta\mathrm{PA_{kin,app}}$). This allows us to
study the internal kinematic misalignment, defined as $|\mathrm{PA_{kin,rec}}-\mathrm{PA_{kin,app}}|-180^\circ$, which should be zero for a regular rotational pattern. Figure~\ref{PA_kin} presents these misalignments for
the stellar (left) and the ionised gas (right) components.

\begin{figure}[!htb]
 \begin{center}
 \includegraphics[width=\linewidth]{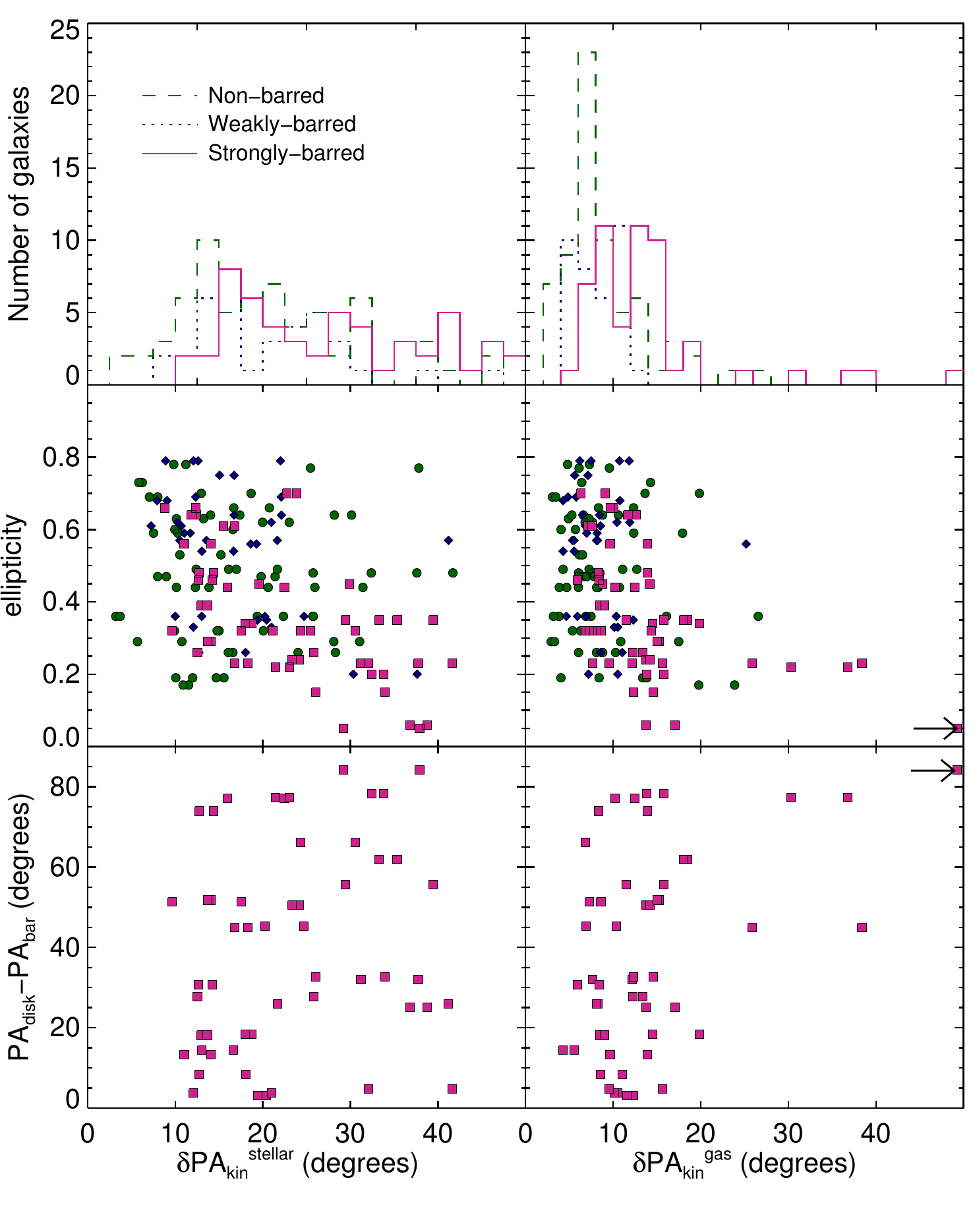}
 \end{center}
   \caption{Top: Histograms of the kinematic departure from a straight line in the velocity maps measured via $\delta$PA$_{\mathrm{kin}}^{\mathrm{stellar}}$ (left) and $\delta$PA$_{\mathrm{kin}}^{\mathrm{gas}}$ (right) separated by  different bar strengths. Dashed-lines represent the non-barred galaxies, dotted-lines correspond to the weakly-barred galaxies and solid-lines shows the distribution of strong barred galaxies. Middle: Apparent flattening of the galaxies via the ellipticity,  as in previous figures, the green-filled circles represent the non-barred galaxies, blue diamonds correspond to weakly-barred galaxies and violet squares represent the barred galaxies. Bottom: Difference between the  PA of the disk (PA$_{\mathrm{disk}}$) and the PA of the bar (PA$_{\mathrm{bar}}$) for strongly-barred galaxies only. Arrows in middle and bottom gas panels indicate the galaxy NGC~171 ($\delta$PA$_{\mathrm{kin}}^{\mathrm{gas}} \sim $ 90 $^{\circ}$ ) }
 \label{Histo_dis} 
 \end{figure}

We find that $\sim$ 92\% of the objects have internal kinematic misalignment smaller
than 15$^\circ$ for both the stellar and the ionised gas. This is for different stellar masses and ellipticities (see Fig.~\ref{PA_kin_e}). Assuming that our objects are a representative sample of isolated galaxies, we suggest that any distortion in the line-of-sight velocity distribution that produces a kinematic misalignment larger than those values is not caused by internal processes. The kinematic deviations from a straight line $\delta$PA$_{\mathrm{kin}}$ range from 5$^\circ$ to  40$^\circ$  in both  components for a large fraction of the sample ($\sim$\,90\%). The mean value of these deviations is twice for the stellar component with respect to the ionized gas ($\sim$ 15$^\circ$ and $\sim$ 7$^\circ$). Their standard deviation is also twice for stars respect to the stellar component ($\sim$ 8$^\circ$ and 4$^\circ$, respectively). In Fig.~\ref{Histo_dis}(top-panel) we plot the distribution of  $\delta$PA$_{\mathrm{kin}}$ for different barred strengths.  The fact that the mean deviation of the stellar component is almost twice the deviation from the ionized gas could be explained by different factors. On the one hand, although we are allow to determine in each spaxel the stellar velocity with relatively small uncertainty, the change of velocity for joint spaxels could be smaller than the spectral resolution leading in a higher scatter in the positions of maximum velocity. On the other hand,  the method used here to derived kinematic properties is entirely based on the symmetry that a pure rotational velocity field should display. Therefore, any deviation translates in a perturbation in any of the parameters we are using to quantify or velocity fields. Small $\delta$PA$_{\mathrm{kin}}$ in the ionized gas indicates that gas lies in a thin disk, while the stellar component present 'dynamical-heated' structures (e.g., dynamical heated disk, pressure-supported bulge, etc).

As we mention above, the barred galaxy \object{NGC\,171}  is the object where is more difficult to trace a symmetric velocity field in its ionised gas component (see Appendix~\ref{sec:maps}), the deviation of the maximmum-velocity positions from a straight line is of the order of $\delta$PA$_{\mathrm{kin}}$ $\sim$90$^\circ$. In Sec.~\ref{sec:Discussion}, we discuss the possible relation of $\delta$PA$_{\mathrm{kin}}$ with the level of bar strength of the non-interacting galaxies.

\subsection{Dependence on the measuring radius}
\label{sec:Impact}

In previous sections, to compare the PA$_{\mathrm{kin}}$ of both components, we used the same distance to average the polar coordinates of positions defined by the lines of nodes (see Sec.~\ref{sec:Robust_Kinematic}). In our data, usually
the stellar velocity map limits this distance where the average is done. For barred galaxies, it is similar (and a few times smaller) than half the length of the bar (see r$\mathrm{_{bar}}$ in table \ref{table_morph}  and r in tables \ref{table_Skin} \ref{table_Gkin}).
For the majority of the galaxies, the ionised gas extends further out for these galaxies. Our results could therefore be biased because of measuring the average kinematic orientation on a distance similar or close to the bar length.

To study the impact of this distance in the global orientation of the
galaxies, we calculated the morpho-kinematic misalignments at different radii in
a sub-sample of 49 low-inclined galaxies ($\epsilon<0.5$ ).  PA$_{\mathrm{kin}}$ is computed 
up to two different radius, r$_{\mathrm{out}}$ and r$_{\mathrm{in}}$ (see table
\ref{table_kin_bar}). The former defines the largest extension of the velocity
map in any component. For barred galaxies r$_{\mathrm{in}}$ defines the bar
length, for non-barred galaxies we define it as half of the r$_{\mathrm{out}}$.
PA$_{\mathrm{kin}}$(r$_{\mathrm{out}}$)  is the average of the polar coordinates
from  r$_{\mathrm{in}}$ to r$_{\mathrm{out}}$, while
PA$_{\mathrm{kin}}$(r$_{\mathrm{in}}$) is the average from the kinematic centre
up to r$_{\mathrm{in}}$.

Left panels of Fig.~\ref{PA_bars} show, for both components, the morpho-kinematic misalignment respect to the orientation of an ellipse at the outermost region of each galaxy (PA$\mathrm{_{morph}^{out}}$ ).. Red symbols correspond to misalignments with PA$_{\mathrm{kin}}$(r$_{\mathrm{in}}$), while blue ones represent misalignments for PA$_{\mathrm{kin}}$(r$_{\mathrm{out}}$ ). Note that in most of the cases PA$\mathrm{_{morph}^{out}}$ is measured at larger radii than  PA$_{\mathrm{kin}}$(r$_{\mathrm{out}}$). Right panels compare the difference between PA$_{\mathrm{kin}}$(r$_{\mathrm{in}}$) and PA$_{\mathrm{kin}}$(r$_{\mathrm{out}}$). Figure~\ref{PA_bars} evidences the similar alignment of the kinematic PA for different morphologies  at the different measuring radius, r$_{\mathrm{in}}$ and r$_{\mathrm{out}}$, in particular for barred galaxies. On the one hand, the comparison of kinematic PAs and PA$\mathrm{_{morph}^{out}}$
(left panels) reveals that 90\% of the objects have differences smaller than
16$^\circ$ (20$^\circ$) for the outer (inner) region in the stellar component.
Differences became smaller for the ionised gas (13$^\circ$ and 20$^\circ$ for
outer and inner regions, respectively). On the other hand, when we compare PA$_{\mathrm{kin}}$ at the inner and outer radii (right panels), we find that 90\% of the galaxies have differences smaller than 20$^\circ$ (16$^\circ$) for the stellar (ionised gas) component. These plots quantify what we observed in velocity fields; at different radii the orientation of the major kinematic PA remains rather constant. Moreover at different radii, PA$_{\mathrm{kin}}$ seems to be aligned with the orientation of the major photometric axis rather than other morphological local features such as bars. We find one outlier in both components: the strongly barred late-type galaxy (\object{NGC\,171}). This galaxy presents the smallest ellipticity in the sample (see Fig.~\ref{PA_morph_e}), making difficult to determine a reliable estimation of PA$\mathrm{_{morph}^{out}}$. From these morpho-kinematic differences as well as the differences presented in Sections \ref{sec:mk_mis} and \ref{sec:int_kin} we suggest that  for this sample of non-interacting galaxies, the stars and the ionised gas follow the potential of the disk. 

\section{The impact of bars in velocity maps}
\label{sec:Discussion}

Simultaneous comparisons of the stellar and ionised gas resolved kinematics in
barred galaxies are rather scarce. \cite{2006MNRAS.369..529F} studied those
cases found in a sample of 24 spiral bulges. The sample included nine barred
galaxies. Kinematic misalignments between the two components for these galaxies
ranged from 1$^{\circ}$ to 38$^{\circ}$. A subsequent detailed study of the galaxy
NGC\,5448 \citep{2005MNRAS.364..773F} revealed a kinematic misalignment between
these two components of 25$^{\circ}$. These differences are explained by the fact
that the morphological PA of the galaxies was measured within the same FoV
probed by their kinematics (i.e. the inner regions of those galaxies). The
PA$\mathrm{_{morph}}$ in those cases are thus biased due to the bar.
Throughout this study we find that the global kinematic orientation (measured by
the major-axis kinematic PA) for the stellar and the ionised gas are aligned for
non-interacting objects in the CALIFA survey (see Fig.~\ref{PA_morph} and
\ref{PA_kin}). This result appears to hold at different radial lengths of the 
galaxies (Fig.~\ref{PA_bars}), even for barred galaxies. To quantify the departures from perfect axisymmetry in the global kinematic orientation, we plot in top panels of Fig.~\ref{Histo_dis} histograms showing the deviations from a perfect straight line observed in the stellar and ionised-gas kinematic maps ($\delta$PA$_{\mathrm{kin}}^{\mathrm{stellar}}$, $\delta$PA$_{\mathrm{kin}}^{\mathrm{gas}}$). As we already note in Sec.~\ref{sec:int_kin} the mean value of these deviations is larger for the stellar component in comparison to the ionized gas. For both components the strongly barred galaxies present slightly larger values respect to the non-barred galaxies. In particular the ionised gas component presents a difference between the distribution of barred and non-barred galaxies. This suggest that although the global kinematic orientation is stable in non-interacting galaxies, in particular for barred galaxies, some other kinematic parameters such as $\delta$PA$_{\mathrm{kin}}^{\mathrm{gas}}$ (at higher spatial and spectral resolution as the present velocity fields ) can be sensitive to the presence of
strong bars. 
\begin{figure}[!htb]
\begin{center}
\includegraphics[width=\linewidth]{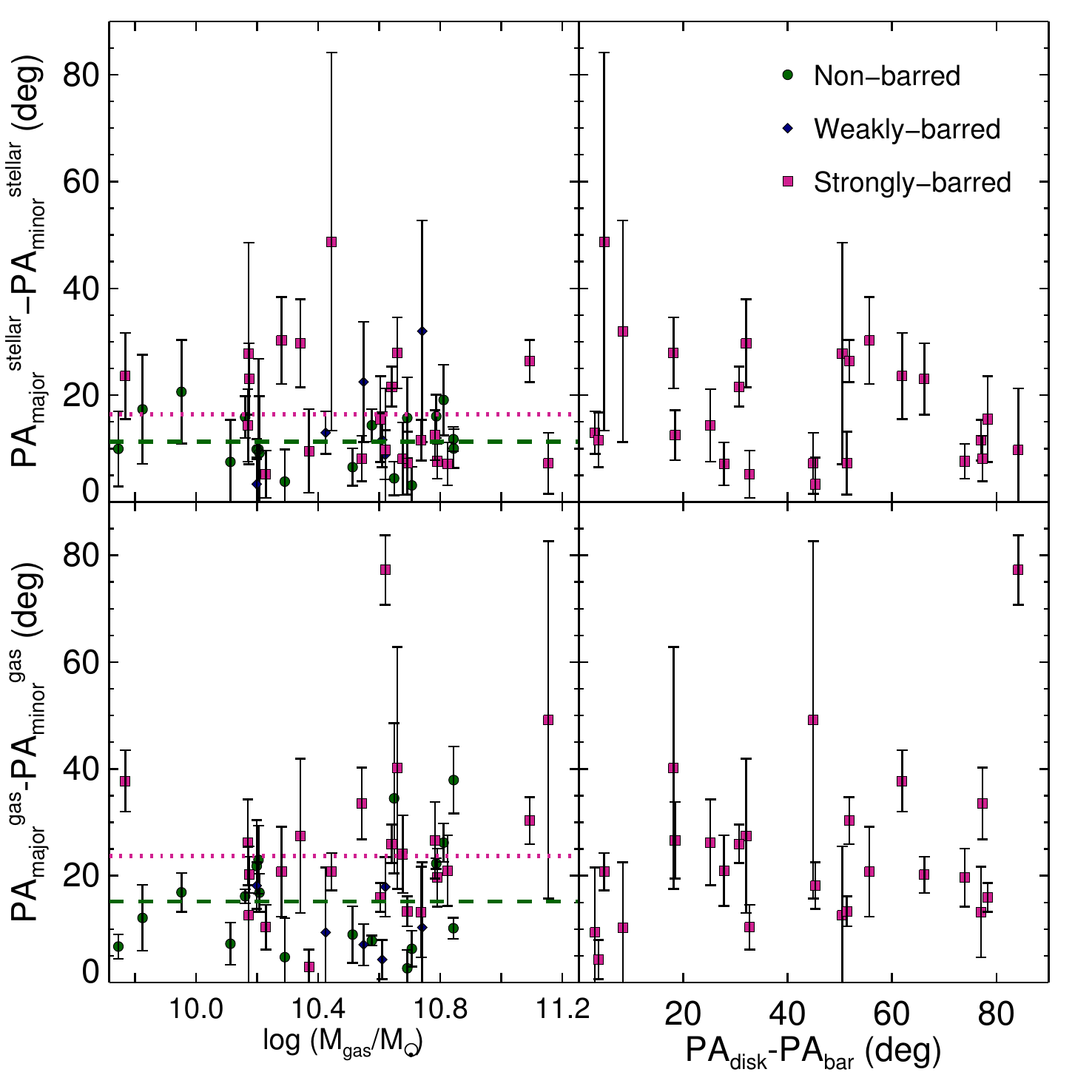}
\end{center}
\caption{ Misalignment between the major and minor kinematic PA for a subsample of low-inclined galaxies($\epsilon$~<~0.5). 0$^{\circ}$ represents perpendicularity between major and minor axes (see section~\ref{sec:Discussion}   for details). Left panels show these misalignments as function of the stellar mass for the stellar (top) and the ionised gas (bottom) components. Right panels present these kinematic  misalignment as function of the morphological alignment of the bar with respect to the disk for barred galaxies. As in previous figures,  the green-filled circles represent the non-barred galaxies, blue diamonds correspond to weakly-barred galaxies and violet squares represent the barred galaxies. Dotted-violet lines represent the mean value for the barred galaxies while dashed-green lines represent the mean value for the non-barred galaxies.}
\label{minor} 
\end{figure}

It is important to note that  there is still a fair fraction of strongly barred
systems with $\delta$PA$_{\mathrm{kin}}$ consistent with those observed in
non-barred galaxies. We have investigated whether this effect is caused by the flattening of the galaxy,or due to the relative orientation of the bar with respect to the disk major axis. This is shown in the middle and lower panels of 
Fig.~\ref{Histo_dis}. The projected deviation of the kinemtic PA ($\delta$PA$_{\mathrm{kin}}$) seems to grow at lower apparent flattening of the galaxy (as a surrogate for inclination). This trend is particular clear for barred galaxies, it may be induced by projections effects. Using the method presented in this study, any kinematic deviation from a straight line would be more easily quantified in  a  nearly face-on velocity field  than in the same field viewed as edge-on. For high inclined galaxies with large kinematic deviations, we may even trace vertical motions rather than radial deviation from a circular velocity field. Note however that other kinematic indicators such as the intrinsic kinematic misalignment do not present a clear trend respect to the ellipticity (see Appendix~\ref{sec:inclination_effects}). As for the orientation of the bar with respect to the disk major axis, the kinematic deviations covers a wide range of values for a given alignment between the bar and disk orientation, indicating that at least for barred galaxies the orientation of the bar with respect to the disk may not affect drastically the scatter in the location of  maximum velocities. 

The velocity maps of barred galaxies often show, however, clear signatures of
perturbations from a symmetric velocity field \citep[e.g.,
][]{2005MNRAS.360.1201H, 2009ApJ...704.1657F}. A characteristic ``S''-shape in
the zero-velocity curve is observed in several of these ionised gas kinematic
maps \citep[e.g.,][]{1980ApJ...242..913P, 2001AJ....121.2540G,
2006MNRAS.365..367E}. For a pure rotating disk galaxy the major and minor axis are perpendicular everywhere \citep[]{1998gaas.book.....B}. To account for the distortion in the zero-curve velocity we estimate the difference between the above kinematic PAs. For each low-inclined galaxy ($\epsilon <$ 0.5) we estimate the largest departure among the four kinematic PAs (approaching and receding major PAs and their corresponding minor PAs). 0$^{\circ}$ represents perpendicularity between major and minor axes. In Fig.~\ref{minor} we represent these departures against the stellar mass (left panels) and the photometric alignment of the bar respect to the disk for barred galaxies  (right panels) for the stellar (top panels) and nebular components (bottom panels).  For the stellar component, almost all the low-inclined galaxies present differences smaller than $\sim$30$^{\circ}$ except for the barred galaxy NGC~6155. Even more, strongly-barred galaxies spread homogeneously in this range as the non-barred objects. The average departure between strongly and non barred galaxies is  similar ($\sim$11$^{\circ}$ and $\sim$15$^{\circ}$, respectively). The ionised gas component display a wide range of differences between the major and minor kinematic PAs. In particular, strongly-barred galaxies present a wider range of  differences than non-barred sample. The average of this difference for the barred galaxies is larger  ($\sim$24$^{\circ}$) respect to the unbarred galaxies ($\sim$15$^{\circ}$). According to numerical simulations, the difference between the major and minor kinematic PA is best seen in velocity fields where the bar makes an angle of 45$^{\circ}$ with  the major axis of the galaxy \citep[e.g., ][]{1984PhR...114..319A, 2012MNRAS.425L..10S}. Due to the moderate sub-sample of barred galaxies in this study and the rather large uncertainty in determine the minor kinematic PA, it is not clear weather larger differences in the kinematic PAs are observed at this specific angle between the bar and the disk PA. Note however, that for the stellar component an increment in this difference seems to be present at PA$_{\mathrm{disk}}$ - PA$_{\mathrm{bar}}$ $\sim$ 50$^{\circ}$. In the ionised gas,  for a given morphological alignment between the bar and the disk we find a broad range of differences between the major and minor kinematic PA. Given the spectral and spatial resolution in our velocities distributions, we consider that the difference between the major and minor kinematic PAs is a sensitive indicator of the presence of a bar in particular for the ionised gas component. Detailed individual studies are required in order to explain the shape and length of the distortion in this velocity curve as well as to determine if the ionized gas is more sensitive component to the bar potential. These issues are  beyond the scope of this study, however our results motivate numerical simulations to address the nature of the kinematics in non-interacting galaxies. As opposed to the previous kinematic indicators (see Fig.~\ref{PA_morph} and~\ref{PA_kin}) this indicator of kinematic distortion varies significantly for non-interacting galaxies making difficult to use it as a discriminator to distinguish the kinematic distortion produced by the merging event or a secular process.   
\\

\section{Conclusions}
\label{sec:conclusions}

We have studied the stellar and ionised-gas kinematics for a sample of 80
non-interacting galaxies in the CALIFA survey. The fraction of barred and
non-barred objects in this sample is similar. We find that for 90\% of the
sample the global orientation of the stellar and ionised gas is fairly aligned
(misalignment smaller than $\sim$ 20$^\circ$) with respect to the global photometric orientation of the galaxy, including barred and non-barred objects. From the method used to measure the major kinematic PA we study the internal kinematic PA misalignments namely, the difference between the receding and approaching kinematic PA. We find intrinsic aligned velocity fields in both components for a large fraction of the sample (intrinsic misalignments smaller than 15$^\circ$ degrees for the stellar and the ionised gas components). 

We also compared the derived kinematic PA for the stellar and the
ionised gas components. We observe a tight alignment (16$^\circ$ degrees of
difference) between both components for the majority of this non-interacting
sample. This result holds even in barred galaxies when the comparison is carried
out either at the radius of the bar or further out, the radius of the disk. From
these results we suggest that the global kinematics in non-interacting galaxies in
both the stellar and the ionised gas components (measured by the major kinematic position angle) seems to be dominated by the mass of the disk, rather than  other morphological component that can induce non-circular
patterns in the observed velocity fields such as bars. Even though locally bars
can redistribute the angular momentum, energy and mass; the global kinematic
pattern that dominates across the galactic disk -including the bar- is
consistent with a regular rotational pattern. 

We will use the results presented in this paper to gauge the kinematic 
distortions caused by external forces in galaxies that undergo a merger event in Barrera-Ballesteros et al. (in prep.).

\section*{Acknowledgements}
We thank the referee for a thorough reading of this work and his/her comments and suggestions. We also thank Lindsay Holmes for allow us to use her DiskFit kinematic modelling for comparison with our results. This study makes use of the data provided by the Calar Alto Legacy Field Area (CALIFA) survey (http://www.califa.caha.es). Based on observations collected at the Centro Astron\'{o}mico Hispano Alem\'{a}n (CAHA) at Calar Alto, operated jointly by the Max-Planck-Institut f\"{u}r Astronomie and the Instituto de Astrof\'{\i}sica de Andalucia (CSIC). CALIFA is the first legacy survey performed at the Calar Alto. The CALIFA collaboration would like to thank to the IAA-CSIC and MPIA-MPG as major partners of the observatory, and CAHA itself, for the unique access to the telescope time and support  in manpower and infrastructures. The CALIFA collaboration also thanks the CAHA staff for the dedication to this project. J.~B.-B. and B.~G.-L. thank the support from the Plan Nacional de I+D+i (PNAYA) funding programs (AYA2012-39408-C02-02) of Spanish Ministry of Economy and Competitiveness (MINECO).
J.~F.-B. acknowledges support from the Ram\'on y Cajal Program, grants
AYA2010-21322-C03-02 from the Spanish Ministry of Economy and Competitiveness
(MINECO). We also acknowledge support from the FP7 Marie Curie Actions of the
European Commission, via the Initial Training Network DAGAL under REA grant
agreement number 289313. R.A.M is funded by the Spanish program of International Campus of Excellence Moncloa (CEI).
L.~V.~M. acknowledges support from the Grant AYA2011-30491-C02-01 co-financed by MICINN and FEDER funds, and the Junta de Andalucia (Spain) grants P08-FQM-4205 and TIC-114.
J.I.P. acknowledges financial support from the Spanish MINECO under grant AYA2010-21887-C04-01 and from Junta de
Andaluc\'{\i}a Excellence Project PEX2011-FQM7058.
J.M.A. acknowledges support from the European Research Council Starting Grant (SEDmorph; P.I. V. Wild)

\bibliographystyle{aa} 
\bibliography{noInter_final}


\begin{appendix}

\section{Kinematic alignments versus apparent ellipticity}
\label{sec:inclination_effects} 

All the kinematic parameters presented in this study are determined on the plane of the sky. Therefore, it is expected that any intrinsic kinematic property would be affected by the inclination of the galaxy with respect to the plane of the sky. In this appendix we show the same kinematic misalignments presented in section \ref{sec:alignments} with respect to the apparent ellipticity as proxy of the apparent inclination of the galaxies.

  \begin{figure}[!htb]
   \includegraphics[width=\linewidth]{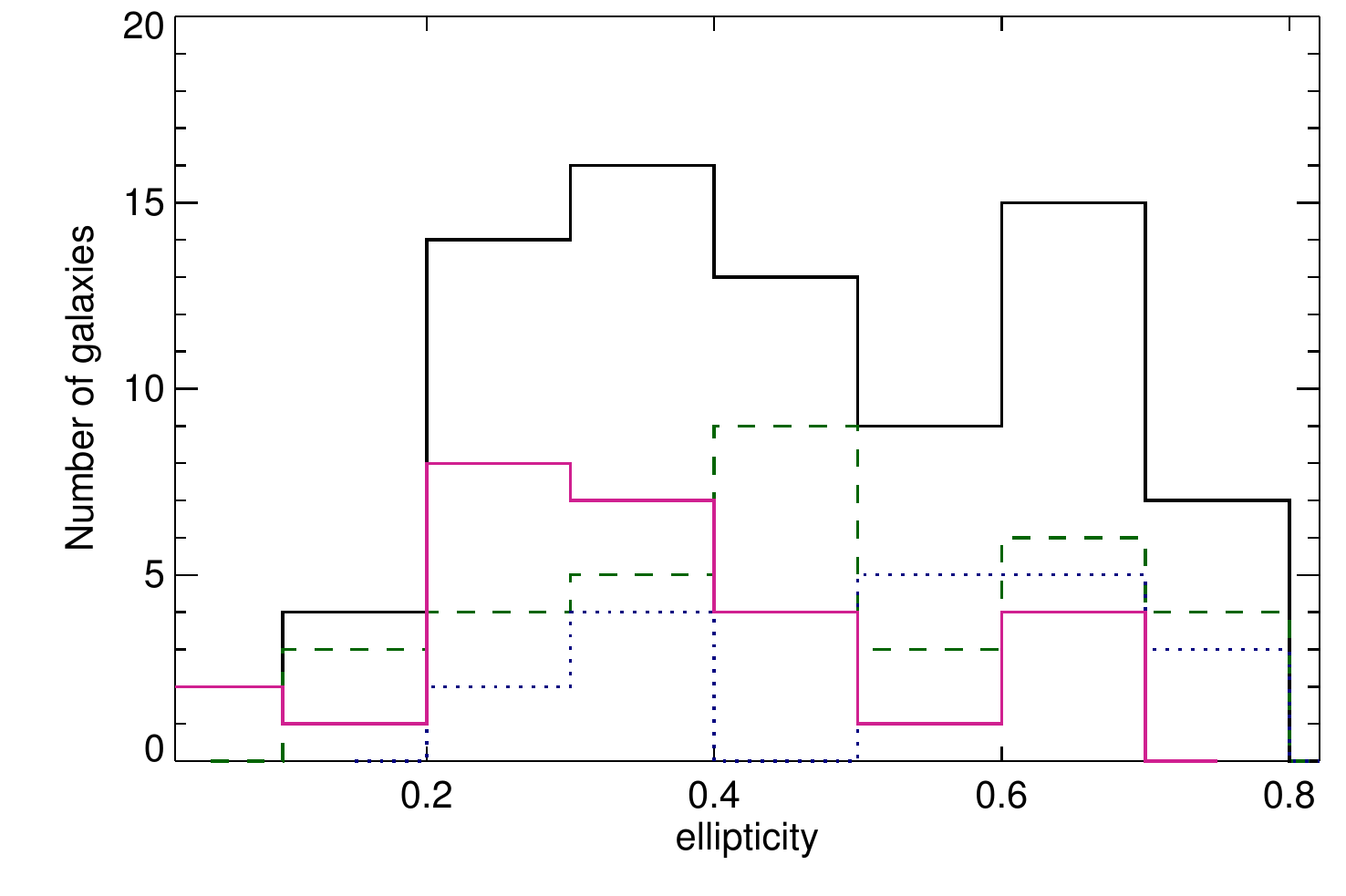}
   \caption{\label{ell_hist} Distribution of the ellipticity as proxy of the inclination for the galaxies used in this study. Dashed-lines represent the non-barred galaxies, dotted-lines correspond to the weakly-barred galaxies and solid-lines shows the distribution of strong barred galaxies.}
  \end{figure}

In Figure \ref{ell_hist} we plot the distribution of the non-interacting galaxies with respect to the ellipticity. We have a fair coverage of galaxies at different inclinations , in particular for the low-inclined ones ($\epsilon$~<~0.5). In order to have consistent  results, we also try to have the same fraction of barred vs unbarred galaxies at different ellipticity bins.

\begin{figure}[!htb]
    \includegraphics[width=\linewidth]{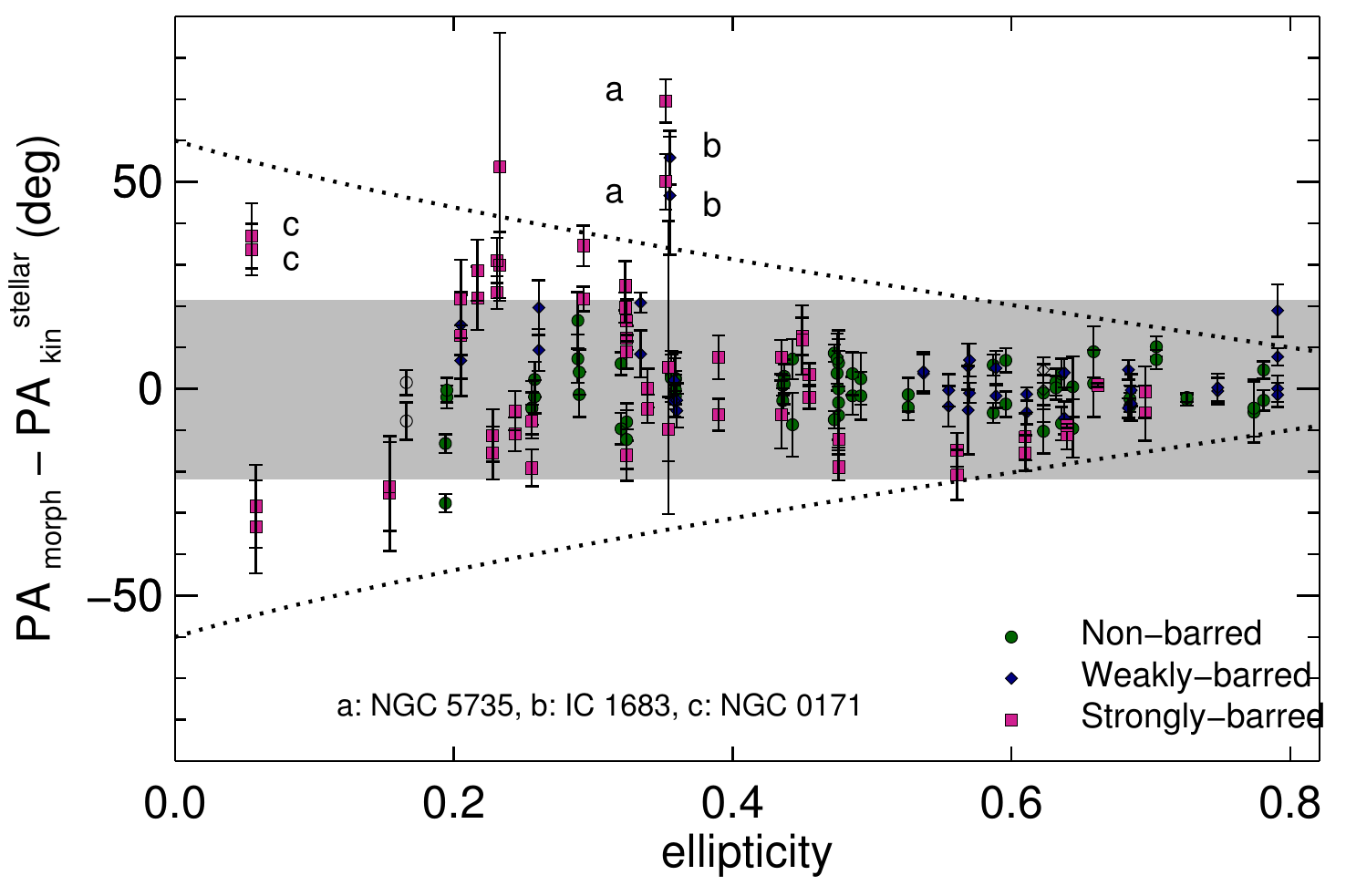}
    \includegraphics[width=\linewidth]{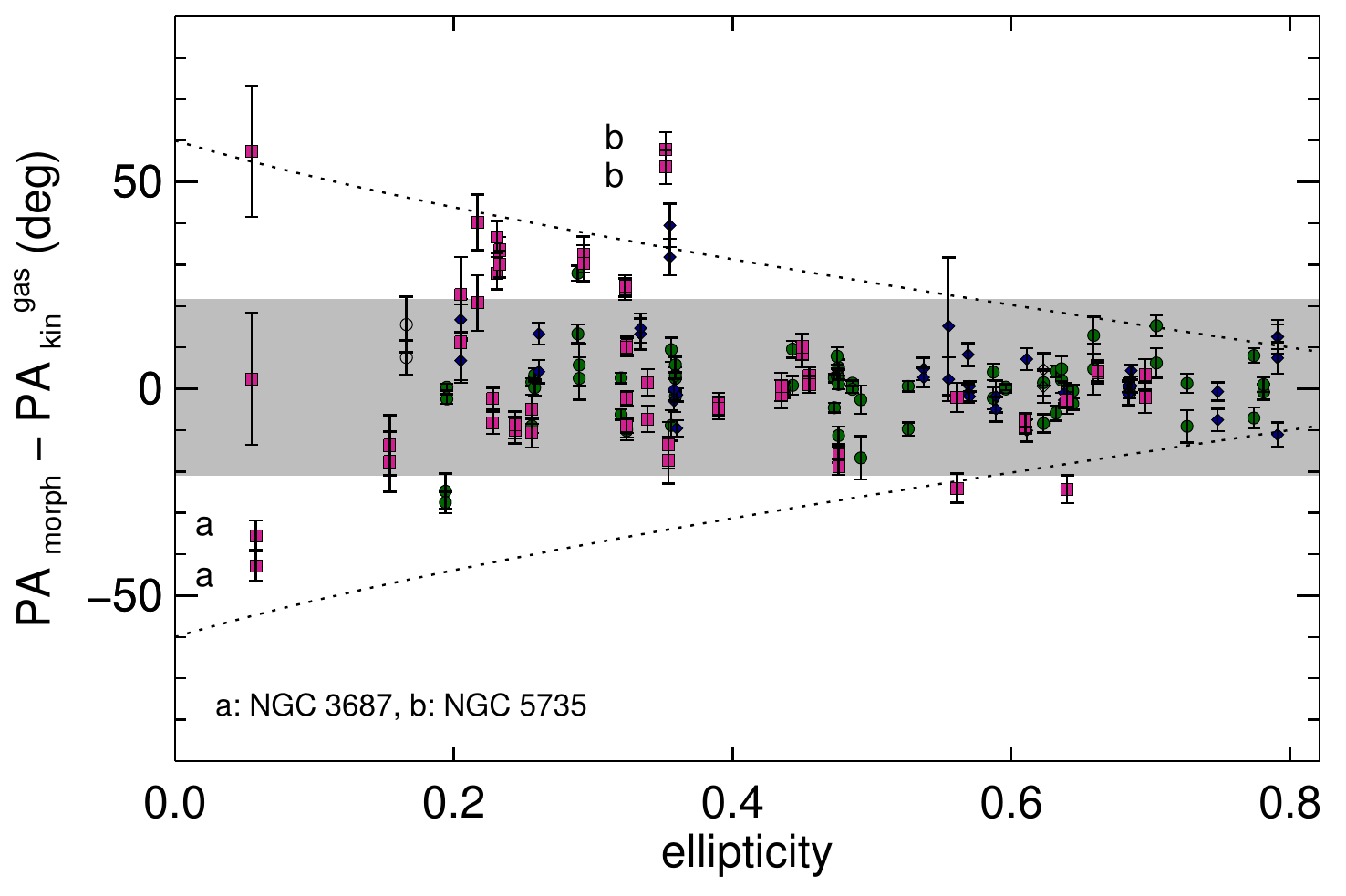}
    \caption{\label{PA_morph_e} Difference between the morphological PA ($\mathrm{PA_{morph}}$) and the kinematic PA ($\mathrm{PA_{kin}}$) against the ellipticity of the outermost isophote of the galaxy for the stars (top panel) and the ionised gas (bottom panel). Both kinematic sides of the velocity maps were compared with $\mathrm{PA_{morph}}$. In each panel, the green-filled circles represent the non-barred galaxies, blue diamonds are weakly-barred galaxies and violet squares correspond to barred galaxies. For early-type galaxies we use empty circles. Labels indicate the 3-$\sigma$ outliers. Dashed lines represent the projection effect of a morpho-kinematic misalignment of 60$^{\circ}$ measured in the plane of the galaxy. In each panel, the grey area represents the 2$\sigma$ dispersion of the sample.}
  \end{figure}

\begin{figure}[!htb]
    \includegraphics[width=\linewidth]{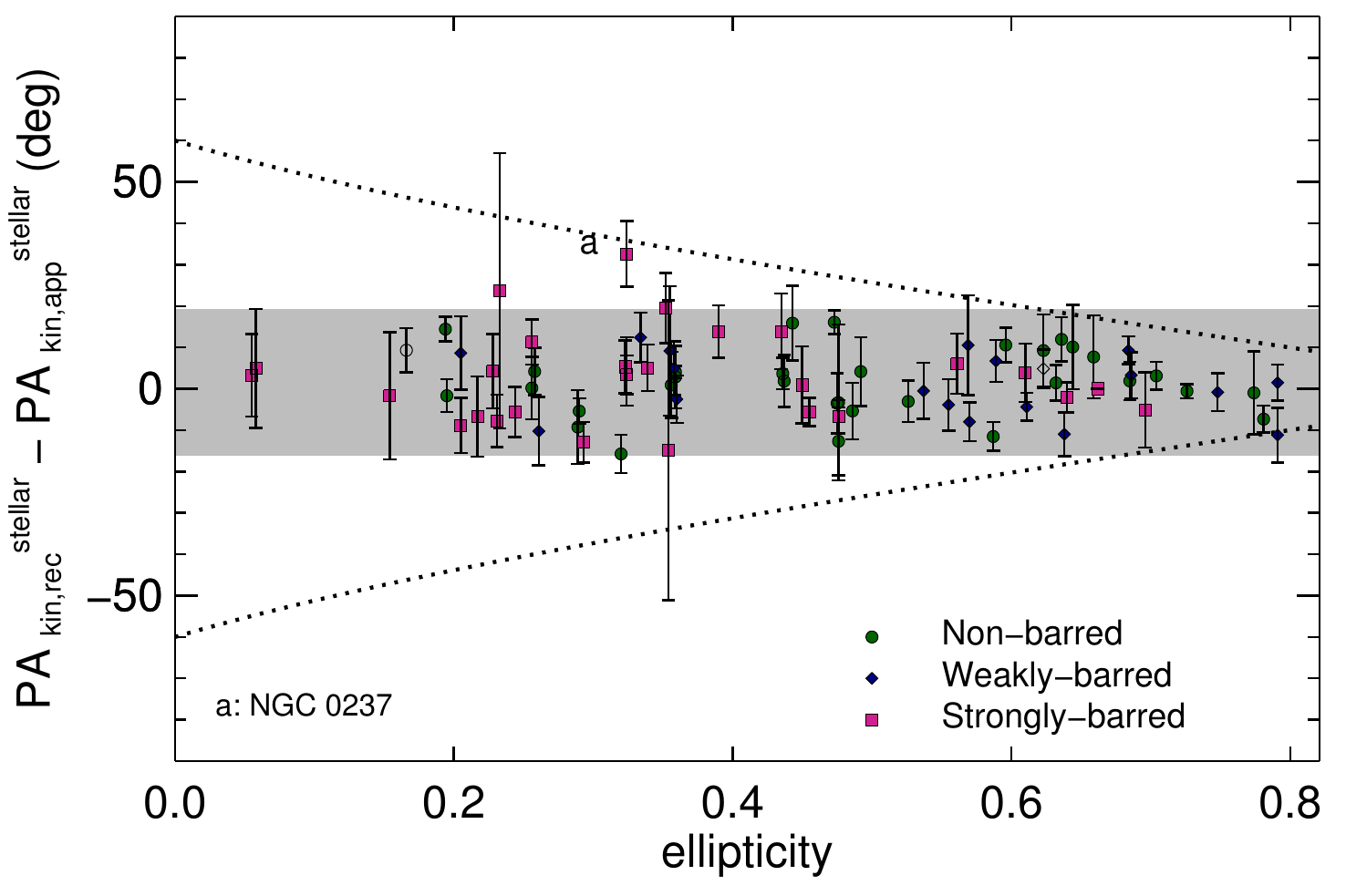}
    \includegraphics[width=\linewidth]{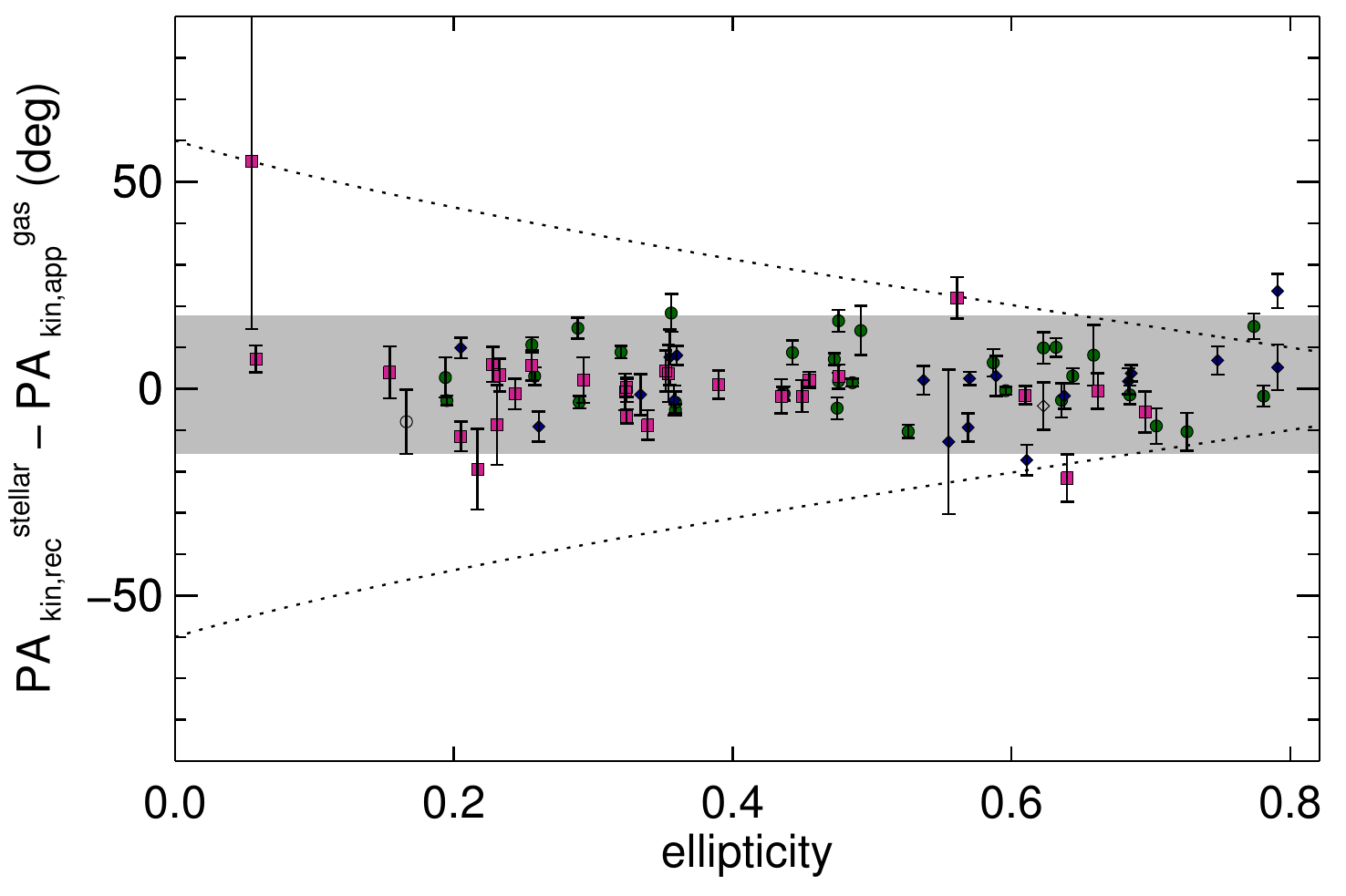}
    \caption{\label{PA_kin_e} Internal kinematic misalignment between the receding PA ($\mathrm{PA_{kin,rec}}$) and approaching ($\mathrm{PA_{kin,app}}$) kinematic position angle against the ellipticity. In each panel the green-filled circles represent the non-barred galaxies, blue diamonds correspond to weakly-barred galaxies and violet squares represent the barred galaxies. Empty circles represent the early-type galaxies. Labels indicate the 3-$\sigma$ outlier. Dashed lines represent the projection effect of a morpho-kinematic misalignment of 60$^{\circ}$ measured in the plane of the galaxy.  In each panel, the grey area represents the 2$\sigma$ dispersion of the sample.}
  \end{figure}

\begin{figure}[!htb]
    \includegraphics[width=\linewidth]{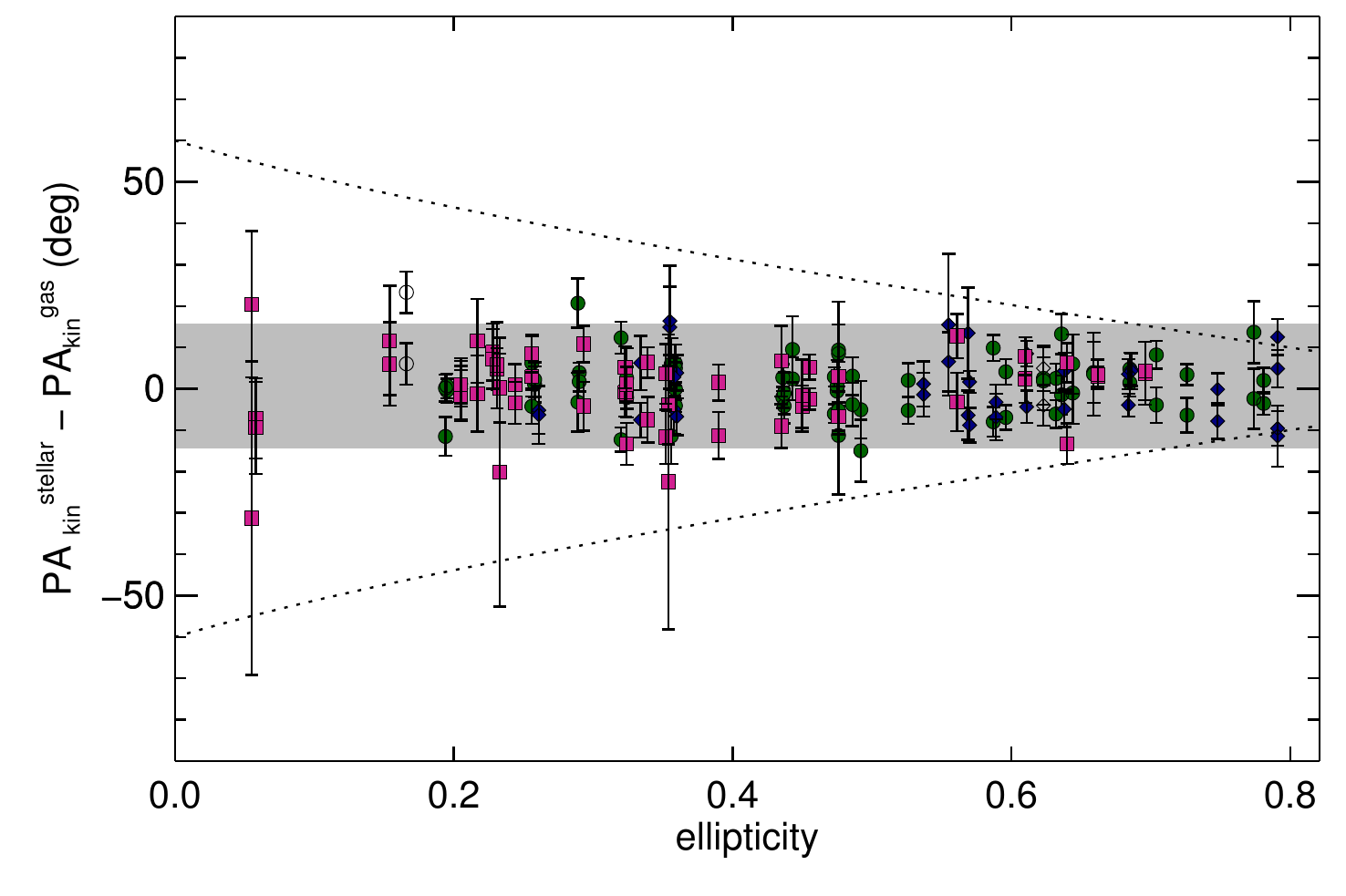}
    \caption{\label{PA_kin_e1} Stellar and ionised gas difference respect the ellipticity. the green-filled circles represent the non-barred galaxies, blue diamonds correspond to weakly-barred galaxies and violet squares represent the barred galaxies. Empty circles represent the early-type galaxies. Dashed lines represent the projection effect of a morpho-kinematic misalignment of 60$^{\circ}$ measured in the plane of the galaxy. In each panel, the grey area represents the 2$\sigma$ dispersion of the sample.}
  \end{figure}

In Figure \ref{PA_morph_e} we plot the morpho-kinematic misalignments explained in section \ref{sec:mk_mis} against the ellipticity for both components as well as the internal kinematic misalignment in each component ( Fig.~\ref{PA_kin_e}) and the comparison of the kinematic PA between the stars and the gas ( Fig.~\ref{PA_kin_e1}).  We find large morpho-kinematic misalignments in galaxies with low ellipticities ($\epsilon$\,<\,0.3, see Fig.~\ref{PA_morph_e}).
 
For the kinematic parameters independent of the morphology (i.e., Figs.~\ref{PA_kin_e} and ~\ref{PA_kin_e1}), we find rather similar values at different ellipticity bins. In each panel of these figures we plot an estimation of the projections effects as function of the ellipticity assuming a face-on misalignment of 60$^{\circ}$ (see Fig.~\ref{PA_morph_e}) in any of the (morpho-) kinematic indicators. This value was chosen to  approximately match the  misalignments found at large ellipticities. Within uncertainties all the galaxies  display internal kinematic misalignments smaller as the ones expected for their inclinations. In other words for these misalignments we do not find large scatters at low ellipticities as expected for a quantity heavily affected by projection effects.

\clearpage
\newpage

\section{Tables}
 \input{tabletex_morph_prop.txt}

 \input{tabletex_Skin_prop.txt}


 \input{tabletex_Gkin_prop.txt}


 \input{tabletex_kin_prop_bar.txt}

\clearpage
\newpage

\section{Stellar and Ionized velocity fields}
\label{sec:maps}
In this appendix we present the velocity field of the 80 non-interacting galaxies.
Each row present one galaxy similar as in Fig.\ref{kin_example}.
From left to right: SDSS r-band image of the galaxy. White solid line represents the photometric PA measured at the same galactocentric distance as the kinematic PA ($\mathrm{PA_{morph}}$, see section \ref{sec:Robust_Kinematic}). Dashed line represents the photometric PA measured at the outermost isophote of the image (PA$\mathrm{_{morph}^{out}}$, see section \ref{sec:Impact}).
Next two panels show the stellar and ionised gas velocity maps, respectively. Green points highlight the locations where the maximum velocity is located at given radius, determined from the position-velocity diagram. Black lines for each kinematic side represent the average kinematic PA (PA$_{\mathrm{kin}}$), while white thin lines along the zero-velocity curve show the average minor kinematic PA. Next panel  shows the distance from the galactic centre versus the maximum for the stellar (blue) and ionised gas (red) components. The curve along  0\,km\,s$^{-1}$ represent the velocities along the zero-velocity curve. Uncertainty in velocity are determined from Monte Carlo simulations. Last panel shows the dependence of the different PA with respect to the radius. $\mathrm{PA_{morph}}$,  $\mathrm{PA_{kin}^{stellar}}$  and  $\mathrm{PA_{kin}^{gas}}$  are represented by open  circles, filled blue diamonds and filled red squares, respectively. For barred galaxies we highlight with dashed lines the length of the bar and its orientation.

 \begin{figure*}[!htb]
 \includegraphics[width=\linewidth]{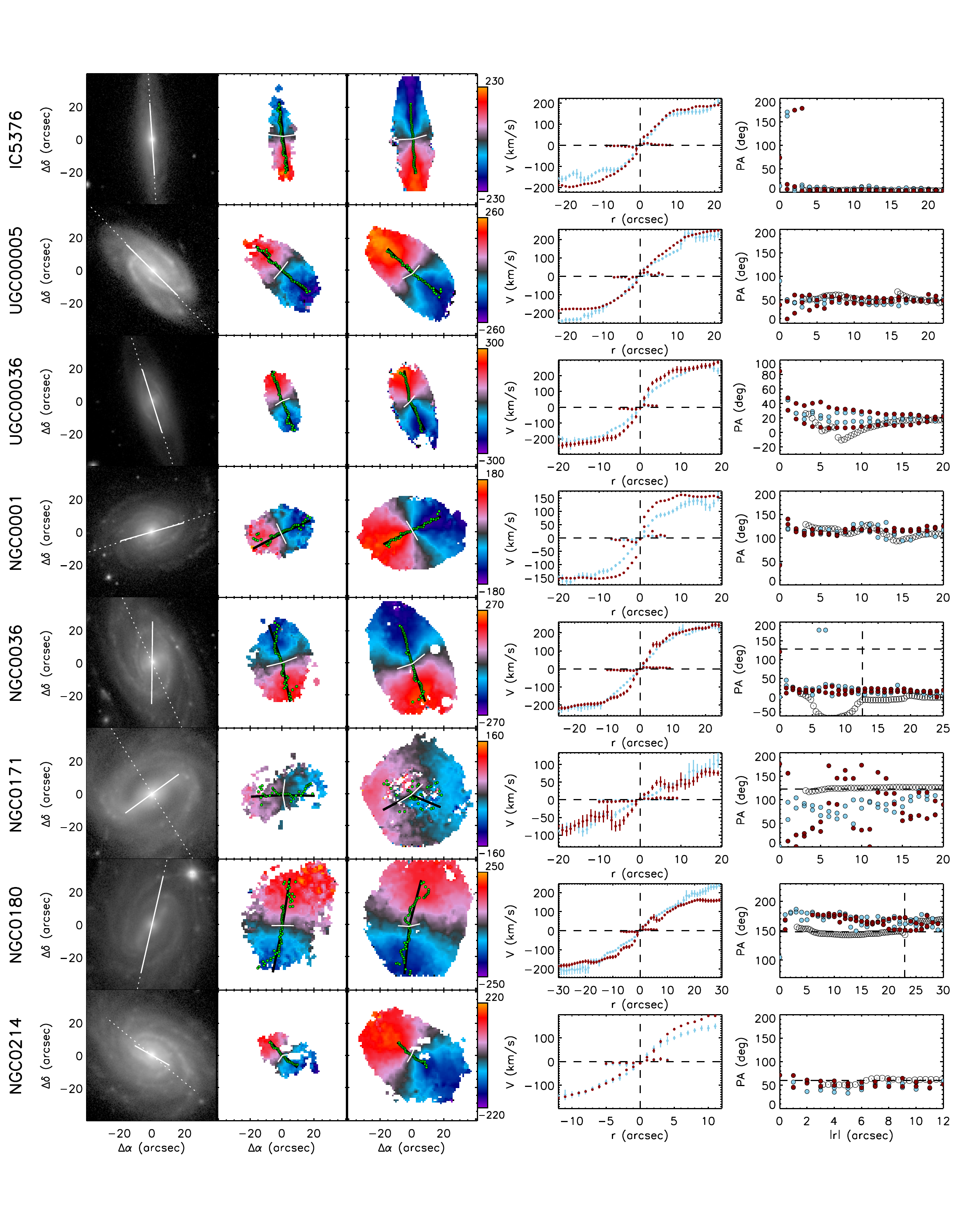}
  \caption{\label{app} Stellar and ionised gas velocity fields for the sample of galaxies used in this study.}
  \end{figure*}

 \begin{figure*}[!htb]
  \addtocounter{figure}{-1}
 \includegraphics[width=\linewidth]{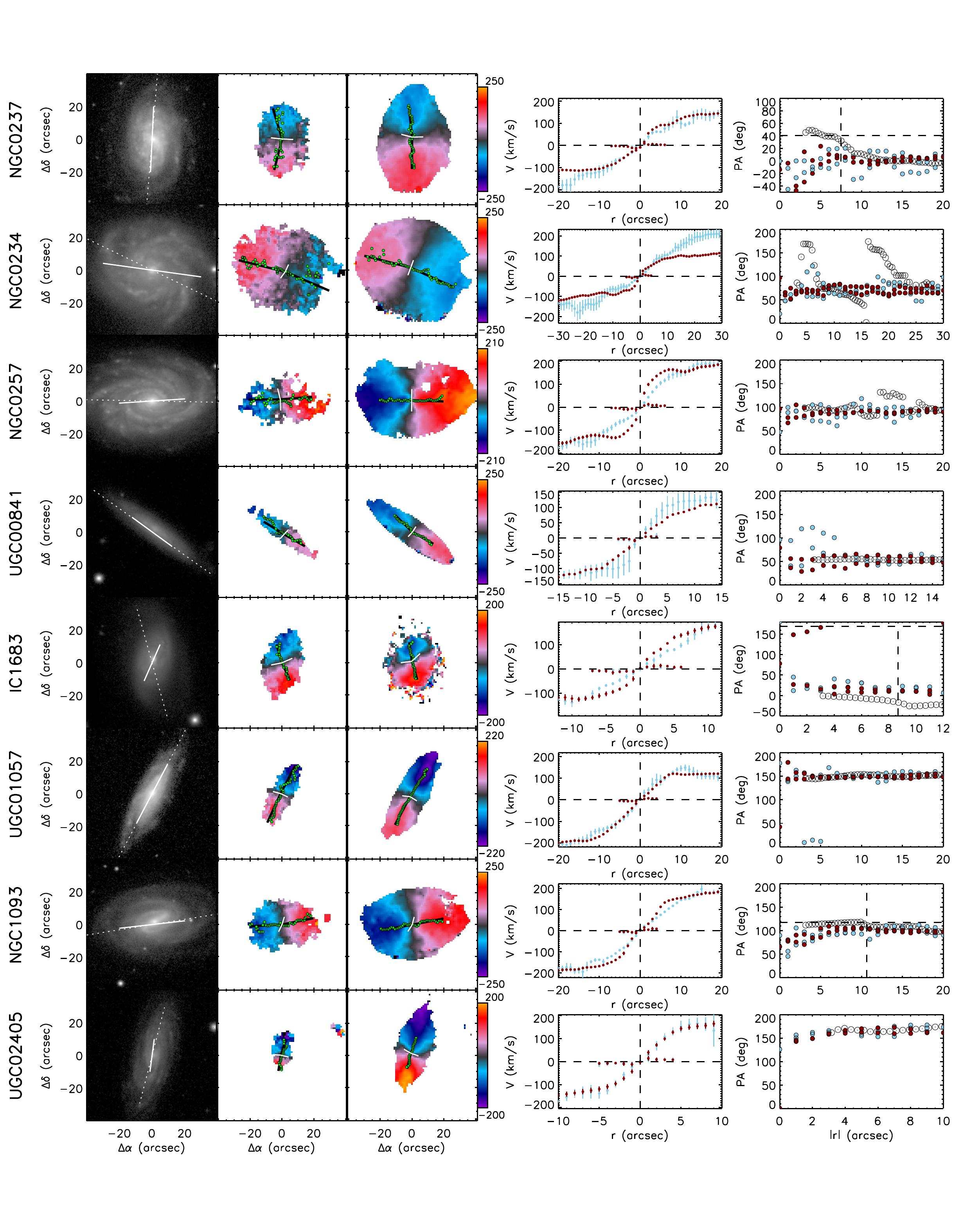}
  \caption{\label{app1} - \textit{continued}}
  \end{figure*}
  
 \begin{figure*}[!htb]
   \addtocounter{figure}{-1}
 \includegraphics[width=\linewidth]{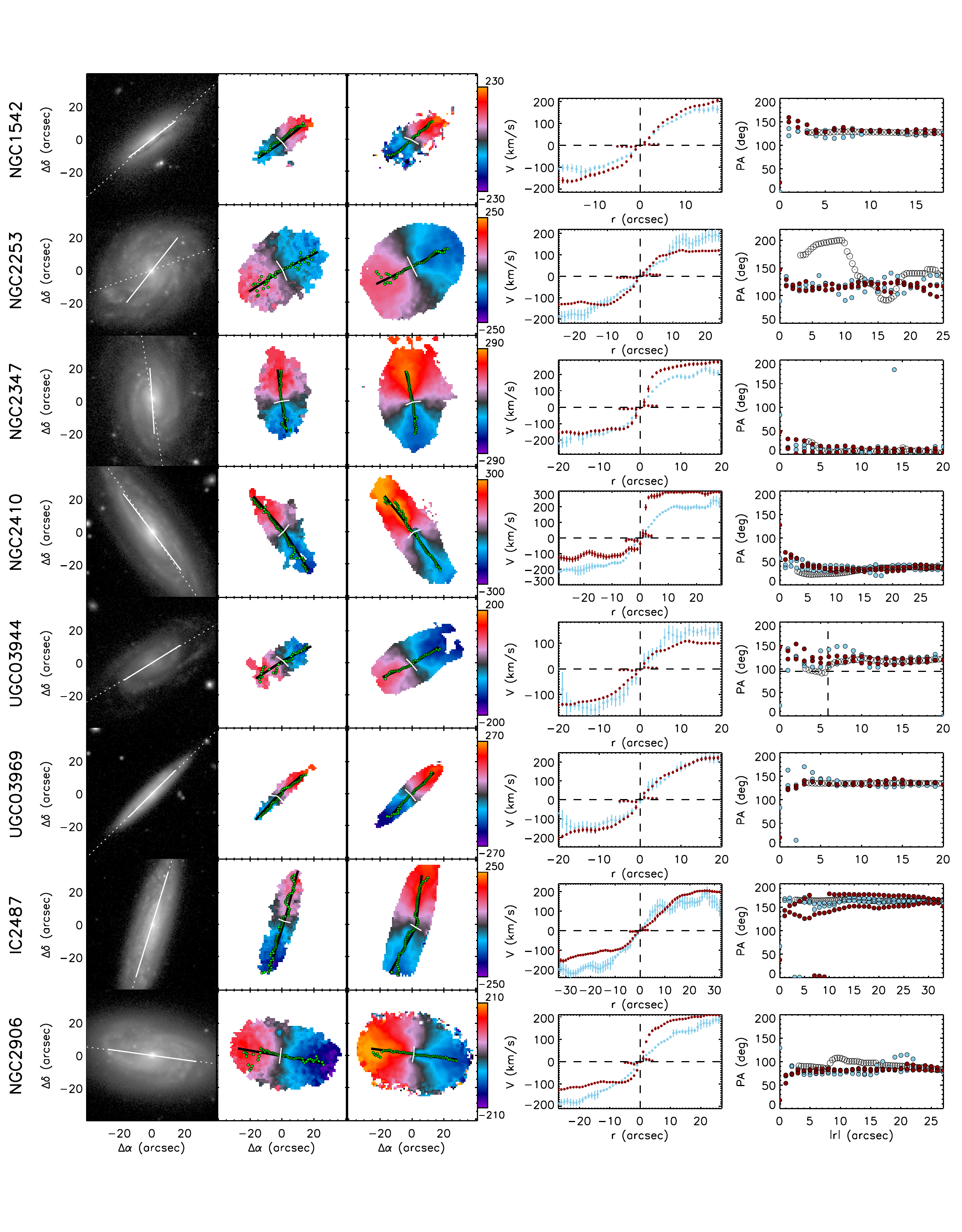}
 \caption{\label{app2} - \textit{continued}}
 \end{figure*}
 
 \begin{figure*}[!htb]
 \includegraphics[width=\linewidth]{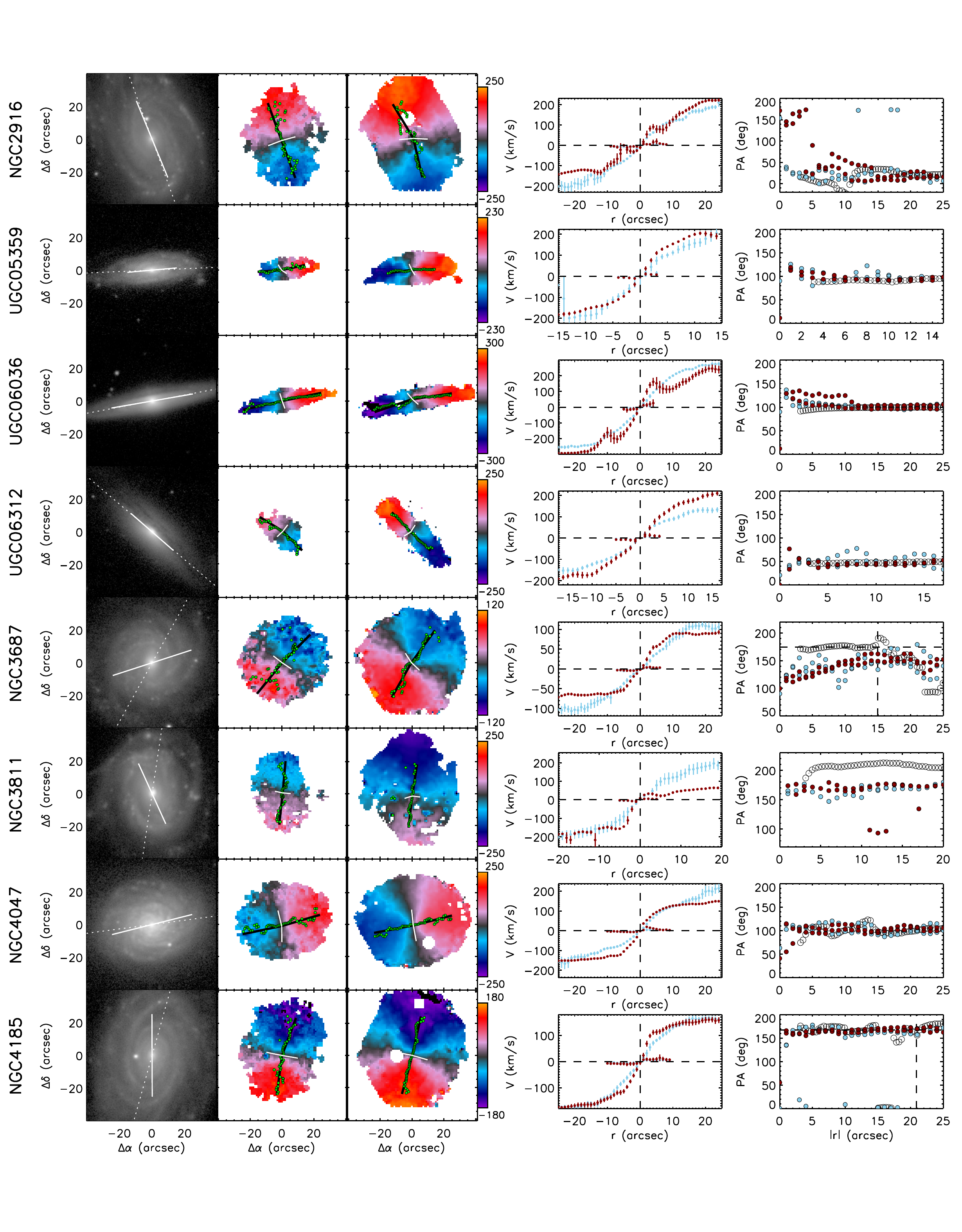}
  \addtocounter{figure}{-1}
  \caption{\label{app3} - \textit{continued}}
  \end{figure*}

  \begin{figure*}[!htb]
  \includegraphics[width=\linewidth]{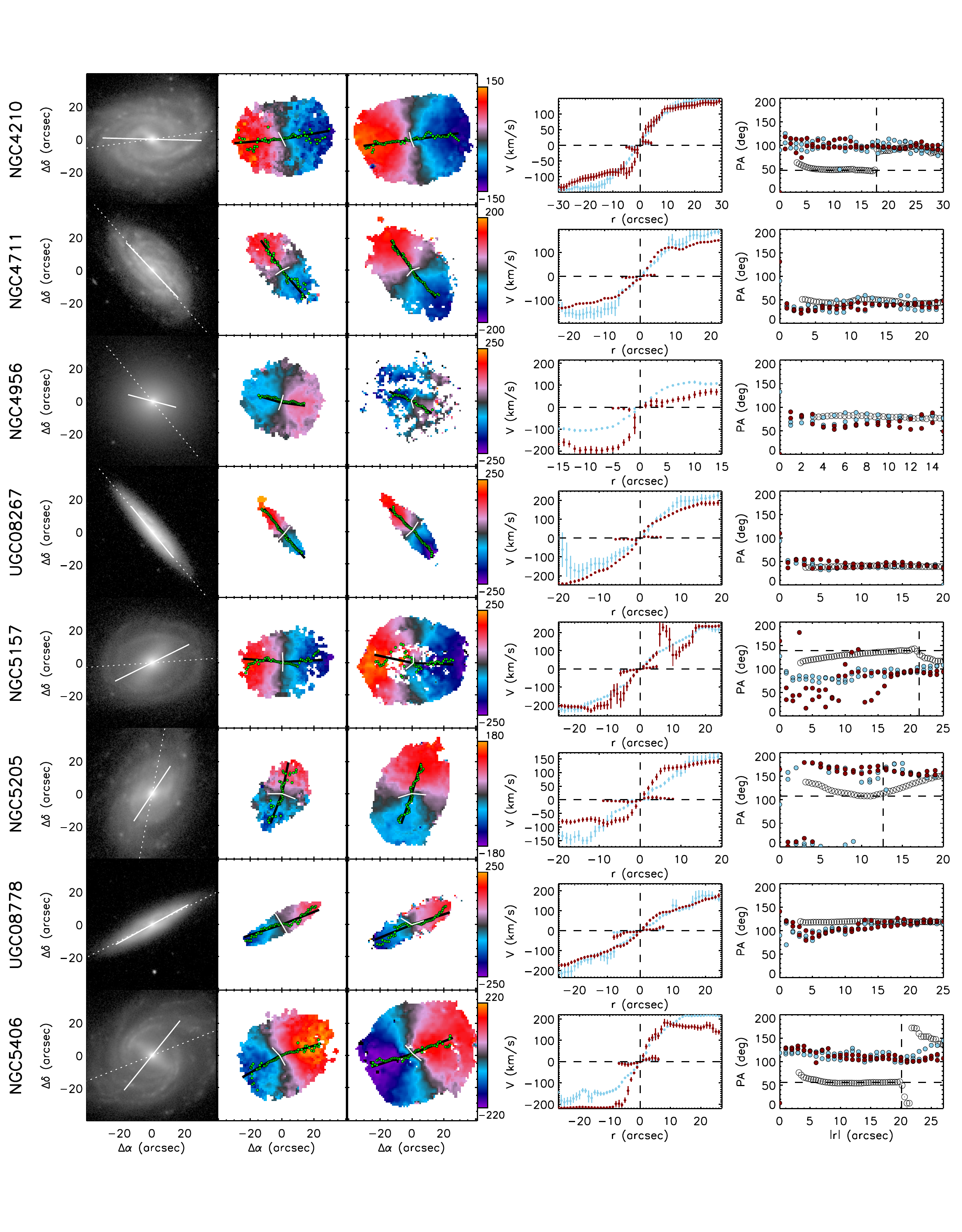}
  \addtocounter{figure}{-1}
  \caption{\label{app4} - \textit{continued}}
  \end{figure*}

  \begin{figure*}[!htb]
  \includegraphics[width=\linewidth]{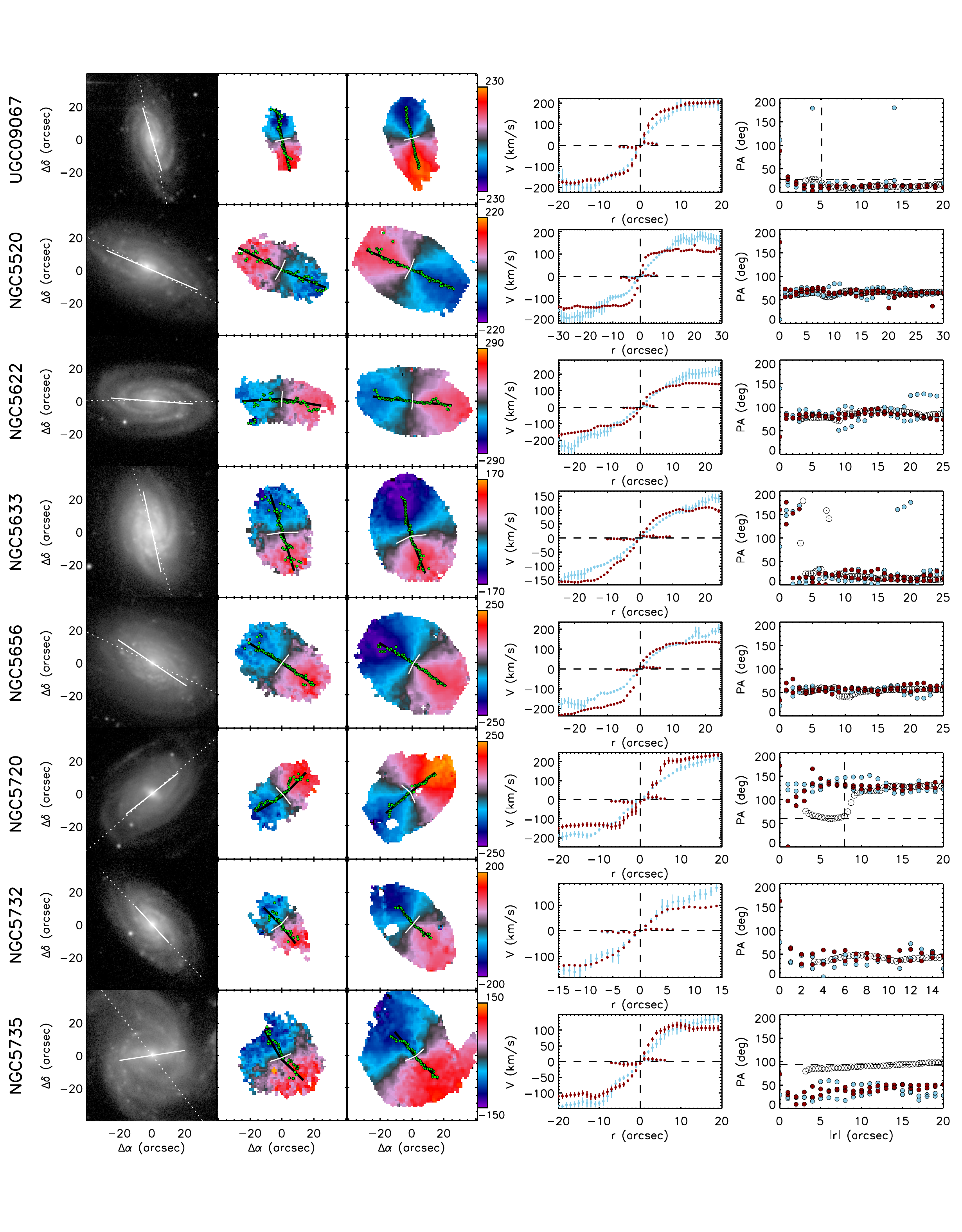}
  \addtocounter{figure}{-1}
  \caption{\label{app5} - \textit{continued}}
  \end{figure*}
 
  \begin{figure*}[!htb]
  \includegraphics[width=\linewidth]{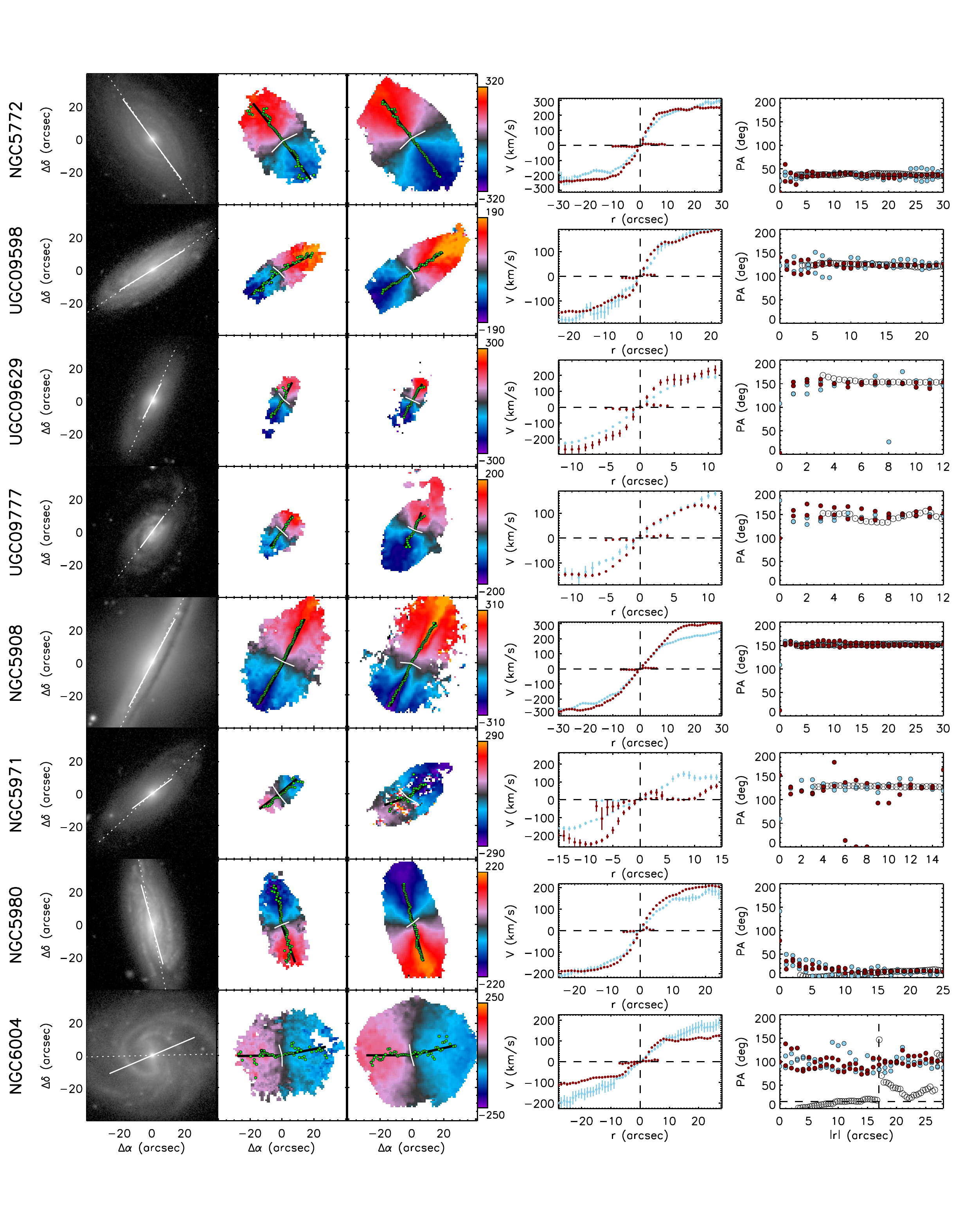}
  \addtocounter{figure}{-1}
  \caption{\label{app6} - \textit{continued}}
  \end{figure*}
 
  \begin{figure*}[!htb]
  \includegraphics[width=\linewidth]{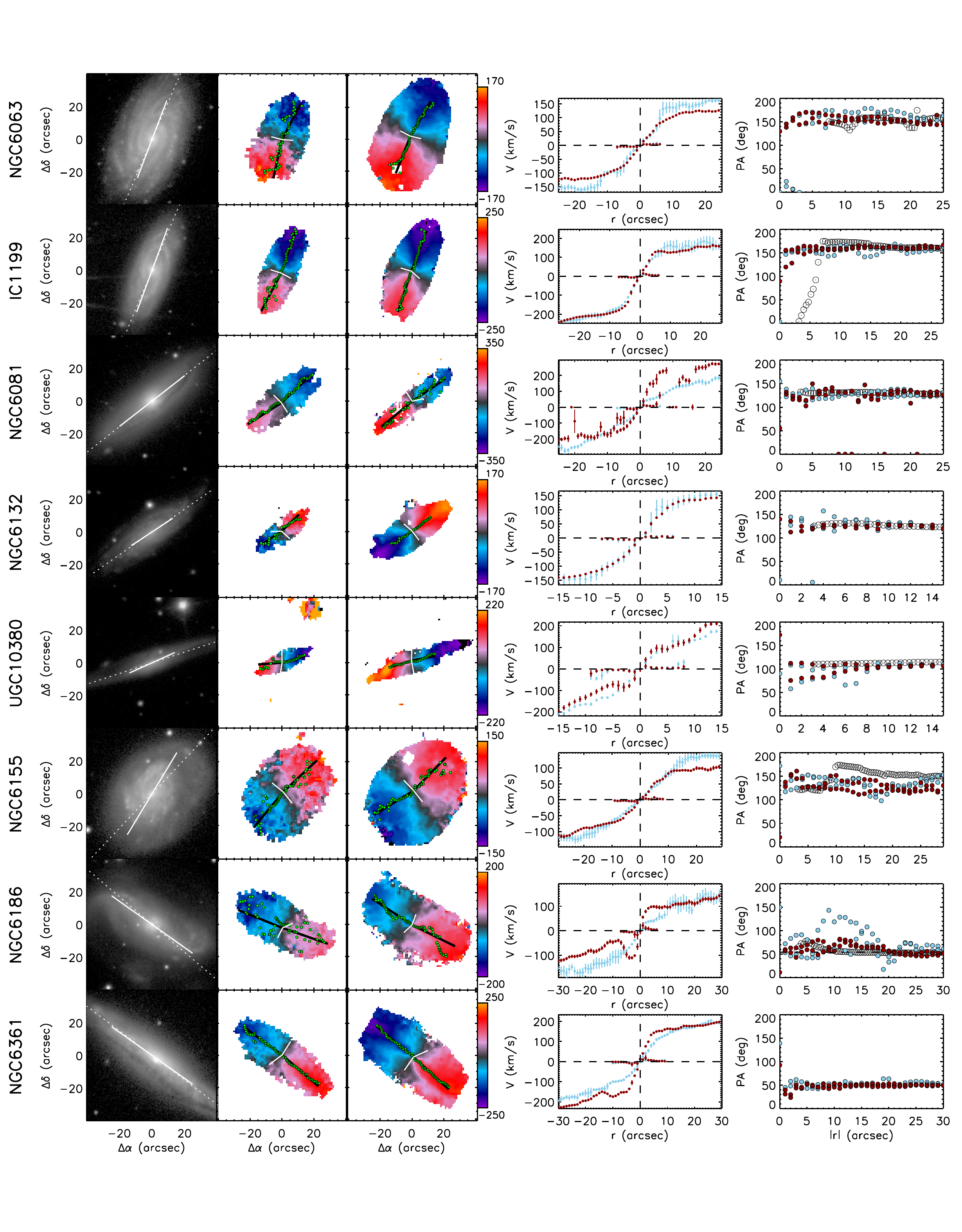}
  \addtocounter{figure}{-1}
  \caption{\label{app7} - \textit{continued}}
  \end{figure*}
 
  \begin{figure*}[!htb]
  \includegraphics[width=\linewidth]{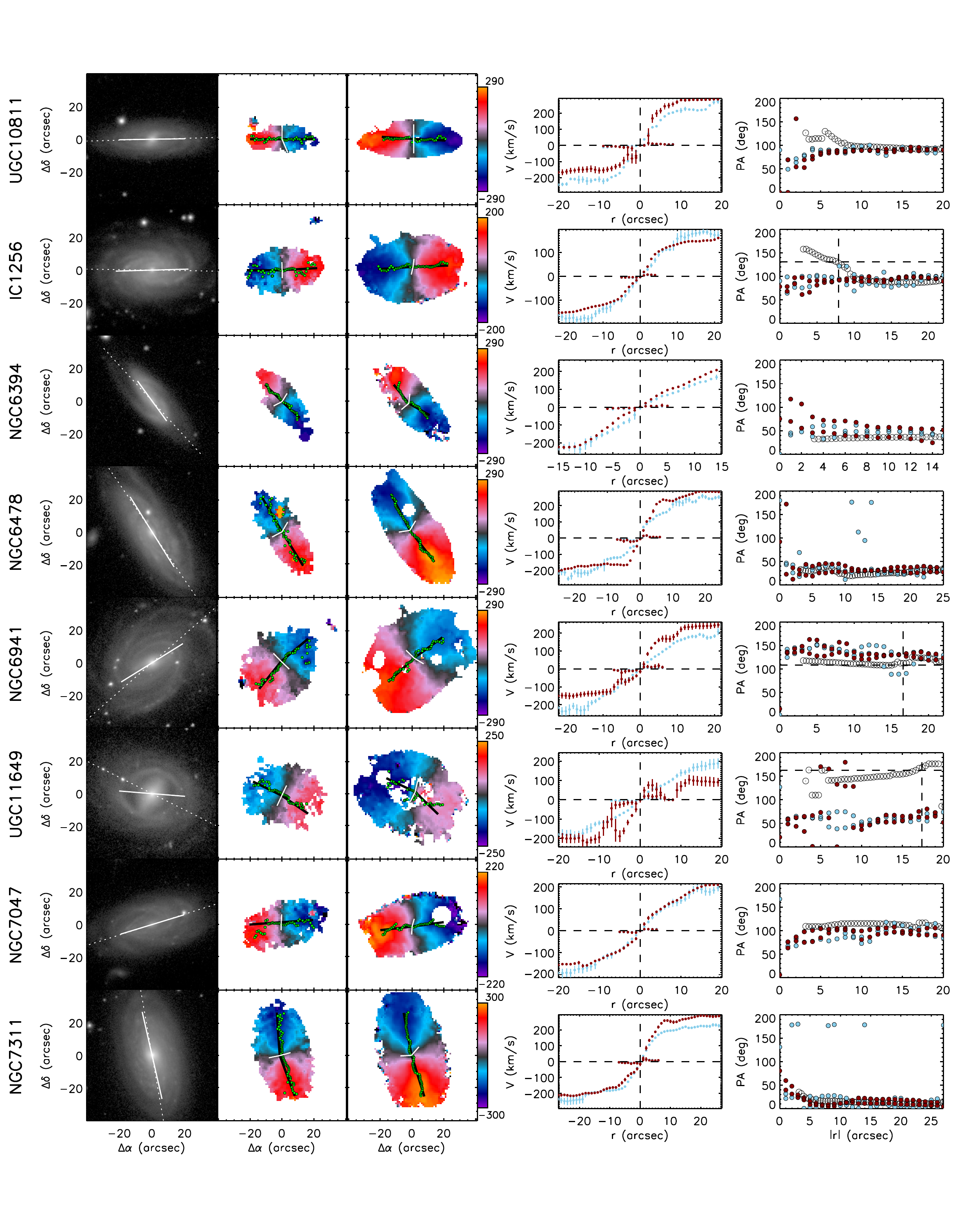}
  \addtocounter{figure}{-1}
  \caption{\label{app8} - \textit{continued}}
  \end{figure*}
 
 \begin{figure*}[!htb]
 \includegraphics[width=\linewidth]{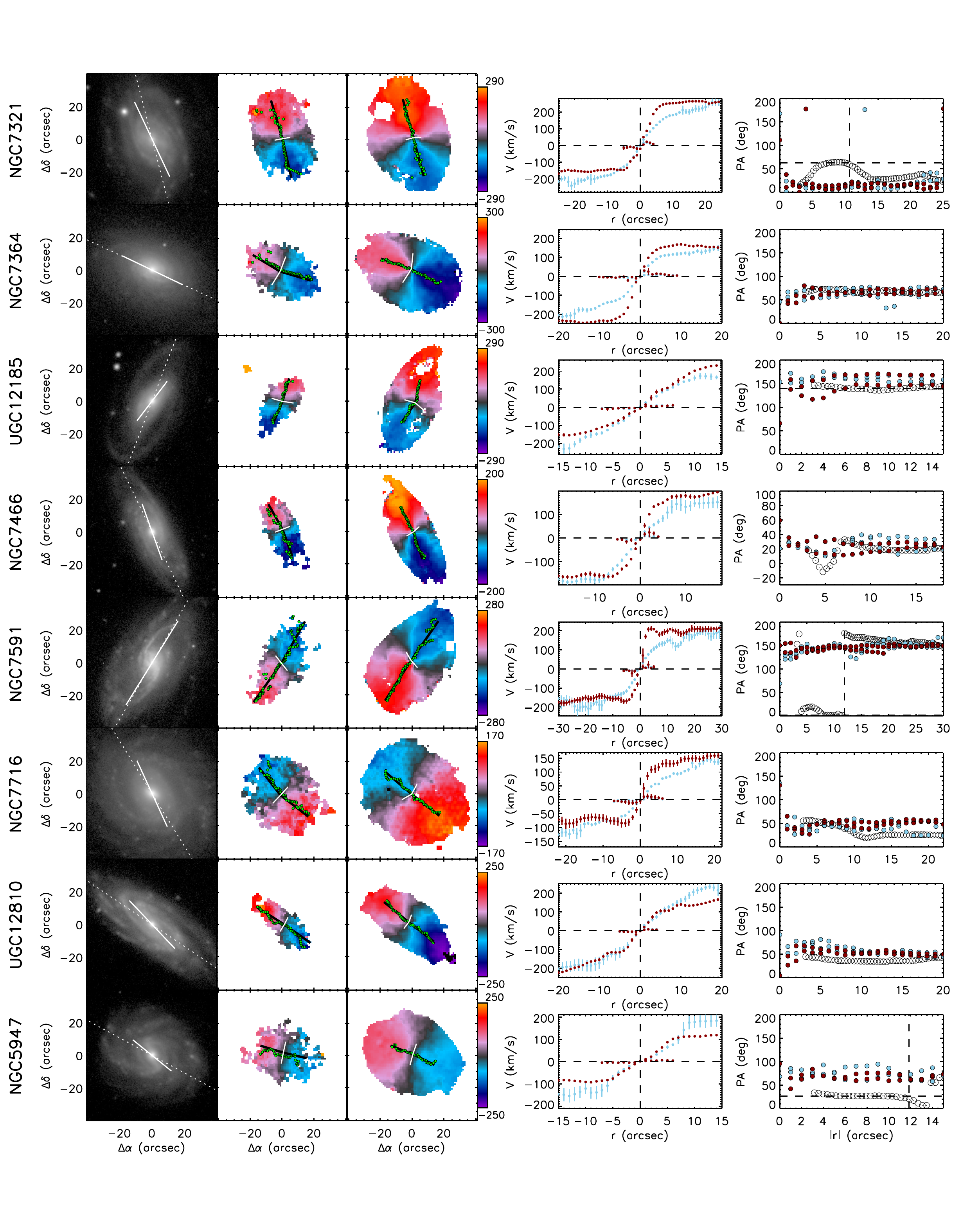}
  \addtocounter{figure}{-1}
  \caption{\label{app9} - \textit{continued}}
  \end{figure*}

\end{appendix}

\end{document}

%% file: tabletex_morph_prop.txt
\begin{table*}[!htb]
\caption{\label{table_morph} Morphological parameters of the sample of non-interacting CALIFA galaxies used in this study.
} 
\begin{center}
\renewcommand{\thefootnote}{\alph{footnote}}
\begin{tabular} {c c c c c c c c c c}
\toprule
 CALIFA id & Name & Morphological&     Bar     & $\epsilon$   & PA$\mathrm{_{morph}^{out}}$ & PA$\mathrm{_{bar}}$ & r$\mathrm{_{bar}}$ & Stellar mass       \\
           &      &     type     &  strength   &              &           ($^{\circ}$)      &   ($^{\circ}$)      &        (arcsec)    & log(M/M$_{\odot}$) \\
    (1)    &  (2) &     (3)      &     (4)     &     (5)      &           (6)               &   (7)               &        (8)         & (9) \\
\midrule
  1 & \object{ IC5376 } & Sb & A & 0.69 & 50 & ... & ... &  10.57 \\
  2 & \object{ UGC00005 } & Sbc & A & 0.53 & 50 & ... & ... &  10.72 \\
  7 & \object{ UGC00036 } & Sab & AB & 0.61 & 50 & ... & ... &  10.88 \\
  8 & \object{ NGC0001 } & Sbc & A & 0.32 & 50 & ... & ... &  10.71 \\
 10 & \object{ NGC0036 } & Sb & B & 0.48 & 50 & 128 &  12 &  10.79 \\
 23 & \object{ NGC0171 } & Sb & B & 0.05 & 50 & 122 &  33 &  10.62 \\
 25 & \object{ NGC0180 } & Sb & B & 0.34 & 50 & 147 &  22 &  10.78 \\
 28 & \object{ NGC0214 } & Sbc & AB & 0.26 & 50 &  59 &  14 &  10.74 \\
 30 & \object{ NGC0237 } & Sc & B & 0.32 & 50 &  40 &   7 &  10.17 \\
 31 & \object{ NGC0234 } & Sc & AB & 0.20 & 50 & ... & ... &  10.55 \\
 33 & \object{ NGC0257 } & Sc & A & 0.36 & 50 & ... & ... &  10.69 \\
 41 & \object{ UGC00841 } & Sbc & A & 0.77 & 50 & ... & ... &   9.71 \\
 43 & \object{ IC1683 } & Sb & AB & 0.35 & 50 & 168 &   8 &  10.43 \\
 53 & \object{ UGC01057 } & Sc & AB & 0.69 & 50 & ... & ... &  10.11 \\
108 & \object{ NGC1093 } & Sbc & B & 0.39 & 50 & 117 &  10 &  10.66 \\
116 & \object{ UGC02405 } & Sbc & A & 0.66 & 50 & ... & ... &  10.40 \\
131 & \object{ NGC1542 } & Sab & AB & 0.59 & 50 & ... & ... &  10.30 \\
147 & \object{ NGC2253 } & Sbc & B & 0.32 & 50 & ... & ... &  10.37 \\
149 & \object{ NGC2347 } & Sbc & AB & 0.36 & 50 & ... & ... &  10.62 \\
151 & \object{ NGC2410 } & Sb & AB & 0.68 & 50 & ... & ... &  10.78 \\
152 & \object{ UGC03944 } & Sbc & AB & 0.57 & 50 &  94 &   5 &   9.87 \\
153 & \object{ UGC03969 } & Sb & A & 0.78 & 50 & ... & ... &  10.61 \\
273 & \object{ IC2487 } & Sc & AB & 0.79 & 50 & ... & ... &  10.29 \\
275 & \object{ NGC2906 } & Sbc & A & 0.44 & 50 & ... & ... &  10.29 \\
277 & \object{ NGC2916 } & Sbc & A & 0.36 & 50 & ... & ... &  10.65 \\
307 & \object{ UGC05359 } & Sb & B & 0.70 & 50 & ... & ... &  10.54 \\
364 & \object{ UGC06036 } & Sa & A & 0.73 & 50 & ... & ... &  11.09 \\
386 & \object{ UGC06312 } & Sab & A & 0.64 & 50 & ... & ... &  10.93 \\
414 & \object{ NGC3687 } & Sb & B & 0.06 & 50 & 174 &  14 &  10.17 \\
436 & \object{ NGC3811 } & Sbc & B & 0.23 & 50 &  22 &  24 &  10.34 \\
489 & \object{ NGC4047 } & Sbc & A & 0.26 & 50 & ... & ... &  10.58 \\
515 & \object{ NGC4185 } & Sbc & AB & 0.33 & 50 & 167 &  20 &  10.61 \\
518 & \object{ NGC4210 } & Sb & B & 0.24 & 50 &  47 &  17 &  10.17 \\
580 & \object{ NGC4711 } & Sbc & A & 0.47 & 50 & ... & ... &  10.20 \\
602 & \object{ NGC4956 } & E1 & A & 0.17 & 50 & ... & ... &  10.99 \\
610 & \object{ UGC08267 } & Sb & AB & 0.75 & 50 & ... & ... &  10.68 \\
624 & \object{ NGC5157 } & Sab & B & 0.23 & 50 & 138 &  21 &  11.15 \\
630 & \object{ NGC5205 } & Sbc & B & 0.35 & 50 & 108 &  12 &   9.77 \\
664 & \object{ UGC08778 } & Sb & A & 0.70 & 50 & ... & ... &  10.15 \\
684 & \object{ NGC5406 } & Sb & B & 0.29 & 50 &  56 &  20 &  11.09 \\
714 & \object{ UGC09067 } & Sbc & AB & 0.54 & 50 &  27 &   5 &  10.36 \\
715 & \object{ NGC5520 } & Sbc & A & 0.49 & 50 & ... & ... &   9.75 \\
743 & \object{ NGC5622 } & Sbc & A & 0.48 & 50 & ... & ... &  10.11 \\
748 & \object{ NGC5633 } & Sbc & A & 0.26 & 50 & ... & ... &  10.16 \\
753 & \object{ NGC5656 } & Sb & A & 0.19 & 50 & ... & ... &  10.51 \\
764 & \object{ NGC5720 } & Sbc & B & 0.44 & 50 &  60 &   7 &  10.74 \\
768 & \object{ NGC5732 } & Sbc & A & 0.48 & 50 & ... & ... &   9.83 \\
771 & \object{ NGC5735 } & Sbc & B & 0.35 & 50 &  94 &  20 &  10.28 \\
777 & \object{ NGC5772 } & Sab & A & 0.44 & 50 & ... & ... &  10.84 \\
779 & \object{ UGC09598 } & Sbc & AB & 0.64 & 50 & ... & ... &  10.45 \\
782 & \object{ UGC09629 } & E7 & AB & 0.62 & 50 & ... & ... &  11.02 \\
\bottomrule
\end{tabular}
\tablefoot{
(1) CALIFA ID number.
(2) name of the galaxy.
(3) morphological type from visual clasification (see Walcher et. al in preparation for details).
(4) bar strength of the galaxy from the same morphological visual clasification (see Walcher et. al. in preparation for details).
(5) and (6) ellipticity ($\epsilon$) and position angle  (PA$\mathrm{_{morph}^{out}}$) from an ellipse fitting at the largest scale isophote of the SDSS r-band image. Both measurements were inferred using the IRAF task \textit{ellipse}.
(7) and (8) position angle (PA$\mathrm{_{bar}}$) and length of the bar (r$\mathrm{_{bar}}$) derived from an ellipse fitting on the SDSS r-band image.
(9) stellar masses (see Walcher et. al in preparation for details).
}
\end{center}
\begin{minipage}{\textwidth}
\end{minipage}
\end{table*} 
   
\begin{table*}[!htb]
\addtocounter{table}{-1}
\caption{continue} \label{table_kin_stellar_int}
\begin{center}
\renewcommand{\thefootnote}{\alph{footnote}}
\begin{tabular} {c c c c c c c c c c}
\toprule
 CALIFA id & Name & Morphological & Barredness & $\epsilon$   & PA$\mathrm{_{morph}^{out}}$ & PA$\mathrm{_{bar}}$ & r$\mathrm{_{bar}}$ & Stellar mass       \\
           &      &     type     &             &              &           ($^{\circ}$)      &   ($^{\circ}$)      &        (arcsec)    & log(M/M$_{\odot}$) \\
    (1)    &  (2) &     (3)      &     (4)     &     (5)      &           (6)               &   (7)               &        (8)         & (9) \\
\midrule
790 & \object{ UGC09777 } & Sbc & A & 0.49 & 50 & ... & ... &  10.21 \\
791 & \object{ NGC5908 } & Sa & A & 0.36 & 50 & ... & ... &  11.12 \\
804 & \object{ NGC5971 } & Sb & AB & 0.56 & 50 & ... & ... &  10.22 \\
810 & \object{ NGC5980 } & Sbc & A & 0.60 & 50 & ... & ... &  10.62 \\
813 & \object{ NGC6004 } & Sbc & B & 0.20 & 50 &  15 &  16 &  10.60 \\
823 & \object{ NGC6063 } & Sbc & A & 0.44 & 50 & ... & ... &   9.95 \\
824 & \object{ IC1199 } & Sb & AB & 0.57 & 50 & ... & ... &  10.20 \\
826 & \object{ NGC6081 } & S0a & A & 0.59 & 50 & ... & ... &  11.04 \\
831 & \object{ NGC6132 } & Sbc & A & 0.64 & 50 & ... & ... &  10.15 \\
834 & \object{ UGC10380 } & Sb & AB & 0.79 & 50 & ... & ... &  10.92 \\
836 & \object{ NGC6155 } & Sc & A & 0.29 & 50 & ... & ... &  10.01 \\
842 & \object{ NGC6186 } & Sb & B & 0.23 & 50 &  54 &  35 &  10.44 \\
853 & \object{ NGC6361 } & Sab & A & 0.29 & 50 & ... & ... &  10.81 \\
854 & \object{ UGC10811 } & Sb & B & 0.66 & 50 & ... & ... &  10.75 \\
856 & \object{ IC1256 } & Sb & AB & 0.36 & 50 & 130 &   7 &  10.20 \\
857 & \object{ NGC6394 } & Sbc & B & 0.64 & 50 & ... & ... &  10.79 \\
862 & \object{ NGC6478 } & Sc & A & 0.63 & 50 & ... & ... &  10.90 \\
869 & \object{ NGC6941 } & Sb & B & 0.26 & 50 & 108 &  16 &  10.82 \\
872 & \object{ UGC11649 } & Sab & B & 0.22 & 50 & 162 &  17 &  10.54 \\
876 & \object{ NGC7047 } & Sbc & B & 0.45 & 50 & 115 & 20 &  10.68 \\
886 & \object{ NGC7311 } & Sa & A & 0.47 & 50 & ... & ... &  10.84 \\
887 & \object{ NGC7321 } & Sbc & B & 0.32 & 50 &  62 &  10 &  10.69 \\
889 & \object{ NGC7364 } & Sab & A & 0.32 & 50 & ... & ... &  10.79 \\
890 & \object{ UGC12185 } & Sb & B & 0.56 & 50 & 139 &  22 &  10.58 \\
896 & \object{ NGC7466 } & Sbc & A & 0.62 & 50 & ... & ... &  10.70 \\
904 & \object{ NGC7591 } & Sbc & B & 0.46 & 50 &   1 &  11 &  10.64 \\
924 & \object{ NGC7716 } & Sb & A & 0.19 & 50 & ... & ... &  10.20 \\
929 & \object{ UGC12810 } & Sbc & B & 0.61 & 50 & ... & ... &  10.61 \\
938 & \object{ NGC5947 } & Sbc & B & 0.15 & 50 &  26 &  11 &  10.23 \\
\bottomrule
\end{tabular}
\end{center}
\begin{minipage}{\textwidth}
\end{minipage}
\end{table*} 
\newpage

%% file: tabletex_Skin_prop.txt
\begin{table*}[!htb]
\tiny
\caption{\label{table_Skin}
 Stellar Kinematic properties of the non-interacting sample selected for this study included in the CALIFA survey.
}
\begin{center}
\renewcommand{\thefootnote}{\alph{footnote}}
\begin{tabular} {c c c c c c c c c c c}
\toprule
    id  &$\Delta\alpha$  & $\Delta\delta$ & V$\mathrm{_{sys}}$  &r$\mathrm{_{max}}$  &  r$\mathrm{_{min}}$  & PA$\mathrm{_{morph}}$ (r$\mathrm{_{max}}$)    &  \multicolumn{2}{c}{PA approaching}                    & \multicolumn{2}{c}{PA receding}                                  \\
        &                &                &                     &                    &                      &                                               &  PA$\mathrm{_{kin}}$  & $\delta$PA$\mathrm{_{kin}}$ & PA$\mathrm{_{kin}}$  & $\delta$PA$\mathrm{_{kin}}$ \\
        &   (arcsec)     &      (arcsec)  &    (km s$^{-1}$)    &      (arcsec)      &      (arcsec)        &           ($^{\circ}$)                        &  ($^{\circ}$)         & ($^{\circ}$)                & ($^{\circ}$)         & ($^{\circ}$)                \\
    (1) &      (2)       &       (3)      &         (4)         &        (5)         &          (6)         &           (7)                                 &          (8)          &            (9)              &          (10)         &       (11)                   \\
\midrule
  1 & \phantom{-}0.6 $\pm$ 0.6 & \phantom{-}1.7 $\pm$ 0.5 &  4963 $\pm$    9 & 22 & -- & \phantom{00}3.7 $\pm$ 0.4 & \phantom{00}6.0 $\pm$  3.3 & \phantom{00}7.1 $\pm$  2.2 & \phantom{00}8.2 $\pm$  3.2 & \phantom{00}8.0 $\pm$  1.8 \\
  2 & ... & ... &  7209 $\pm$    1 & 22 & -- & \phantom{0}45.2 $\pm$ 0.9 & \phantom{0}49.9 $\pm$  2.7 & \phantom{0}10.5 $\pm$  2.7 & \phantom{0}45.0 $\pm$  0.2 & \phantom{0}15.2 $\pm$  2.7 \\
  7 & ... & ... &  6238 $\pm$    1 & 20 & -- & \phantom{0}17.4 $\pm$ 0.2 & \phantom{0}22.7 $\pm$  2.8 & \phantom{0}10.7 $\pm$  2.3 & \phantom{0}18.8 $\pm$  1.6 & \phantom{00}7.3 $\pm$  1.8 \\
  8 & -0.6 $\pm$ 0.5 & -1.1 $\pm$ 0.6 &  4515 $\pm$    9 & 20 &  7 & 106.0 $\pm$ 1.1 & 114.6 $\pm$  4.2 & \phantom{0}15.0 $\pm$  4.9 & 118.2 $\pm$  6.2 & \phantom{0}20.1 $\pm$  3.7 \\
 10 & ... & ... &  5939 $\pm$    1 & 25 & 10 & 179.3 $\pm$ 0.6 & \phantom{0}19.4 $\pm$  2.7 & \phantom{0}14.4 $\pm$  2.1 & \phantom{0}12.0 $\pm$  2.3 & \phantom{0}12.7 $\pm$  2.4 \\
 23 & \phantom{-}0.4 $\pm$ 0.5 & -1.0 $\pm$ 0.6 &  3857 $\pm$    1 & 20 & 10 & 126.0 $\pm$ 0.7 & \phantom{0}88.8 $\pm$  7.2 & \phantom{0}29.2 $\pm$  7.6 & \phantom{0}92.0 $\pm$  4.9 & \phantom{0}37.9 $\pm$ 10.2 \\
 25 & \phantom{-}0.4 $\pm$ 0.5 & -0.7 $\pm$ 0.4 &  5204 $\pm$    6 & 30 &  7 & 167.1 $\pm$ 1.0 & 168.1 $\pm$  4.0 & \phantom{0}18.7 $\pm$  3.0 & 172.0 $\pm$  3.8 & \phantom{0}18.0 $\pm$  3.4 \\
 28 & \phantom{-}0.2 $\pm$ 0.5 & \phantom{-}0.2 $\pm$ 0.7 &  4500 $\pm$   12 & 12 &  5 & \phantom{0}60.4 $\pm$ 0.8 & \phantom{0}51.3 $\pm$  5.4 & \phantom{0}12.7 $\pm$  4.7 & \phantom{0}40.2 $\pm$  6.3 & \phantom{0}18.0 $\pm$  4.9 \\
 30 & ... & ... &  4142 $\pm$    1 & 20 &  7 & 177.3 $\pm$ 0.8 & \phantom{0}17.0 $\pm$  5.8 & \phantom{0}24.3 $\pm$  5.8 & 165.3 $\pm$  5.2 & \phantom{0}30.6 $\pm$  7.4 \\
 31 & \phantom{-}1.9 $\pm$ 0.7 & -1.1 $\pm$ 0.8 &  4391 $\pm$    6 & 30 &  5 & \phantom{0}82.2 $\pm$ 4.9 & \phantom{0}66.5 $\pm$  7.3 & \phantom{0}37.6 $\pm$  5.5 & \phantom{0}76.4 $\pm$  6.7 & \phantom{0}30.3 $\pm$  9.4 \\
 33 & -1.1 $\pm$ 0.7 & \phantom{-}1.2 $\pm$ 0.8 &  5183 $\pm$   12 & 20 &  7 & \phantom{0}93.7 $\pm$ 1.5 & \phantom{0}91.0 $\pm$  6.3 & \phantom{0}25.7 $\pm$  5.6 & \phantom{0}94.4 $\pm$  4.8 & \phantom{0}25.8 $\pm$  7.6 \\
 41 & \phantom{-}1.3 $\pm$ 0.5 & -1.4 $\pm$ 0.6 &  5524 $\pm$    9 & 15 & -- & \phantom{0}53.9 $\pm$ 0.2 & \phantom{0}58.6 $\pm$  7.1 & \phantom{0}37.8 $\pm$  9.9 & \phantom{0}60.3 $\pm$  5.1 & \phantom{0}25.4 $\pm$  7.5 \\
 43 & ... & ... &  4839 $\pm$    9 & 12 &  7 & 156.4 $\pm$ 1.0 & \phantom{0}19.0 $\pm$  9.7 & \phantom{0}20.4 $\pm$  8.1 & \phantom{0}31.8 $\pm$  7.2 & \phantom{0}19.4 $\pm$  6.4 \\
 53 & -0.8 $\pm$ 0.5 & \phantom{-}0.1 $\pm$ 0.6 &  6294 $\pm$   12 & 20 & -- & 152.2 $\pm$ 0.5 & 152.6 $\pm$  4.7 & \phantom{0}22.1 $\pm$  4.7 & 155.1 $\pm$  3.8 & \phantom{0}12.4 $\pm$  2.9 \\
108 & ... & ... &  5224 $\pm$    1 & 20 &  5 & \phantom{0}96.5 $\pm$ 1.2 & \phantom{0}92.9 $\pm$  3.0 & \phantom{0}13.0 $\pm$  5.0 & 101.4 $\pm$  1.8 & \phantom{0}13.7 $\pm$  4.2 \\
116 & -0.8 $\pm$ 0.4 & \phantom{-}0.1 $\pm$ 0.8 &  7676 $\pm$   21 & 10 & -- & 172.4 $\pm$ 1.4 & 164.5 $\pm$  6.3 & \phantom{0}20.8 $\pm$  7.8 & 171.7 $\pm$  7.2 & \phantom{0}16.7 $\pm$  7.1 \\
131 & \phantom{-}0.2 $\pm$ 0.5 & -1.8 $\pm$ 0.6 &  3689 $\pm$   10 & 18 & -- & 128.0 $\pm$ 0.4 & 123.4 $\pm$  3.6 & \phantom{0}11.7 $\pm$  2.6 & 129.2 $\pm$  2.8 & \phantom{0}11.0 $\pm$  2.7 \\
147 & ... & ... &  3549 $\pm$    0 & 25 &  7 & 141.7 $\pm$ 1.9 & 117.0 $\pm$  5.7 & \phantom{0}25.5 $\pm$  3.5 & 122.1 $\pm$  3.3 & \phantom{0}21.1 $\pm$  5.0 \\
149 & -0.3 $\pm$ 0.5 & -0.0 $\pm$ 0.4 &  4396 $\pm$   12 & 20 &  5 & \phantom{00}3.9 $\pm$ 0.8 & \phantom{00}8.5 $\pm$  4.3 & \phantom{0}13.0 $\pm$  3.1 & \phantom{00}7.2 $\pm$  4.0 & \phantom{0}10.0 $\pm$  3.7 \\
151 & \phantom{-}1.4 $\pm$ 0.5 & \phantom{-}0.5 $\pm$ 0.5 &  4662 $\pm$    9 & 29 & -- & \phantom{0}37.0 $\pm$ 0.3 & \phantom{0}32.4 $\pm$  2.5 & \phantom{00}9.1 $\pm$  1.8 & \phantom{0}41.7 $\pm$  2.3 & \phantom{00}7.9 $\pm$  1.4 \\
152 & \phantom{-}0.6 $\pm$ 0.6 & \phantom{-}1.3 $\pm$ 0.7 &  3881 $\pm$    9 & 20 & -- & 121.9 $\pm$ 0.5 & 116.9 $\pm$  5.9 & \phantom{0}21.7 $\pm$  5.3 & 125.2 $\pm$ 10.0 & \phantom{0}41.2 $\pm$ 14.8 \\
153 & -3.1 $\pm$ 0.7 & -1.6 $\pm$ 0.8 &  7984 $\pm$   18 & 20 & -- & 134.9 $\pm$ 0.3 & 137.7 $\pm$  2.2 & \phantom{0}11.2 $\pm$  3.0 & 130.2 $\pm$  2.4 & \phantom{00}9.8 $\pm$  3.8 \\
273 & \phantom{-}1.3 $\pm$ 0.7 & \phantom{-}1.9 $\pm$ 0.8 &  4326 $\pm$   11 & 33 & -- & 162.9 $\pm$ 0.2 & 162.6 $\pm$  2.7 & \phantom{0}12.6 $\pm$  2.1 & 164.5 $\pm$  2.5 & \phantom{0}12.1 $\pm$  2.3 \\
275 & -1.0 $\pm$ 0.4 & -0.3 $\pm$ 0.4 &  2169 $\pm$    5 & 27 &  5 & \phantom{0}82.1 $\pm$ 0.6 & \phantom{0}79.6 $\pm$  4.6 & \phantom{0}13.9 $\pm$  2.6 & \phantom{0}80.2 $\pm$  4.9 & \phantom{0}20.6 $\pm$  4.6 \\
277 & -0.3 $\pm$ 0.4 & -0.8 $\pm$ 0.5 &  3690 $\pm$    6 & 25 &  9 & \phantom{0}22.3 $\pm$ 0.7 & \phantom{0}19.7 $\pm$  5.9 & \phantom{0}22.4 $\pm$  3.4 & \phantom{0}19.6 $\pm$  6.4 & \phantom{0}19.4 $\pm$  2.9 \\
307 & ... & ... &  8344 $\pm$   16 & 15 & -- & \phantom{0}95.2 $\pm$ 0.5 & 101.8 $\pm$  6.3 & \phantom{0}23.9 $\pm$ 13.3 & \phantom{0}95.5 $\pm$  5.6 & \phantom{0}22.7 $\pm$  7.2 \\
364 & ... & ... &  6474 $\pm$    1 & 25 & -- & \phantom{0}99.8 $\pm$ 0.4 & 102.6 $\pm$  1.4 & \phantom{00}6.2 $\pm$  0.8 & 101.8 $\pm$  0.8 & \phantom{00}5.9 $\pm$  1.0 \\
386 & \phantom{-}0.5 $\pm$ 0.6 & \phantom{-}0.4 $\pm$ 0.5 &  6297 $\pm$   10 & 17 & -- & \phantom{0}48.9 $\pm$ 0.9 & \phantom{0}44.9 $\pm$  3.9 & \phantom{0}14.0 $\pm$  3.2 & \phantom{0}58.1 $\pm$  4.7 & \phantom{0}17.4 $\pm$  3.9 \\
414 & ... & ... &  2521 $\pm$    0 & 25 &  7 & 107.9 $\pm$ 3.2 & 135.6 $\pm$ 12.8 & \phantom{0}36.8 $\pm$  7.8 & 141.5 $\pm$ 11.0 & \phantom{0}38.7 $\pm$  6.4 \\
436 & \phantom{-}0.3 $\pm$ 0.7 & \phantom{-}0.5 $\pm$ 0.7 &  3106 $\pm$    9 & 20 &  5 & \phantom{0}24.3 $\pm$ 1.1 & 145.1 $\pm$ 39.0 & \phantom{0}31.2 $\pm$  8.0 & 173.2 $\pm$  5.9 & \phantom{0}37.8 $\pm$ 11.5 \\
489 & -0.2 $\pm$ 0.4 & -0.9 $\pm$ 0.7 &  3392 $\pm$    5 & 25 & 10 & 104.1 $\pm$ 1.0 & 102.4 $\pm$  3.7 & \phantom{0}16.6 $\pm$  3.9 & 106.6 $\pm$  3.8 & \phantom{0}16.1 $\pm$  3.4 \\
515 & ... & ... &  3868 $\pm$    1 & 25 & 10 & \phantom{00}0.2 $\pm$ 1.1 & 159.5 $\pm$  2.3 & \phantom{0}12.1 $\pm$  3.0 & 171.8 $\pm$  5.4 & \phantom{0}21.0 $\pm$  3.1 \\
518 & ... & ... &  2712 $\pm$    0 & 30 &  5 & \phantom{0}88.5 $\pm$ 1.0 & \phantom{0}99.2 $\pm$  4.4 & \phantom{0}24.2 $\pm$  3.5 & \phantom{0}93.0 $\pm$  3.5 & \phantom{0}23.4 $\pm$  4.4 \\
580 & ... & ... &  4082 $\pm$    1 & 23 &  5 & \phantom{0}43.3 $\pm$ 0.5 & \phantom{0}40.6 $\pm$  4.4 & \phantom{0}21.4 $\pm$  3.9 & \phantom{0}36.1 $\pm$  5.1 & \phantom{0}19.8 $\pm$  3.7 \\
602 & -0.4 $\pm$ 0.2 & -1.0 $\pm$ 0.3 &  4737 $\pm$    5 & 15 & -- & \phantom{0}75.2 $\pm$ 0.8 & \phantom{0}72.9 $\pm$  3.2 & \phantom{0}11.5 $\pm$  3.0 & \phantom{0}82.4 $\pm$  4.4 & \phantom{0}10.9 $\pm$  4.7 \\
610 & \phantom{-}1.4 $\pm$ 0.8 & -1.2 $\pm$ 1.0 &  7125 $\pm$   19 & 20 & -- & \phantom{0}39.4 $\pm$ 0.3 & \phantom{0}40.2 $\pm$  3.2 & \phantom{0}16.7 $\pm$  5.0 & \phantom{0}38.7 $\pm$  3.6 & \phantom{0}15.1 $\pm$  4.5 \\
624 & \phantom{-}0.5 $\pm$ 0.4 & \phantom{-}0.5 $\pm$ 0.3 &  7255 $\pm$    7 & 25 &  6 & 116.4 $\pm$ 1.5 & \phantom{0}92.9 $\pm$  3.5 & \phantom{0}18.3 $\pm$  2.8 & \phantom{0}85.2 $\pm$  4.8 & \phantom{0}16.8 $\pm$  3.2 \\
630 & ... & ... &  1768 $\pm$    1 & 20 &  9 & 146.4 $\pm$ 0.7 & 157.6 $\pm$  8.0 & \phantom{0}35.3 $\pm$  8.0 & 170.1 $\pm$  7.6 & \phantom{0}33.3 $\pm$  6.5 \\
664 & ... & ... &  3220 $\pm$    0 & 25 & -- & 118.7 $\pm$ 0.2 & 109.4 $\pm$  2.6 & \phantom{0}13.0 $\pm$  2.0 & 111.5 $\pm$  2.9 & \phantom{0}18.7 $\pm$  2.7 \\
684 & -1.5 $\pm$ 0.3 & -0.2 $\pm$ 0.3 &  5331 $\pm$    6 & 27 &  7 & 141.2 $\pm$ 2.1 & 119.9 $\pm$  2.3 & \phantom{0}14.1 $\pm$  2.5 & 107.5 $\pm$  4.6 & \phantom{0}13.7 $\pm$  2.6 \\
714 & \phantom{-}0.4 $\pm$ 0.7 & -0.4 $\pm$ 0.6 &  7793 $\pm$   18 & 20 & -- & \phantom{0}16.7 $\pm$ 0.5 & \phantom{0}12.7 $\pm$  5.0 & \phantom{0}13.0 $\pm$  2.9 & \phantom{0}13.0 $\pm$  4.3 & \phantom{0}16.7 $\pm$  4.1 \\
715 & -0.7 $\pm$ 0.8 & \phantom{-}0.0 $\pm$ 0.7 &  1885 $\pm$    9 & 30 &  7 & \phantom{0}66.5 $\pm$ 0.4 & \phantom{0}68.7 $\pm$  3.9 & \phantom{0}17.0 $\pm$  3.4 & \phantom{0}62.4 $\pm$  5.0 & \phantom{0}21.7 $\pm$  3.6 \\
743 & -0.1 $\pm$ 0.6 & \phantom{-}1.3 $\pm$ 0.7 &  3868 $\pm$    8 & 25 &  5 & \phantom{0}85.9 $\pm$ 0.5 & \phantom{0}92.1 $\pm$  7.5 & \phantom{0}32.4 $\pm$  5.2 & \phantom{0}81.0 $\pm$  6.4 & \phantom{0}25.7 $\pm$  6.5 \\
748 & \phantom{-}0.3 $\pm$ 0.7 & -0.4 $\pm$ 0.4 &  2355 $\pm$    4 & 25 & 10 & 192.5 $\pm$ 0.5 & \phantom{0}17.8 $\pm$  6.2 & \phantom{0}24.0 $\pm$  4.8 & \phantom{0}17.6 $\pm$  4.1 & \phantom{0}28.3 $\pm$  6.3 \\
753 & -0.3 $\pm$ 0.4 & -0.5 $\pm$ 0.5 &  3175 $\pm$    4 & 25 &  7 & \phantom{0}56.1 $\pm$ 0.6 & \phantom{0}57.4 $\pm$  3.0 & \phantom{0}15.6 $\pm$  2.8 & \phantom{0}57.0 $\pm$  3.2 & \phantom{0}10.1 $\pm$  1.8 \\
764 & \phantom{-}1.0 $\pm$ 0.7 & -1.1 $\pm$ 0.8 &  7704 $\pm$   11 & 20 &  7 & 128.4 $\pm$ 0.7 & 121.3 $\pm$  4.0 & \phantom{0}16.0 $\pm$  4.8 & 136.1 $\pm$  7.8 & \phantom{0}22.5 $\pm$  4.2 \\
768 & -1.0 $\pm$ 0.5 & -0.2 $\pm$ 0.6 &  3735 $\pm$    8 & 15 &  7 & \phantom{0}42.4 $\pm$ 0.9 & \phantom{0}50.0 $\pm$ 20.6 & \phantom{0}37.6 $\pm$ 12.1 & \phantom{0}41.7 $\pm$ 13.8 & \phantom{0}41.7 $\pm$  7.4 \\
771 & -1.1 $\pm$ 0.8 & -1.8 $\pm$ 0.9 &  3762 $\pm$    9 & 20 &  7 & \phantom{0}98.7 $\pm$ 0.9 & \phantom{0}29.5 $\pm$  6.6 & \phantom{0}29.5 $\pm$  7.6 & \phantom{0}46.5 $\pm$  9.7 & \phantom{0}39.5 $\pm$ 14.3 \\
777 & \phantom{-}0.5 $\pm$ 0.4 & -1.0 $\pm$ 0.4 &  4836 $\pm$    9 & 30 & 10 & \phantom{0}36.1 $\pm$ 0.4 & \phantom{0}35.7 $\pm$  2.0 & \phantom{0}10.1 $\pm$  1.6 & \phantom{0}38.4 $\pm$  3.1 & \phantom{0}12.3 $\pm$  1.9 \\
779 & -1.8 $\pm$ 0.5 & -1.1 $\pm$ 0.6 &  5527 $\pm$    8 & 23 & -- & 121.8 $\pm$ 0.3 & 129.4 $\pm$  4.4 & \phantom{0}22.1 $\pm$  5.0 & 117.9 $\pm$  3.7 & \phantom{0}16.7 $\pm$  5.7 \\
782 & \phantom{-}0.9 $\pm$ 0.3 & \phantom{-}0.9 $\pm$ 0.3 &  7803 $\pm$   10 & 12 & -- & 152.9 $\pm$ 0.5 & 147.6 $\pm$  3.0 & \phantom{0}10.3 $\pm$  3.3 & 152.9 $\pm$  3.6 & \phantom{0}21.0 $\pm$  8.2 \\
\bottomrule
\end{tabular}
\tablefoot{
(1) CALIFA identifier.
(2) and (3) Location of the kinematic centre. The value listed is the gradient peak (GP) with respect to the optical nucleus (ON) if not, ON is used as kinematic centre (see section \ref{sec:Robust_Kinematic} for details).
(4) Systemic velocity derived from integrated the velocities in a 2.7$\arcsec$ aperture centred in the kinematic centre.
(5) Radius used to average the polar coordinates of the positions from the lines of nodes (see section \ref{sec:Robust_Kinematic} for details).
(6) Radius used to average the polar coordinates of the positions from the cero-velocity line (see section \ref{sec:Robust_Kinematic} for details).
(7) Morphological PA inferred by fitting an elipse to an isophote at radius r$\mathrm{_{max}}$ in the r-band SDSS image.
(8) Kinematic position angle at radius r for the  approaching side (see section \ref{sec:Robust_Kinematic} for details).
(9) Standard deviation of the kinematic position angle of the  approaching side (see section \ref{sec:Robust_Kinematic} for details).
(10) Kinematic position angle at radius r for the  receding side.
(11) Standard deviation of the kinematic position angle of the  receding side.
 Values in parenthesis for each row represent the errors obtained from Monte Carlo simulations.
}
\end{center}
\begin{minipage}{\textwidth}
\end{minipage}
\end{table*} 
   
\begin{table*}[!htb]
\addtocounter{table}{-1}
\tiny
\caption{continue  \ref{table_Skin} }
\begin{center}
\renewcommand{\thefootnote}{\alph{footnote}}
\begin{tabular} {c c c c c c c c c c c}
\toprule
    id  &$\Delta\alpha$  & $\Delta\delta$ & V$\mathrm{_{sys}}$  &r$\mathrm{_{max}}$  &  r$\mathrm{_{min}}$  & PA$\mathrm{_{morph}}$ (r$\mathrm{_{max}}$)    &  \multicolumn{2}{c}{PA approaching}                    & \multicolumn{2}{c}{PA receding}             \\
        &                &                &                     &                    &                      &                                               &  PA$\mathrm{_{kin}}$  & $\delta$PA$\mathrm{_{kin}}$ & PA$\mathrm{_{kin}}$  & $\delta$PA$\mathrm{_{kin}}$ \\
        &   (arcsec)     &      (arcsec)  &    (km s$^{-1}$)    &      (arcsec)      &      (arcsec)        &           ($^{\circ}$)                        &  ($^{\circ}$)         & ($^{\circ}$)                & ($^{\circ}$)         & ($^{\circ}$)                \\
    (1) &      (2)       &       (3)      &         (4)         &        (5)         &          (6)         &           (7)                                 &          (8)          &            (9)              &          (10)         &       (11)                   \\
\midrule
790 & ... & ... &  4677 $\pm$    1 & 12 &  5 & 143.6 $\pm$ 1.3 & 142.4 $\pm$  5.6 & \phantom{0}16.1 $\pm$  4.3 & 144.5 $\pm$  5.9 & \phantom{0}12.4 $\pm$  4.5 \\
791 & \phantom{-}1.2 $\pm$ 0.3 & \phantom{-}0.4 $\pm$ 0.3 &  3316 $\pm$    5 & 30 & -- & 152.3 $\pm$ 0.3 & 149.7 $\pm$  1.2 & \phantom{00}3.2 $\pm$  0.7 & 154.5 $\pm$  1.3 & \phantom{00}3.7 $\pm$  0.6 \\
804 & -2.2 $\pm$ 0.5 & -0.5 $\pm$ 0.5 &  3388 $\pm$    7 & 15 & -- & 126.8 $\pm$ 0.5 & 130.5 $\pm$  4.4 & \phantom{0}19.3 $\pm$  4.8 & 127.7 $\pm$  4.7 & \phantom{0}18.6 $\pm$  5.8 \\
810 & -0.3 $\pm$ 0.5 & -0.1 $\pm$ 0.5 &  4117 $\pm$    9 & 25 & -- & \phantom{0}15.3 $\pm$ 0.5 & \phantom{00}8.6 $\pm$  3.0 & \phantom{0}10.0 $\pm$  2.4 & \phantom{0}19.6 $\pm$  3.1 & \phantom{0}16.6 $\pm$  3.0 \\
813 & ... & ... &  3836 $\pm$    1 & 28 &  7 & 113.1 $\pm$ 8.8 & 101.1 $\pm$  5.5 & \phantom{0}33.8 $\pm$  5.3 & \phantom{0}90.0 $\pm$  4.4 & \phantom{0}32.4 $\pm$  6.1 \\
823 & ... & ... &  2838 $\pm$    1 & 25 &  7 & 159.2 $\pm$ 0.6 & 152.2 $\pm$  4.5 & \phantom{0}31.4 $\pm$  6.5 & 168.5 $\pm$  7.8 & \phantom{0}26.0 $\pm$  4.3 \\
824 & -0.4 $\pm$ 0.5 & -1.7 $\pm$ 0.6 &  4730 $\pm$   14 & 27 & -- & 159.8 $\pm$ 0.2 & 161.6 $\pm$  3.0 & \phantom{0}10.5 $\pm$  1.8 & 152.7 $\pm$  4.0 & \phantom{0}13.6 $\pm$  2.2 \\
826 & -0.4 $\pm$ 0.5 & -1.2 $\pm$ 0.5 &  5070 $\pm$    9 & 25 & -- & 127.9 $\pm$ 0.3 & 133.4 $\pm$  2.3 & \phantom{0}10.3 $\pm$  2.2 & 122.5 $\pm$  2.2 & \phantom{00}7.5 $\pm$  1.3 \\
831 & -0.3 $\pm$ 0.8 & \phantom{-}0.4 $\pm$ 0.7 &  4957 $\pm$   12 & 15 & -- & 123.4 $\pm$ 0.6 & 123.8 $\pm$  9.4 & \phantom{0}30.1 $\pm$  9.9 & 133.8 $\pm$  7.1 & \phantom{0}28.2 $\pm$  8.4 \\
834 & ... & ... &  8694 $\pm$    1 & 15 & -- & 114.2 $\pm$ 0.5 & 106.3 $\pm$  1.9 & \phantom{00}8.9 $\pm$  2.6 & \phantom{0}95.0 $\pm$  5.8 & \phantom{0}22.0 $\pm$  3.4 \\
836 & \phantom{-}0.6 $\pm$ 0.8 & \phantom{-}0.8 $\pm$ 0.8 &  2424 $\pm$    4 & 29 & -- & 148.9 $\pm$ 0.9 & 140.5 $\pm$  5.3 & \phantom{0}28.1 $\pm$  7.3 & 133.4 $\pm$  6.1 & \phantom{0}31.1 $\pm$  5.0 \\
842 & \phantom{-}0.3 $\pm$ 0.5 & -2.2 $\pm$ 0.5 &  2970 $\pm$    6 & 30 &  7 & \phantom{0}54.7 $\pm$ 0.4 & \phantom{0}66.9 $\pm$  5.6 & \phantom{0}41.6 $\pm$  4.5 & \phantom{0}70.4 $\pm$  5.3 & \phantom{0}32.0 $\pm$  4.6 \\
853 & ... & ... &  3788 $\pm$    1 & 30 & 10 & \phantom{0}53.0 $\pm$ 5.0 & \phantom{0}53.6 $\pm$  2.9 & \phantom{0}10.8 $\pm$  2.1 & \phantom{0}49.5 $\pm$  2.2 & \phantom{00}5.7 $\pm$  1.4 \\
854 & ... & ... &  8625 $\pm$    1 & 20 & -- & \phantom{0}90.9 $\pm$ 0.4 & \phantom{0}90.0 $\pm$  4.3 & \phantom{00}8.8 $\pm$  3.0 & \phantom{0}90.0 $\pm$  4.5 & \phantom{0}12.3 $\pm$  2.9 \\
856 & ... & ... &  4696 $\pm$    1 & 22 &  5 & \phantom{0}91.6 $\pm$ 0.7 & \phantom{0}90.0 $\pm$  4.3 & \phantom{0}24.7 $\pm$  7.1 & \phantom{0}93.7 $\pm$  3.7 & \phantom{0}20.2 $\pm$  3.9 \\
857 & ... & ... &  8453 $\pm$    1 & 15 & -- & \phantom{0}36.1 $\pm$ 0.4 & \phantom{0}47.8 $\pm$  3.9 & \phantom{0}12.4 $\pm$  3.5 & \phantom{0}45.0 $\pm$  0.2 & \phantom{0}11.9 $\pm$  2.8 \\
862 & ... & ... &  6704 $\pm$    1 & 25 & -- & \phantom{0}32.1 $\pm$ 0.4 & \phantom{0}30.2 $\pm$  3.4 & \phantom{0}13.2 $\pm$  2.1 & \phantom{0}31.7 $\pm$  2.2 & \phantom{0}10.1 $\pm$  1.9 \\
869 & -0.6 $\pm$ 0.5 & \phantom{-}0.9 $\pm$ 0.5 &  6157 $\pm$    7 & 22 &  7 & 121.8 $\pm$ 1.8 & 130.8 $\pm$  4.2 & \phantom{0}25.8 $\pm$  3.8 & 141.4 $\pm$  3.9 & \phantom{0}12.5 $\pm$  2.8 \\
872 & -0.2 $\pm$ 0.7 & -1.3 $\pm$ 0.7 &  3767 $\pm$   11 & 20 &  7 & \phantom{0}85.4 $\pm$ 3.0 & \phantom{0}64.2 $\pm$  6.3 & \phantom{0}21.5 $\pm$  3.8 & \phantom{0}56.3 $\pm$  6.9 & \phantom{0}23.1 $\pm$  4.3 \\
876 & -0.6 $\pm$ 0.8 & \phantom{-}1.4 $\pm$ 0.8 &  5740 $\pm$   13 & 20 &  5 & 106.9 $\pm$ 1.3 & \phantom{0}94.2 $\pm$  4.6 & \phantom{0}19.6 $\pm$  5.0 & \phantom{0}95.6 $\pm$  9.2 & \phantom{0}29.9 $\pm$  5.4 \\
886 & -1.4 $\pm$ 0.3 & -0.1 $\pm$ 0.2 &  4488 $\pm$    5 & 27 &  7 & \phantom{0}12.8 $\pm$ 0.3 & \phantom{00}4.2 $\pm$  1.8 & \phantom{00}8.0 $\pm$  1.6 & \phantom{0}20.0 $\pm$  2.2 & \phantom{00}9.0 $\pm$  2.0 \\
887 & \phantom{-}0.3 $\pm$ 0.4 & \phantom{-}0.4 $\pm$ 0.3 &  7065 $\pm$    9 & 25 &  5 & \phantom{0}25.2 $\pm$ 1.4 & \phantom{0}12.6 $\pm$  2.9 & \phantom{00}9.7 $\pm$  2.1 & \phantom{0}16.7 $\pm$  4.4 & \phantom{0}17.6 $\pm$  2.5 \\
889 & -1.2 $\pm$ 0.4 & -0.9 $\pm$ 0.4 &  4852 $\pm$    7 & 20 & 10 & \phantom{0}64.4 $\pm$ 0.7 & \phantom{0}73.4 $\pm$  3.9 & \phantom{0} 9.9 $\pm$  2.1 & \phantom{0}58.6 $\pm$  2.9 & \phantom{0}14.8 $\pm$  3.1 \\
890 & \phantom{-}0.4 $\pm$ 0.5 & -0.1 $\pm$ 0.4 &  6541 $\pm$   10 & 15 & -- & 142.6 $\pm$ 0.5 & 157.5 $\pm$  4.2 & \phantom{0}11.0 $\pm$  2.6 & 162.8 $\pm$  5.4 & \phantom{0}14.1 $\pm$  2.8 \\
896 & \phantom{-}0.3 $\pm$ 0.5 & \phantom{-}1.7 $\pm$ 0.6 &  7439 $\pm$   11 & 18 & -- & \phantom{0}19.6 $\pm$ 0.5 & \phantom{0}18.7 $\pm$  6.9 & \phantom{0}20.0 $\pm$  5.5 & \phantom{0}29.6 $\pm$  5.6 & \phantom{0}23.0 $\pm$  4.6 \\
904 & -0.5 $\pm$ 0.5 & -0.0 $\pm$ 0.5 &  4913 $\pm$   13 & 30 &  6 & 148.2 $\pm$ 1.1 & 149.8 $\pm$  2.5 & \phantom{0}14.2 $\pm$  2.6 & 145.1 $\pm$  2.6 & \phantom{0}12.7 $\pm$  2.6 \\
924 & -1.0 $\pm$ 0.0 & -1.0 $\pm$ 0.0 &  2571 $\pm$    1 & 22 &  7 & \phantom{0}24.6 $\pm$ 0.8 & \phantom{0}37.4 $\pm$  1.8 & \phantom{0}12.0 $\pm$  1.7 & \phantom{0}51.4 $\pm$  3.2 & \phantom{0}14.7 $\pm$  3.7 \\
929 & \phantom{-}1.4 $\pm$ 0.8 & \phantom{-}0.1 $\pm$ 0.7 &  7977 $\pm$   17 & 20 & -- & \phantom{0}43.4 $\pm$ 0.8 & \phantom{0}54.9 $\pm$  4.6 & \phantom{0}16.8 $\pm$  3.6 & \phantom{0}58.2 $\pm$  4.0 & \phantom{0}15.5 $\pm$  3.6 \\
938 & \phantom{-}1.0 $\pm$ 0.6 & \phantom{-}1.7 $\pm$ 0.5 &  5896 $\pm$    4 & 15 &  7 & \phantom{0}51.1 $\pm$ 5.9 & \phantom{0}75.3 $\pm$ 10.7 & \phantom{0}34.0 $\pm$  6.3 & \phantom{0}73.2 $\pm$  8.1 & \phantom{0}26.1 $\pm$  5.4 \\
\bottomrule
\end{tabular}
\end{center}
\begin{minipage}{\textwidth}
\end{minipage}
\end{table*}

%% file: tabletex_Gkin_prop.txt
\begin{table*}[!htb]
\tiny
\caption{\label{table_Gkin}
Ionized gas kinematic properties of the non-interacting sample selected for this study included in the CALIFA survey.
}
\begin{center}
\renewcommand{\thefootnote}{\alph{footnote}}
\begin{tabular} {c c c c c c c c c c c}
\toprule
    id  &$\Delta\alpha$  & $\Delta\delta$ & V$\mathrm{_{sys}}$  &r$\mathrm{_{max}}$  &  r$\mathrm{_{min}}$  & PA$\mathrm{_{morph}}$ (r$\mathrm{_{max}}$)    &  \multicolumn{2}{c}{PA approaching}                    & \multicolumn{2}{c}{PA receding}                                  \\
        &                &                &                     &                    &                      &                                               &  PA$\mathrm{_{kin}}$  & $\delta$PA$\mathrm{_{kin}}$ & PA$\mathrm{_{kin}}$  & $\delta$PA$\mathrm{_{kin}}$ \\
        &   (arcsec)     &      (arcsec)  &    (km s$^{-1}$)    &      (arcsec)      &      (arcsec)        &           ($^{\circ}$)                        &  ($^{\circ}$)         & ($^{\circ}$)                & ($^{\circ}$)         & ($^{\circ}$)                \\
    (1) &      (2)       &       (3)      &         (4)         &        (5)         &          (6)         &           (7)                                 &          (8)          &            (9)              &          (10)         &       (11)                   \\
\midrule
  1 & ... & ... &  4990 $\pm$    2 & 22 & ... & \phantom{00}3.7 $\pm$ 0.4 & \phantom{00}4.5 $\pm$  1.2 & \phantom{00}3.1 $\pm$  1.0 & \phantom{00}3.5 $\pm$  1.6 & \phantom{00}3.5 $\pm$  1.3 \\
  2 & \phantom{-}0.2 $\pm$ 0.1 & -2.3 $\pm$ 0.1 &  7247 $\pm$    1 & 22 & ... & \phantom{0}45.2 $\pm$ 0.9 & \phantom{0}55.0 $\pm$  1.4 & \phantom{00}6.1 $\pm$  0.7 & \phantom{0}44.6 $\pm$  1.0 & \phantom{00}6.5 $\pm$  1.5 \\
  7 & -1.4 $\pm$ 0.3 & -0.5 $\pm$ 0.4 &  6299 $\pm$   18 & 20 & ... & \phantom{0}17.4 $\pm$ 0.2 & \phantom{0}27.8 $\pm$  2.7 & \phantom{00}8.6 $\pm$  2.1 & \phantom{0}10.3 $\pm$  2.4 & \phantom{00}6.8 $\pm$  1.7 \\
  8 & \phantom{-}0.1 $\pm$ 0.1 & \phantom{-}0.9 $\pm$ 0.1 &  4534 $\pm$    0 & 20 &  7 & 106.0 $\pm$ 1.1 & 116.2 $\pm$  2.1 & \phantom{00}7.4 $\pm$  1.0 & 115.7 $\pm$  1.7 & \phantom{00}7.3 $\pm$  1.0 \\
 10 & \phantom{-}0.1 $\pm$ 0.2 & -0.7 $\pm$ 0.3 &  5974 $\pm$    8 & 25 & 10 & 179.3 $\pm$ 0.6 & \phantom{0}16.4 $\pm$  1.4 & \phantom{00}8.3 $\pm$  1.5 & \phantom{0}18.9 $\pm$  2.3 & \phantom{0}13.9 $\pm$  2.5 \\
 23 & -0.5 $\pm$ 0.6 & -0.9 $\pm$ 0.6 &  3859 $\pm$    6 & 20 & 10 & 126.0 $\pm$ 0.7 & \phantom{0}69.9 $\pm$ 16.1 & \phantom{0}91.4 $\pm$  9.2 & 108.2 $\pm$ 33.0 & \phantom{0}91.4 $\pm$ 21.9 \\
 25 & -1.8 $\pm$ 0.5 & -1.2 $\pm$ 0.6 &  5239 $\pm$   10 & 30 &  7 & 167.1 $\pm$ 1.0 & 174.6 $\pm$  2.4 & \phantom{0}14.5 $\pm$  3.2 & 165.6 $\pm$  1.8 & \phantom{0}19.8 $\pm$  2.7 \\
 28 & -0.4 $\pm$ 0.2 & \phantom{-}0.0 $\pm$ 0.3 &  4486 $\pm$    4 & 12 &  5 & \phantom{0}60.4 $\pm$ 0.8 & \phantom{0}56.2 $\pm$  2.3 & \phantom{0}11.0 $\pm$  2.6 & \phantom{0}47.3 $\pm$  2.4 & \phantom{00}8.6 $\pm$  2.5 \\
 30 & \phantom{-}0.1 $\pm$ 0.1 & \phantom{-}1.4 $\pm$ 0.1 &  4111 $\pm$    2 & 20 &  7 & 177.3 $\pm$ 0.8 & \phantom{00}5.7 $\pm$  1.9 & \phantom{00}6.8 $\pm$  1.7 & 179.6 $\pm$  0.6 & \phantom{00}6.9 $\pm$  3.0 \\
 31 & -0.8 $\pm$ 0.2 & \phantom{-}1.5 $\pm$ 0.2 &  4438 $\pm$    2 & 30 &  5 & \phantom{0}82.2 $\pm$ 4.9 & \phantom{0}65.4 $\pm$  1.1 & \phantom{00}7.2 $\pm$  1.3 & \phantom{0}75.9 $\pm$  1.9 & \phantom{0}10.5 $\pm$  2.2 \\
 33 & -0.2 $\pm$ 0.2 & \phantom{-}0.5 $\pm$ 0.2 &  5213 $\pm$    6 & 20 &  7 & \phantom{0}93.7 $\pm$ 1.5 & \phantom{0}91.5 $\pm$  2.1 & \phantom{00}7.5 $\pm$  1.1 & \phantom{0}87.7 $\pm$  1.2 & \phantom{00}6.9 $\pm$  1.3 \\
 41 & ... & ... &  5547 $\pm$    0 & 15 & ... & \phantom{0}53.9 $\pm$ 0.2 & \phantom{0}45.9 $\pm$  2.2 & \phantom{00}9.6 $\pm$  2.4 & \phantom{0}60.6 $\pm$  2.2 & \phantom{00}6.1 $\pm$  1.3 \\
 43 & -0.1 $\pm$ 0.3 & \phantom{-}1.0 $\pm$ 0.4 &  4811 $\pm$    7 & 12 &  7 & 156.4 $\pm$ 1.0 & \phantom{00}7.9 $\pm$  4.0 & \phantom{0}12.3 $\pm$  3.8 & \phantom{0}16.1 $\pm$  4.5 & \phantom{0}11.5 $\pm$  4.3 \\
 53 & -1.2 $\pm$ 0.3 & -1.8 $\pm$ 0.3 &  6356 $\pm$    4 & 20 & ... & 152.2 $\pm$ 0.5 & 148.0 $\pm$  1.4 & \phantom{00}4.8 $\pm$  1.0 & 151.7 $\pm$  1.6 & \phantom{00}5.8 $\pm$  1.2 \\
108 & \phantom{-}0.1 $\pm$ 0.2 & -0.1 $\pm$ 0.3 &  5235 $\pm$    2 & 20 &  5 & \phantom{0}96.5 $\pm$ 1.2 & 100.6 $\pm$  2.2 & \phantom{00}8.5 $\pm$  2.0 & 100.4 $\pm$  2.3 & \phantom{00}9.0 $\pm$  2.4 \\
116 & -0.7 $\pm$ 0.4 & \phantom{-}1.4 $\pm$ 0.4 &  7696 $\pm$   12 & 10 & ... & 172.4 $\pm$ 1.4 & 159.9 $\pm$  4.9 & \phantom{00}8.3 $\pm$  2.7 & 167.3 $\pm$  7.0 & \phantom{0}12.3 $\pm$  5.0 \\
131 & -0.1 $\pm$ 0.5 & -0.4 $\pm$ 0.5 &  3693 $\pm$    7 & 18 & ... & 128.0 $\pm$ 0.4 & 130.3 $\pm$  3.5 & \phantom{00}8.2 $\pm$  2.5 & 132.9 $\pm$  2.9 & \phantom{00}7.0 $\pm$  1.9 \\
147 & \phantom{-}0.1 $\pm$ 0.1 & \phantom{-}2.2 $\pm$ 0.2 &  3568 $\pm$    2 & 25 &  7 & 141.7 $\pm$ 1.9 & 117.7 $\pm$  1.4 & \phantom{00}8.1 $\pm$  1.4 & 116.4 $\pm$  4.3 & \phantom{0}14.3 $\pm$  2.6 \\
149 & \phantom{-}0.6 $\pm$ 0.2 & -0.6 $\pm$ 0.2 &  4359 $\pm$   10 & 20 &  5 & \phantom{00}3.9 $\pm$ 0.8 & \phantom{00}5.3 $\pm$  1.7 & \phantom{00}4.7 $\pm$  0.9 & \phantom{0}13.8 $\pm$  1.7 & \phantom{00}5.9 $\pm$  1.5 \\
151 & \phantom{-}0.7 $\pm$ 0.3 & -0.3 $\pm$ 0.2 &  4574 $\pm$   18 & 29 & ... & \phantom{0}37.0 $\pm$ 0.3 & \phantom{0}36.6 $\pm$  1.3 & \phantom{00}4.3 $\pm$  0.8 & \phantom{0}37.3 $\pm$  3.0 & \phantom{0}10.8 $\pm$  3.7 \\
152 & -0.1 $\pm$ 0.3 & -1.1 $\pm$ 0.3 &  3916 $\pm$    4 & 20 & ... & 121.9 $\pm$ 0.5 & 122.8 $\pm$  1.9 & \phantom{00}8.2 $\pm$  2.2 & 113.9 $\pm$  2.2 & \phantom{00}8.1 $\pm$  1.8 \\
153 & -0.6 $\pm$ 0.4 & -0.7 $\pm$ 0.3 &  8081 $\pm$   10 & 20 & ... & 134.9 $\pm$ 0.3 & 135.7 $\pm$  2.1 & \phantom{00}7.3 $\pm$  1.8 & 134.0 $\pm$  1.9 & \phantom{00}4.8 $\pm$  1.1 \\
273 & \phantom{-}2.4 $\pm$ 0.9 & -1.9 $\pm$ 0.6 &  4270 $\pm$    5 & 33 & ... & 162.9 $\pm$ 0.2 & 150.0 $\pm$  2.9 & \phantom{0}10.7 $\pm$  2.1 & 174.1 $\pm$  2.6 & \phantom{00}6.2 $\pm$  1.5 \\
275 & \phantom{-}1.2 $\pm$ 0.2 & \phantom{-}0.6 $\pm$ 0.1 &  2086 $\pm$    6 & 27 &  5 & \phantom{0}82.1 $\pm$ 0.6 & \phantom{0}83.2 $\pm$  0.9 & \phantom{00}3.9 $\pm$  0.8 & \phantom{0}82.3 $\pm$  1.3 & \phantom{00}8.1 $\pm$  1.9 \\
277 & \phantom{-}0.9 $\pm$ 0.4 & \phantom{-}0.4 $\pm$ 0.6 &  3649 $\pm$    6 & 25 &  9 & \phantom{0}22.3 $\pm$ 0.7 & \phantom{0}12.8 $\pm$  2.5 & \phantom{0}16.1 $\pm$  2.8 & \phantom{0}32.5 $\pm$  4.0 & \phantom{0}26.6 $\pm$  3.1 \\
307 & -0.2 $\pm$ 0.3 & \phantom{-}0.2 $\pm$ 0.3 &  8439 $\pm$   10 & 15 & ... & \phantom{0}95.2 $\pm$ 0.5 & \phantom{0}97.2 $\pm$  4.3 & \phantom{00}9.1 $\pm$  3.4 & \phantom{0}92.0 $\pm$  3.2 & \phantom{00}6.3 $\pm$  2.5 \\
364 & \phantom{-}0.2 $\pm$ 0.6 & \phantom{-}1.1 $\pm$ 0.6 &  6582 $\pm$   22 & 25 & ... & \phantom{0}99.8 $\pm$ 0.4 & 108.6 $\pm$  4.3 & \phantom{0}14.3 $\pm$  2.7 & \phantom{0}98.6 $\pm$  2.8 & \phantom{00}6.4 $\pm$  1.7 \\
386 & -0.7 $\pm$ 0.5 & -0.1 $\pm$ 0.6 &  6332 $\pm$   10 & 17 & ... & \phantom{0}48.9 $\pm$ 0.9 & \phantom{0}46.6 $\pm$  3.4 & \phantom{00}8.8 $\pm$  2.6 & \phantom{0}44.2 $\pm$  2.9 & \phantom{0}10.6 $\pm$  2.6 \\
414 & -0.4 $\pm$ 0.3 & \phantom{-}0.1 $\pm$ 0.3 &  2497 $\pm$    3 & 25 &  7 & 107.9 $\pm$ 3.2 & 143.4 $\pm$  2.3 & \phantom{0}13.8 $\pm$  2.6 & 150.5 $\pm$  2.8 & \phantom{0}17.1 $\pm$  3.7 \\
436 & \phantom{-}0.1 $\pm$ 0.3 & -1.5 $\pm$ 0.3 &  3158 $\pm$    6 & 20 &  5 & \phantom{0}24.3 $\pm$ 1.1 & 171.5 $\pm$  3.1 & \phantom{0}12.1 $\pm$  4.7 & 173.7 $\pm$  2.7 & \phantom{00}7.7 $\pm$  1.9 \\
489 & \phantom{-}0.8 $\pm$ 0.1 & -0.6 $\pm$ 0.2 &  3409 $\pm$    1 & 25 & 10 & 104.1 $\pm$ 1.0 & 100.8 $\pm$  1.5 & \phantom{00}8.8 $\pm$  1.3 & 104.2 $\pm$  2.0 & \phantom{00}8.1 $\pm$  1.2 \\
515 & \phantom{-}0.1 $\pm$ 0.4 & \phantom{-}0.1 $\pm$ 0.3 &  3866 $\pm$    7 & 25 & 10 & \phantom{00}0.2 $\pm$ 1.1 & 166.6 $\pm$  3.3 & \phantom{0}10.5 $\pm$  1.9 & 165.4 $\pm$  3.2 & \phantom{0}10.1 $\pm$  1.8 \\
518 & -0.6 $\pm$ 0.5 & -0.5 $\pm$ 0.2 &  2690 $\pm$   11 & 30 &  5 & \phantom{0}88.5 $\pm$ 1.0 & \phantom{0}98.1 $\pm$  2.9 & \phantom{0}13.7 $\pm$  1.9 & \phantom{0}97.3 $\pm$  2.5 & \phantom{0}14.2 $\pm$  3.1 \\
580 & \phantom{-}0.0 $\pm$ 0.2 & -0.0 $\pm$ 0.2 &  4063 $\pm$    2 & 23 &  5 & \phantom{0}43.3 $\pm$ 0.5 & \phantom{0}40.1 $\pm$  1.5 & \phantom{00}6.6 $\pm$  1.6 & \phantom{0}35.5 $\pm$  1.6 & \phantom{00}7.8 $\pm$  1.2 \\
602 & ... & ... &  4799 $\pm$   12 & 15 & ... & \phantom{0}75.2 $\pm$ 0.8 & \phantom{0}67.0 $\pm$  3.7 & \phantom{0}23.9 $\pm$  6.9 & \phantom{0}59.3 $\pm$  6.5 & \phantom{0}19.8 $\pm$  4.8 \\
610 & \phantom{-}0.5 $\pm$ 0.4 & \phantom{-}1.1 $\pm$ 0.5 &  7223 $\pm$   12 & 20 & ... & \phantom{0}39.4 $\pm$ 0.3 & \phantom{0}39.9 $\pm$  2.0 & \phantom{00}5.6 $\pm$  1.3 & \phantom{0}47.0 $\pm$  2.6 & \phantom{00}7.1 $\pm$  1.5 \\
624 & \phantom{-}0.9 $\pm$ 0.6 & -0.3 $\pm$ 0.4 &  7309 $\pm$    7 & 25 &  6 & 116.4 $\pm$ 1.5 & \phantom{0}87.4 $\pm$  4.6 & \phantom{0}25.9 $\pm$  5.1 & \phantom{0}78.1 $\pm$  7.9 & \phantom{0}38.4 $\pm$  4.2 \\
630 & -0.1 $\pm$ 0.4 & \phantom{-}0.0 $\pm$ 0.5 &  1727 $\pm$    7 & 20 &  9 & 146.4 $\pm$ 0.7 & 159.5 $\pm$  4.0 & \phantom{0}18.5 $\pm$  4.4 & 163.9 $\pm$  3.7 & \phantom{0}18.1 $\pm$  4.3 \\
664 & ... & ... &  3204 $\pm$    5 & 25 & ... & 118.7 $\pm$ 0.2 & 113.3 $\pm$  3.2 & \phantom{0}19.8 $\pm$  3.3 & 103.6 $\pm$  3.2 & \phantom{0}13.6 $\pm$  3.5 \\
684 & \phantom{-}0.4 $\pm$ 0.4 & \phantom{-}1.2 $\pm$ 0.4 &  5439 $\pm$   10 & 27 &  7 & 141.2 $\pm$ 2.1 & 109.5 $\pm$  3.8 & \phantom{0}15.3 $\pm$  4.5 & 111.3 $\pm$  4.2 & \phantom{0}15.0 $\pm$  2.0 \\
714 & -0.2 $\pm$ 0.2 & \phantom{-}0.4 $\pm$ 0.2 &  7831 $\pm$   10 & 20 & ... & \phantom{0}16.7 $\pm$ 0.5 & \phantom{0}12.1 $\pm$  2.4 & \phantom{00}5.6 $\pm$  1.3 & \phantom{0}13.5 $\pm$  1.9 & \phantom{00}4.3 $\pm$  1.0 \\
715 & -1.0 $\pm$ 0.1 & \phantom{-}0.7 $\pm$ 0.0 &  1884 $\pm$    1 & 30 &  7 & \phantom{0}66.5 $\pm$ 0.4 & \phantom{0}64.9 $\pm$  0.5 & \phantom{00}4.3 $\pm$  0.6 & \phantom{0}66.7 $\pm$  0.8 & \phantom{00}7.3 $\pm$  1.6 \\
743 & \phantom{-}0.3 $\pm$ 0.2 & -0.3 $\pm$ 0.1 &  3865 $\pm$    5 & 25 &  5 & \phantom{0}85.9 $\pm$ 0.5 & \phantom{0}83.0 $\pm$  1.6 & \phantom{00}6.0 $\pm$  0.9 & \phantom{0}84.9 $\pm$  1.3 & \phantom{00}8.5 $\pm$  1.0 \\
748 & -0.9 $\pm$ 0.1 & -2.7 $\pm$ 0.1 &  2350 $\pm$    2 & 25 & 10 & 192.5 $\pm$ 0.5 & \phantom{0}10.7 $\pm$  0.7 & \phantom{00}6.1 $\pm$  1.0 & \phantom{0}21.4 $\pm$  1.4 & \phantom{0}10.3 $\pm$  1.4 \\
753 & \phantom{-}0.8 $\pm$ 0.1 & -0.6 $\pm$ 0.1 &  3222 $\pm$    4 & 25 &  7 & \phantom{0}56.1 $\pm$ 0.6 & \phantom{0}58.4 $\pm$  0.7 & \phantom{00}8.4 $\pm$  0.7 & \phantom{0}55.7 $\pm$  0.6 & \phantom{00}4.0 $\pm$  0.4 \\
764 & -0.9 $\pm$ 0.3 & \phantom{-}0.7 $\pm$ 0.2 &  7726 $\pm$   12 & 20 &  7 & 128.4 $\pm$ 0.7 & 129.4 $\pm$  2.8 & \phantom{0}10.2 $\pm$  2.4 & 128.4 $\pm$  2.3 & \phantom{0}12.4 $\pm$  3.9 \\
768 & \phantom{-}0.4 $\pm$ 0.2 & -0.0 $\pm$ 0.2 &  3765 $\pm$    3 & 15 &  7 & \phantom{0}42.4 $\pm$ 0.9 & \phantom{0}36.8 $\pm$  1.5 & \phantom{00}8.5 $\pm$  1.8 & \phantom{0}53.7 $\pm$  1.8 & \phantom{00}7.7 $\pm$  1.8 \\
771 & \phantom{-}2.2 $\pm$ 0.3 & -0.5 $\pm$ 0.4 &  3741 $\pm$    9 & 20 &  7 & \phantom{0}98.7 $\pm$ 0.9 & \phantom{0}40.7 $\pm$  3.9 & \phantom{0}15.8 $\pm$  2.7 & \phantom{0}44.3 $\pm$  2.8 & \phantom{0}11.5 $\pm$  2.9 \\
777 & -0.5 $\pm$ 0.2 & \phantom{-}0.6 $\pm$ 0.2 &  4877 $\pm$    9 & 30 & 10 & \phantom{0}36.1 $\pm$ 0.4 & \phantom{0}37.4 $\pm$  1.0 & \phantom{00}4.7 $\pm$  0.9 & \phantom{0}36.4 $\pm$  1.2 & \phantom{00}6.0 $\pm$  1.0 \\
779 & -0.1 $\pm$ 0.3 & -1.1 $\pm$ 0.3 &  5559 $\pm$    5 & 23 & ... & 121.8 $\pm$ 0.3 & 124.9 $\pm$  2.1 & \phantom{00}6.5 $\pm$  1.3 & 122.5 $\pm$  2.0 & \phantom{00}8.4 $\pm$  1.7 \\
782 & \phantom{-}0.3 $\pm$ 0.5 & \phantom{-}0.5 $\pm$ 0.5 &  7878 $\pm$   29 & 12 & ... & 152.9 $\pm$ 0.5 & 153.0 $\pm$  3.6 & \phantom{0}10.4 $\pm$  2.9 & 149.0 $\pm$  4.6 & \phantom{0}11.9 $\pm$  3.1 \\
\bottomrule
\end{tabular}
\tablefoot{
(1) CALIFA identifier.
(2) and (3) Location of the kinematic centre. The value listed is the gradient peak (GP) with respect to the optical nucleus (ON) if not, ON is used as kinematic centre (see section \ref{sec:Robust_Kinematic} for details).
(4) systemic velocity derived from integrated the velocities in a 2.7$\arcsec$ aperture centred in the kinematic centre.
(5) radius used to average the polar coordinates of the positions from the lines of nodes (see section \ref{sec:Robust_Kinematic} for details).
(6) radius used to average the polar coordinates of the positions from the cero-velocity line (see section \ref{sec:Robust_Kinematic} for details).
(7) morphological PA inferred by fitting an elipse to an isophote at radius r$\mathrm{_{max}}$ in the r-band SDSS image.
(8) kinematic position angle at radius r for the  approaching side (see section \ref{sec:Robust_Kinematic} for details).
(9) standard deviation of the kinematic position angle of the  approaching side (see section \ref{sec:Robust_Kinematic} for details).
(10) kinematic position angle at radius r for the  receding side.
(11) standard deviation of the kinematic position angle of the  receding side.
 Values in parenthesis for each row represent the errors obtained from Monte Carlo simulations.
}
\end{center}
\begin{minipage}{\textwidth}
\end{minipage}
\end{table*} 
   
\begin{table*}[!htb]
\addtocounter{table}{-1}
\tiny
\caption{continue \ref{table_Gkin} }
\begin{center}
\renewcommand{\thefootnote}{\alph{footnote}}
\begin{tabular} {c c c c c c c c c c c}
\toprule
    id  &$\Delta\alpha$  & $\Delta\delta$ & V$\mathrm{_{sys}}$  &r$\mathrm{_{max}}$  &  r$\mathrm{_{min}}$  & PA$\mathrm{_{morph}}$ (r$\mathrm{_{max}}$)    &  \multicolumn{2}{c}{PA approaching}                    & \multicolumn{2}{c}{PA receding}             \\
        &                &                &                     &                    &                      &                                               &  PA$\mathrm{_{kin}}$  & $\delta$PA$\mathrm{_{kin}}$ & PA$\mathrm{_{kin}}$  & $\delta$PA$\mathrm{_{kin}}$ \\
        &   (arcsec)     &      (arcsec)  &    (km s$^{-1}$)    &      (arcsec)      &      (arcsec)        &           ($^{\circ}$)                        &  ($^{\circ}$)         & ($^{\circ}$)                & ($^{\circ}$)         & ($^{\circ}$)                \\
    (1) &      (2)       &       (3)      &         (4)         &        (5)         &          (6)         &           (7)                                 &          (8)          &            (9)              &          (10)         &       (11)                   \\
\midrule
790 & \phantom{-}1.8 $\pm$ 0.3 & \phantom{-}0.9 $\pm$ 0.3 &  4717 $\pm$    6 & 12 &  5 & 143.6 $\pm$ 1.3 & 145.5 $\pm$  3.7 & \phantom{0}11.1 $\pm$  1.9 & 160.7 $\pm$  5.9 & \phantom{0}12.7 $\pm$  3.3 \\
791 & -0.5 $\pm$ 0.2 & -0.3 $\pm$ 0.2 &  3276 $\pm$    4 & 30 & ... & 152.3 $\pm$ 0.3 & 154.1 $\pm$  1.1 & \phantom{00}3.8 $\pm$  0.7 & 148.9 $\pm$  1.0 & \phantom{00}3.2 $\pm$  0.4 \\
804 & -3.1 $\pm$ 0.4 & -2.4 $\pm$ 0.5 &  3486 $\pm$    7 & 15 & ... & 126.8 $\pm$ 0.5 & 124.7 $\pm$  4.7 & \phantom{0}25.2 $\pm$  6.7 & 109.5 $\pm$ 15.9 & \phantom{0}71.6 $\pm$ 12.6 \\
810 & \phantom{-}0.7 $\pm$ 0.1 & \phantom{-}1.1 $\pm$ 0.1 &  4093 $\pm$    1 & 25 & ... & \phantom{0}15.3 $\pm$ 0.5 & \phantom{0}15.4 $\pm$  0.7 & \phantom{00}4.1 $\pm$  0.8 & \phantom{0}15.1 $\pm$  0.6 & \phantom{00}5.7 $\pm$  0.8 \\
813 & -0.6 $\pm$ 0.1 & \phantom{-}0.0 $\pm$ 0.2 &  3818 $\pm$    1 & 28 &  7 & 113.1 $\pm$ 8.8 & 101.9 $\pm$  2.0 & \phantom{0}15.8 $\pm$  2.7 & \phantom{0}90.5 $\pm$  3.0 & \phantom{0}13.8 $\pm$  3.8 \\
823 & -1.0 $\pm$ 0.2 & \phantom{-}1.9 $\pm$ 0.3 &  2825 $\pm$    2 & 25 &  7 & 159.2 $\pm$ 0.6 & 149.6 $\pm$  2.1 & \phantom{00}8.3 $\pm$  1.4 & 158.6 $\pm$  2.1 & \phantom{0}10.7 $\pm$  2.0 \\
824 & -0.4 $\pm$ 0.1 & -0.6 $\pm$ 0.2 &  4732 $\pm$    3 & 27 & ... & 159.8 $\pm$ 0.2 & 159.3 $\pm$  1.0 & \phantom{00}5.4 $\pm$  1.2 & 161.7 $\pm$  1.2 & \phantom{00}5.5 $\pm$  1.1 \\
826 & ... & ... &  5014 $\pm$    2 & 25 & ... & 127.9 $\pm$ 0.3 & 123.9 $\pm$  1.8 & \phantom{0}12.4 $\pm$  2.2 & 129.9 $\pm$  2.2 & \phantom{0}17.9 $\pm$  3.3 \\
831 & \phantom{-}0.2 $\pm$ 0.1 & \phantom{-}0.5 $\pm$ 0.2 &  4968 $\pm$    4 & 15 & ... & 123.4 $\pm$ 0.6 & 124.2 $\pm$  1.5 & \phantom{00}6.6 $\pm$  1.2 & 127.1 $\pm$  1.2 & \phantom{00}5.3 $\pm$  0.9 \\
834 & -0.4 $\pm$ 0.5 & \phantom{-}1.3 $\pm$ 0.5 &  8764 $\pm$   14 & 15 & ... & 114.2 $\pm$ 0.5 & 101.9 $\pm$  3.7 & \phantom{0}11.8 $\pm$  4.3 & 106.8 $\pm$  3.7 & \phantom{00}7.5 $\pm$  2.1 \\
836 & \phantom{-}1.1 $\pm$ 0.2 & \phantom{-}1.9 $\pm$ 0.2 &  2407 $\pm$    0 & 29 & ... & 148.9 $\pm$ 0.9 & 120.9 $\pm$  1.7 & \phantom{0}10.9 $\pm$  1.9 & 135.6 $\pm$  2.2 & \phantom{0}17.5 $\pm$  2.6 \\
842 & \phantom{-}0.0 $\pm$ 0.3 & -0.2 $\pm$ 0.3 &  2955 $\pm$    6 & 30 &  7 & \phantom{0}54.7 $\pm$ 0.4 & \phantom{0}56.8 $\pm$  2.4 & \phantom{00}9.6 $\pm$  1.5 & \phantom{0}63.2 $\pm$  2.6 & \phantom{0}15.6 $\pm$  1.8 \\
853 & \phantom{-}0.8 $\pm$ 0.2 & -1.0 $\pm$ 0.2 &  3809 $\pm$    3 & 30 & 10 & \phantom{0}53.0 $\pm$ 5.0 & \phantom{0}50.3 $\pm$  0.9 & \phantom{00}2.9 $\pm$  0.3 & \phantom{0}47.3 $\pm$  1.2 & \phantom{00}3.3 $\pm$  0.4 \\
854 & \phantom{-}0.9 $\pm$ 0.4 & \phantom{-}0.2 $\pm$ 0.2 &  8654 $\pm$   16 & 20 & ... & \phantom{0}90.9 $\pm$ 0.4 & \phantom{0}87.4 $\pm$  2.3 & \phantom{0}10.1 $\pm$  2.8 & \phantom{0}86.2 $\pm$  3.1 & \phantom{00}9.7 $\pm$  2.9 \\
856 & \phantom{-}0.7 $\pm$ 0.1 & \phantom{-}2.1 $\pm$ 0.2 &  4700 $\pm$    3 & 22 &  5 & \phantom{0}91.6 $\pm$ 0.7 & \phantom{0}94.3 $\pm$  2.6 & \phantom{0}10.4 $\pm$  2.5 & \phantom{0}91.8 $\pm$  2.6 & \phantom{00}6.9 $\pm$  1.8 \\
857 & -2.0 $\pm$ 0.6 & -1.5 $\pm$ 0.4 &  8539 $\pm$    8 & 15 & ... & \phantom{0}36.1 $\pm$ 0.4 & \phantom{0}60.5 $\pm$  3.7 & \phantom{0}12.6 $\pm$  5.6 & \phantom{0}38.2 $\pm$  4.3 & \phantom{0}11.7 $\pm$  3.1 \\
862 & -1.3 $\pm$ 0.2 & \phantom{-}1.0 $\pm$ 0.2 &  6720 $\pm$    4 & 25 & ... & \phantom{0}32.1 $\pm$ 0.4 & \phantom{0}27.8 $\pm$  1.2 & \phantom{00}4.9 $\pm$  1.0 & \phantom{0}37.8 $\pm$  1.7 & \phantom{00}7.3 $\pm$  0.7 \\
869 & \phantom{-}1.3 $\pm$ 0.6 & \phantom{-}3.0 $\pm$ 0.4 &  6139 $\pm$   18 & 22 &  7 & 121.8 $\pm$ 1.8 & 125.9 $\pm$  3.0 & \phantom{0}13.4 $\pm$  3.7 & 133.0 $\pm$  2.7 & \phantom{0}12.3 $\pm$  2.5 \\
872 & \phantom{-}2.0 $\pm$ 0.7 & \phantom{-}1.1 $\pm$ 0.6 &  3813 $\pm$   22 & 20 &  7 & \phantom{0}85.4 $\pm$ 3.0 & \phantom{0}64.3 $\pm$  5.3 & \phantom{0}36.8 $\pm$  3.2 & \phantom{0}45.3 $\pm$  6.7 & \phantom{0}30.3 $\pm$  5.8 \\
876 & \phantom{-}0.1 $\pm$ 0.3 & -1.2 $\pm$ 0.2 &  5743 $\pm$    4 & 20 &  5 & 106.9 $\pm$ 1.3 & \phantom{0}98.1 $\pm$  2.4 & \phantom{00}8.8 $\pm$  1.7 & \phantom{0}96.7 $\pm$  3.2 & \phantom{0}14.2 $\pm$  3.8 \\
886 & -0.3 $\pm$ 0.1 & \phantom{-}0.1 $\pm$ 0.2 &  4455 $\pm$    5 & 27 &  7 & \phantom{0}12.8 $\pm$ 0.3 & \phantom{0}10.3 $\pm$  1.2 & \phantom{00}6.0 $\pm$  1.4 & \phantom{0}17.2 $\pm$  1.1 & \phantom{00}7.0 $\pm$  1.2 \\
887 & \phantom{-}0.8 $\pm$ 0.1 & \phantom{-}0.9 $\pm$ 0.1 &  7056 $\pm$    3 & 25 &  5 & \phantom{0}25.2 $\pm$ 1.4 & \phantom{0}15.3 $\pm$  1.8 & \phantom{00}8.6 $\pm$  1.2 & \phantom{0}15.0 $\pm$  1.6 & \phantom{00}7.3 $\pm$  1.5 \\
889 & -0.5 $\pm$ 0.0 & \phantom{-}0.5 $\pm$ 0.1 &  4883 $\pm$    0 & 20 & 10 & \phantom{0}64.4 $\pm$ 0.7 & \phantom{0}61.7 $\pm$  1.2 & \phantom{00}6.3 $\pm$  0.9 & \phantom{0}70.7 $\pm$  1.2 & \phantom{00}5.4 $\pm$  1.2 \\
890 & \phantom{-}2.1 $\pm$ 0.4 & -1.2 $\pm$ 0.4 &  6533 $\pm$    5 & 15 & ... & 142.6 $\pm$ 0.5 & 144.5 $\pm$  3.0 & \phantom{0}13.9 $\pm$  4.0 & 167.8 $\pm$  4.2 & \phantom{00}9.7 $\pm$  3.8 \\
896 & \phantom{-}0.9 $\pm$ 0.2 & \phantom{-}0.0 $\pm$ 0.3 &  7461 $\pm$    6 & 18 & ... & \phantom{0}19.6 $\pm$ 0.5 & \phantom{0}18.2 $\pm$  2.7 & \phantom{00}7.7 $\pm$  1.4 & \phantom{0}28.1 $\pm$  1.8 & \phantom{00}6.8 $\pm$  1.7 \\
904 & -0.6 $\pm$ 0.1 & \phantom{-}1.5 $\pm$ 0.1 &  4892 $\pm$    0 & 30 &  6 & 148.2 $\pm$ 1.1 & 145.1 $\pm$  1.5 & \phantom{00}8.4 $\pm$  1.1 & 147.2 $\pm$  0.7 & \phantom{00}5.9 $\pm$  1.1 \\
924 & -0.2 $\pm$ 0.4 & -0.3 $\pm$ 0.4 &  2525 $\pm$   12 & 22 &  7 & \phantom{0}24.6 $\pm$ 0.8 & \phantom{0}49.3 $\pm$  3.9 & \phantom{0}13.4 $\pm$  2.9 & \phantom{0}52.0 $\pm$  2.7 & \phantom{0}13.8 $\pm$  3.7 \\
929 & -2.0 $\pm$ 0.3 & \phantom{-}1.4 $\pm$ 0.3 &  8103 $\pm$    6 & 20 & ... & \phantom{0}43.4 $\pm$ 0.8 & \phantom{0}52.4 $\pm$  1.3 & \phantom{00}7.1 $\pm$  1.5 & \phantom{0}50.7 $\pm$  1.9 & \phantom{00}7.6 $\pm$  1.8 \\
938 & \phantom{-}0.6 $\pm$ 0.2 & \phantom{-}0.7 $\pm$ 0.2 &  5913 $\pm$    3 & 15 &  7 & \phantom{0}51.1 $\pm$ 5.9 & \phantom{0}64.6 $\pm$  3.9 & \phantom{0}14.6 $\pm$  3.4 & \phantom{0}68.4 $\pm$  5.2 & \phantom{0}12.3 $\pm$  3.4 \\
\bottomrule
\end{tabular}
\end{center}
\begin{minipage}{\textwidth}
\end{minipage}
\end{table*}

%% file: tabletex_kin_prop_bar.txt
\begin{table*}[!htb]
\tiny
\caption{\label{table_kin_bar}
Kinematic orientation measured at different radius for a sample of low inclined galaxies included in our sample of non-interacting galaxies.
}
\begin{center}
\renewcommand{\thefootnote}{\alph{footnote}}
\begin{tabular} {c c c c c c c c c c c c c}
\toprule
id &                           &                            &                         &                          & \multicolumn{4}{c}{Stars}                                                              & \multicolumn{4}{c}{ Ionized gas}                                  \\
\cmidrule(r){6-9}  \cmidrule(r){10-13} \\
   &                           &                            &                         &                          & \multicolumn{2}{c}{r$\mathrm{_{in}}$}     &  \multicolumn{2}{c}{r$\mathrm{_{out}}$}    & \multicolumn{2}{c}{r$\mathrm{_{in}}$} &  \multicolumn{2}{c}{r$\mathrm{_{out}}$} \\
\cmidrule(r){6-7} \cmidrule(r){8-9}  \cmidrule(r){10-11} \cmidrule(r){12-13} \\
   & r$\mathrm{_{in}^{stars}}$ & r$\mathrm{_{out}^{stars}}$ & r$\mathrm{_{in}^{gas}}$ & r$\mathrm{_{out}^{gas}}$ & PA$\mathrm{_{app}}$ & PA$\mathrm{_{rec}}$ &  PA$\mathrm{_{app}}$ & PA$\mathrm{_{rec}}$ & PA$\mathrm{_{app}}$ & PA$\mathrm{_{rec}}$ &  PA$\mathrm{_{app}}$ & PA$\mathrm{_{rec}}$ \\
   &       (arcsec)            &           (arcsec)         &       (arcsec)          &          (arcsec)        &    (degrees)        &    (degrees)        &    (degrees)         &    (degrees)        &    (degrees)        &    (degrees)        &    (degrees)         &    (degrees)         \\
(1)&       (2)                 &           (3)              &       (4)               &          (5)             &    (6)              &    (7)              &    (8)               &    (9)              &    (10)              &    (11)             &    (12)              &    (13)              \\
\midrule
  8 & 10 & 20 & 14 & 28 & 114 & 117 & 114 & 112 & 114 & 114 & 114 & 117 \\
 10 & 12 & 25 & 12 & 27 &  27 &  15 &  12 &  12 &  18 &  13 &  15 &  17 \\
 23 & 15 & 25 & 27 & 30 &  81 &  84 & 103 &  89 &  77 & 102 &  76 &  76 \\
 25 & 22 & 30 & 22 & 30 & 170 & 166 & 155 & 171 & 170 & 166 & 168 & 155 \\
 28 & 14 & 17 & 14 & 28 &  52 &  42 &  72 &  52 &  55 &  48 &  49 &  51 \\
 30 &  7 & 20 &  7 & 30 & 183 & 162 & 189 & 168 &  12 & 162 &   6 &  91 \\
 31 & 15 & 30 & 16 & 32 &  57 &  74 &  72 &  64 &  67 &  72 &  65 &  73 \\
 33 & 10 & 20 & 16 & 32 &  93 &  90 &  88 &  95 &  91 &  86 &  90 &  95 \\
 43 &  8 & 17 &  8 & 17 &  22 &  29 &   9 &  20 &   7 &  19 &  99 &  10 \\
108 & 10 & 20 & 10 & 27 &  86 &  90 &  96 & 107 &  94 &  95 & 100 &  97 \\
147 & 10 & 20 & 10 & 30 & 117 & 116 & 112 & 125 & 117 & 115 & 117 & 115 \\
149 & 12 & 25 & 12 & 28 &  10 &  10 &   6 &   4 &   5 &  18 &   5 &  12 \\
275 & 13 & 27 & 16 & 32 &  78 &  76 &  79 &  94 &  83 &  80 &  82 &  86 \\
277 & 12 & 25 & 16 & 30 &  15 &  20 &  23 &  13 &  49 &  46 &  12 &  21 \\
414 & 14 & 25 & 14 & 30 & 137 & 136 & 147 & 144 & 133 & 137 & 146 & 152 \\
436 & 14 & 25 & 23 & 25 & 164 & 181 & 178 & 175 & 172 & 173 & 178 & 178 \\
489 & 13 & 27 & 16 & 32 & 102 & 110 &  95 & 102 &  98 & 106 & 105 &  99 \\
515 & 15 & 25 & 16 & 32 & 162 & 168 & 163 & 172 & 164 & 162 & 166 & 169 \\
518 & 17 & 30 & 17 & 30 & 105 &  93 &  90 &  98 & 104 &  97 &  95 &  98 \\
580 & 11 & 23 & 13 & 27 &  35 &  37 &  47 &  33 &  35 &  31 &  42 &  35 \\
602 & 11 & 22 &  5 & 10 &  68 &  81 &  75 &  84 &  71 &  80 &  65 &  63 \\
624 & 21 & 25 & 21 & 30 &  90 &  82 & 105 & 100 &  74 &  70 &  92 &  89 \\
630 & 12 & 22 & 12 & 32 & 160 &  97 & 155 & 158 & 167 & 163 & 159 & 157 \\
684 & 15 & 27 & 20 & 30 & 119 & 111 & 121 & 105 & 114 & 114 & 107 & 108 \\
715 & 15 & 30 & 16 & 32 &  72 &  59 &  66 &  61 &  65 &  67 &  64 &  58 \\
743 & 12 & 25 & 16 & 32 &  80 &  84 &  94 &  78 &  81 &  84 &  81 &  83 \\
748 & 12 & 25 & 16 & 30 &  12 &  20 &  18 &  19 &  12 &  24 &  13 &  22 \\
753 & 12 & 25 & 16 & 32 &  54 &  57 &  55 &  54 &  53 &  57 &  62 &  56 \\
764 &  7 & 20 &  7 & 30 & 132 & 124 & 119 & 138 & 120 & 113 & 130 & 128 \\
768 &  7 & 15 & 14 & 28 &  38 &  25 &  42 &  42 &  35 &  53 &  38 &  47 \\
771 & 10 & 20 & 21 & 30 &  21 &  33 &  35 &  38 &  39 &  41 &  48 &  33 \\
777 & 15 & 30 & 18 & 35 &  36 &  38 &  32 &  38 &  38 &  34 &  36 &  34 \\
790 &  6 & 12 & 10 & 20 & 137 & 143 & 144 & 146 & 144 & 162 & 149 & 155 \\
791 & 15 & 30 & 15 & 30 & 151 & 154 & 149 & 154 & 156 & 149 & 153 & 149 \\
813 & 16 & 28 & 16 & 35 & 106 &  88 &  93 &  88 & 101 &  83 &  99 &  95 \\
823 & 12 & 25 & 16 & 32 & 156 & 165 & 160 & 161 & 149 & 162 & 152 & 154 \\
836 & 14 & 29 & 14 & 29 & 144 & 135 & 131 & 129 & 127 & 138 & 119 & 129 \\
842 & 18 & 30 & 15 & 30 &  90 &  73 &  57 &  60 &  58 &  70 &  55 &  52 \\
853 & 15 & 30 & 15 & 30 &  53 &  46 &  56 &  50 &  49 &  45 &  50 &  48 \\
856 &  7 & 22 &  7 & 30 &  84 &  98 &  95 &  89 &  76 &  88 &  94 &  93 \\
869 & 16 & 22 & 16 & 30 & 130 & 142 & 116 & 132 & 134 & 136 & 124 & 133 \\
872 & 17 & 20 & 17 & 30 &  63 &  55 &  75 &  65 &  61 &  39 &  73 &  71 \\
876 & 10 & 20 & 15 & 25 &  94 &  79 &  96 & 110 & 102 &  89 &  92 & 101 \\
886 & 13 & 27 & 16 & 32 &   2 &  24 &   4 &  18 &  15 &  16 &   7 &  15 \\
887 & 10 & 25 & 10 & 35 &  10 &  10 &  15 &  24 &  11 &  15 &  14 &  13 \\
889 & 11 & 22 & 11 & 28 &  71 &  58 &  74 &  58 &  57 &  66 &  63 &  69 \\
904 & 11 & 30 & 11 & 30 & 143 & 138 & 152 & 146 & 142 & 143 & 145 & 148 \\
924 & 11 & 22 & 16 & 32 &  39 &  48 &  33 &  51 &  43 &  53 &  46 &  48 \\
938 & 11 & 20 & 11 & 30 &  78 &  72 &  67 &  74 &  64 &  69 &  66 &  66 \\
\bottomrule
\end{tabular}
\tablefoot{
(1) CALIFA identifier.
(2) and (3) internal and external radii used to measured the kinematic PA in the stellar component (see section \ref{sec:Impact} for details).
(4) and (5) internal and external radii used to measured the kinematic PA in the ionized gas component.
(6) and (7) internal approaching and receding kinematic position angles for the stellar component (see section \ref{sec:Impact} for details).
(8) and (9) external approaching and receding kinematic position angles for the stellar component.
(10) and (11) internal approaching and receding kinematic position angles for the ionized gas.
(12) and (13) external approaching and receding kinematic position angles for the ionized gas.
}
\end{center}
\begin{minipage}{\textwidth}
\end{minipage}
\end{table*} 